\documentclass[trackchanges]{aastex631}
\usepackage{amsmath}
\usepackage{hyperref}

\usepackage{longtable}
\usepackage{xspace}
\usepackage{cancel}

\newcommand{\DN}{$D_N$\xspace}

\newcommand{\Rs}{$R_s$\xspace}
\newcommand{\nusix}{$\nu_6$\xspace}
\newcommand{\multinest}{\texttt{multinest}\xspace}

\newcommand{\alphas}{$\alpha_{s}\xspace$}
\newcommand{\tcol}{$t_{col}$\xspace}

\shorttitle{A NEO model calibrated to Earth Impactors}
\shortauthors{Deam et al.}
\graphicspath{{./}{figures/}}

\begin{document}

\title{A Near-Earth Object Model Calibrated to Earth Impactors}

\author[0000-0003-1955-628X]{Sophie E. Deam}
\affiliation{Space Science and Technology Centre, Curtin University, GPO Box U1987, Perth WA 6845, Australia}
\affiliation{International Centre for Radio Astronomy Research, Curtin University, GPO Box U1987, Perth WA 6845, Australia}

\author[0000-0001-9226-1870]{Hadrien A. R. Devillepoix}
\affiliation{Space Science and Technology Centre, Curtin University, GPO Box U1987, Perth WA 6845, Australia}
\affiliation{International Centre for Radio Astronomy Research, Curtin University, GPO Box U1987, Perth WA 6845, Australia}

\author[0000-0002-4547-4301]{David Nesvorn\'y}
\affiliation{Solar System Science \& Exploration Division, Southwest Research Institute, 1301 Walnut St., Suite 400,  Boulder, CO 80302, USA}

\author[0000-0003-4766-2098]{Patrick M. Shober}
\affiliation{NASA Astromaterials Research and Exploration Science Division, Johnson Space Center, Houston TX 77058, USA}

\author[0000-0003-2702-673X]{Eleanor K. Sansom}
\affiliation{International Centre for Radio Astronomy Research, Curtin University, GPO Box U1987, Perth WA 6845, Australia}

\author{Jim Albers} 
\affiliation{SETI Institute, 339 Bernardo Ave, Mountain View, CA 94043, USA} 

\author{Eric Anderson}
\altaffiliation{posthumous} 
\affiliation {Astronomical Society of Victoria, GPO Box 1059, Melbourne VIC 3001 Australia}

\author[0000-0001-6285-9847]{Zouhair Benkhaldoun} 
\affiliation{Department of Applied Physics and Astronomy, and Sharjah Academy for Astronomy, Space Sciences and Technology, University of Sharjah, P.O. Box 27272 Sharjah United Arab Emirates.} 
\affiliation{High Energy Physics, Astrophysics and Geosciences Laboratory (LPHEAG), Faculty of Sciences Semlalia , Marrakesh B.P. 2390, Morocco} 
\affiliation{Oukaimeden Observatory, Cadi Ayyad University, 40273, Marrakesh, Morocco.} 

\author[0000-0001-6130-7039]{Peter G. Brown} 
\affiliation{Department of Physics and Astronomy, University of Western Ontario, London, Ontario, N6A 3K7, Canada} 
\affiliation{Institute for Earth and Space Exploration, University of Western Ontario, London, Ontario N6A 5B8, Canada} 

\author[0000-0002-7150-4092]{Luke Daly} 
\affiliation{School of Geographical and Earth Sciences, University of Glasgow, Glasgow, UK.}  
\affiliation{Australian Centre for Microscopy and Microanalysis, The University of Sydney, Sydney, New South Wales, Australia.} 
\affiliation{Department of Materials, University of Oxford, Oxford, UK.} 

\author{George DiBattista} 
\affiliation{ }

\author[0000-0001-8792-246X]{Hasnaa Chennaoui Aoudjehane}
\affiliation{GeoPen Laboratory, Faculty of Science Ain Chock, Hassan II University of Casablanca, km 8 Route d'El Jadida Morocco} 

\author[0000-0001-5210-4002]{Christopher D. K. Herd}
\affiliation{Department of Earth and Atmospheric Sciences, 1-26 Earth Sciences Building, University of Alberta, Edmonton, AB T6G 2E3, Canada} 

\author{Tom Herring} 
\affiliation{Jack C. Davis Observatory, Western Nevada College, 2269 Vanpatten Ave, Carson City, NV 89703, USA} 

\author[0000-0002-1160-7970]{Jonathan Horner} 
\affiliation{University of Southern Queensland, Centre for Astrophysics, West Street, Toowoomba, QLD 4350 Australia}

\author[0000-0003-4735-225X]{Peter Jenniskens} 
\affiliation{SETI Institute, 339 Bernardo Ave, Mountain View, CA 94041, USA} 

\author{Derek C. Poulton} 
\affiliation {Astronomical Society of Victoria, GPO Box 1059, Melbourne VIC 3001 Australia} 

\author[0000-0001-7165-2215]{Martin D. Suttle} 
\affiliation{School of Physical Sciences, The Open University, Walton Hall, Milton Keynes, MK7 6AA, UK} 

\author[0000-0002-3020-4296]{Anna Zappatini} 
\affiliation{Institute of Geological Science, University of Bern, Baltzerstrasse 1+3, 3012 Bern, Switzerland}

\correspondingauthor{Sophie E. Deam} 
\email{sophdeam@gmail.com}

\begin{abstract}

The population of Earth-impacting meteoroids and its size-dependent orbital elements are key to understanding the origin of meteorites and informing on planetary defence efforts. Outstanding questions include the role of collisions in depleting meteoroids on highly evolved orbits and the relative importance of delivery resonances. Those depend on size, with current dynamical models considering only asteroids larger than 10~m in diameter.  Based on 1,202 sporadic meteoroids observed by the Global Fireball Observatory, we created a debiased model of the near-Earth meteoroid population in the 10~g - 150~kg in size (approximately 1~cm - 0.5~m) as they dynamically evolved from the main asteroid belt onto Earth-crossing orbits. The observed impact population is best matched with a collisional half-life decreasing from 3~Myr for meteoroids of 0.6~kg (7~cm) or higher, to 1~Myr below this size, extending to the model lower bound of 10~g. Placing our results in context with near-Earth object models for larger sizes, we find that the inner main belt continues to dominate feeding the small 1~m to 10~m diameter population primarily via the \nusix secular resonance and the 3:1J mean motion resonance. We also evaluated the potential significance of physical processes other than collisions on Earth-impacting meteoroids, such as low-perihelion disruptions from thermal stresses. 

\end{abstract}

\section{Introduction} \label{sec:intro}

Earth experiences a bombardment of interplanetary material as it orbits the Sun, from a continuous
stream of dust particles $\mu$m to mm in size, through the less frequent interception of cm- to m-sized meteoroids, to relatively rare encounters with $>$\,10~m diameter asteroids \citep{Ryabova2019_meteoroidsbook}. 
As this material spans many orders of magnitude, it spans two important regimes: objects that are not easily detectable and those that are hazardous. 
Such limitation and potential risk are a motivation to model the dynamics and evolution of material in the inner Solar System.

Astronomers have been able to discover smaller and smaller near-Earth objects (NEOs) through decades of dedicated surveying.
Here we consider any object with a perihelion distance $q<$1.3~au to be an NEO, and for an overview of NEOs in the context of other small bodies in the Solar System, see \citet{Horner:2020}.
Current estimates of the total number of expected NEOs indicate that we have discovered 90-95\% of objects with an absolute magnitude of $H$ $<$ 17.75 ($\gtrsim$ 1 kilometre in diameter for an albedo of 0.14) \citep{HarrisChodas:2021, Nesvorny:2024NEOMOD2, Deienno:2025}. 
The discovery completeness estimates reduce significantly to only 26-34\% for NEOs with H $<$ 22.75 (diameter approximately $>$ 100~m).
Fortunately, new facilities and surveys such as the NEO Surveyor mission \citep{Mainzer:2023} and the Vera C. Rubin Observatory will discover many more small NEOs \citep{Kurlander:2025} and improve the discovery completeness for such objects. 
NEOs observed with traditional telescopes are at their brightest and most observable when they are very close to Earth - especially if they are on a collision course with the Earth.
Since the discovery of the first such imminent impactor 2008 TC$_{3}$ \citep{Jenniskens:2009}, there have been 11 instances of NEO discovery before impact (\citealp{Farnocchia:2016_2014AA, Jenniskens:2021_2018LA}; 2019 MO\footnote{https://minorplanetcenter.net/mpec/K19/K19M72.html}; \citealp{Geng:2023_2022EB5, Kareta:2024_2022WJ1, Egal:2025_2023CX1, Spurny:2024_2024BX1, Ingebretsen:2025_2024RW1}; 2024 UQ\footnote{https://minorplanetcenter.net/mpec/K24/K24U49.html}; \citealp{Gianotto:2025_2024XA1}) which have all ranged between 0.5~m -- 10~m in size. 
That said, such objects can remain hidden in the Sun's glare and evade detection even during very close encounters, as was seen in the case of the 2013 Chelyabinsk impact \citep{Borovicka:2013}.
This demonstrates that the metre-sized NEOs are the current lower limit for telescopic observability in asteroids.

Fortunately, NEOs become visible in the form of fireballs and meteors if they impact the Earth. 
Dedicated camera networks such as the Global Fireball Observatory (GFO) \citep{Devillepoix:2020}, European Fireball Network \citep{Borovicka:2022paper1}, FRIPON \citep{Colas:2020}, and the Global Meteor Network \citep{Vida:2021} have been recording the night sky for years to capture images of ablating interplanetary material\footnote{A detailed description and historical overview of all networks of meteor cameras worldwide is beyond the scope of this work; for such an overview, we direct the interested reader to \citet{Jenniskens:2025} and references therein.}.
These observatories can estimate the pre-impact orbit and mass of a meteoroid, 
and measure its trajectory to predict a fall location in order to recover a meteorite.
Since the impact frequency of meteoroids decreases with increasing size, these networks capture relatively few meteoroids $\geq$10~cm, instilling the need for long-duration, large sky-area surveys such as the GFO to observe a significant number of such meteoroids.
In contrast, telescopic surveys of NEOs are biased \textit{against} the smallest objects in their observable range, and therefore record fewer observations of that end of the population, overall leading to the least amount of observations of NEOs 10~cm -- 10~m in size.

This scarcely observed population (10~cm -- 10~m) is critical because it covers the region across which object become large enough to transition from being harmless spectacles to being potentially hazardous upon impact.
Most authors typically consider the dividing line between those two regimes to lie around impact diameters of tens of meters \citep{Boslough:2008, Shustov:2017, Silber:2018, Svetsov:2019}.
A specific example of an object that was large enough to be hazardous is the $\sim$19~m asteroid that fell over Chelyabinsk and created an airburst containing the energy of 500 kilotonnes of TNT \citep{Brown:2013, Popova:2013}, the shock wave from which shattered glass and injured many people. 
Impactors causing the scale of damage seen from the Chelyabinsk airburst are estimated to be relatively rare \citep{Gi:2018}.
This rarity, combined with the paucity of decameter-scale NEO observations, has led to an order-of-magnitude uncertainty in the estimated impact frequency of 10 m NEOs -- from 2–3 years up to 20–40 years -- as discussed in \citet{ChowBrown:2025}.

Understanding this population will also inform our understanding of meteorite delivery processes \citep{Borovicka:2015}. 
Meteorites tell us about the chemical makeup and formation processes of the terrestrial planets \citep{BurbineObrien:2004, Piani:2020} and the composition of the Solar System \citep{Marty:2024}.
Knowing the dynamic origin of such meteorites provides greater context to their previous passage through the Solar System (e.g. \citealp{King:2022}) to potentially link their composition and source asteroid family \citep{Binzel:2015, Jenniskens:2025}.

Much of our understanding of the evolution of NEOs larger than 10~m comes from models that trace an object's origin in the main asteroid belt and simulate their subsequent dynamical and physical evolution until they reproduce the characteristics of the observed population.
Many such models, from the innovative \citet{Bottke:2002} and \citet{Greenstreet:2012}, to the more recent models of \citet{Granvik:2018} and \citet{Nesvorny:2023} (hereafter G18 and NEOMOD respectively), the subsequent NEOMOD2 and NEOMOD3 models \citep{Nesvorny:2024NEOMOD2, Nesvorny:2024NEOMOD3}, and the most recent addition of \citet{Deienno:2025} function in a similar way to each other, which we will describe here using the same notation and language as Table~1 in G18.
Each model begins with a synthetic population of asteroids residing in the main belt or cometary reservoirs.
Asteroid fragments, created through collisions \citep{Dohnanyi:1969, Durda:1998, Bottke:2005_2ndpaper_Linking...}, evolve under non-gravitational forces such as Yarkovsky drift and YORP evolution towards potential escape routes \citep{Farinella:1998, MorbidelliGladman:1998, Bottke:2002_asteroids3}.
Each of the models mentioned above mimics this process to varying degrees; NEOMOD simply places particles within a resonance with the same orbital distribution as asteroids neighbouring the resonances, while G18 computes the drift of particles into the escape routes. 
Main belt escape routes - which we will interchangeably call source regions - are locations within the belt where gravitational forces, such as the mean motion and secular resonances with the Jupiter and Saturn respectively, perturb an asteroid’s orbit and cause it to migrate from the main belt into near-Earth space ($q<1.3$~au)\citep{Wetherill:1985, Granvik:2017}, and possibly onto Earth crossing orbits.
All of the above mentioned models integrate particles under these gravitational perturbations through time to map each escape route to a subsequent distribution of time spent on particular orbits in near-Earth space which we call residence time distributions, \Rs.
The models then combine the \Rs to match debiased NEO observations and create a model NEO population.
At this stage, the models represent only the dynamical evolution of NEOs, and subsequent differences to the observed population indicates where physical processes have modified the NEO population along its evolution.

A wealth of information about the physical processes acting on NEOs $>$10~m have been identified and refined with these models.
First of all, the main-belt sampling of escape routes was found to be size-dependent \citep{Granvik:2018}. 
Second, as initially shown by \citet{Granvik:2016}, asteroids that pass close to the Sun appear to completely disrupt.
The removal of low-$q$ particles was implemented in the G18 and NEOMOD models to produce a better match between the main-belt derived distributions and the NEO population. 
They also found that the disruption distance is size dependent. 
Extrapolation of the size-distance relation of \citet{Granvik:2016} to objects of 1~metre in diameter implies a disruption distance of $q$=0.4~au. 
It has not been directly tested whether this phenomenon also effects meteoroids of this size, though fireball observations do show a paucity at such perihelia \citep{Wiegert:2020}.
Finally, \citet{Granvik:2024} have found a NEO source from around the Earth and Venus created by tidally disrupted material.

The extrapolation of these size-dependent features or introduction of new processes to smaller meteoroid sizes is an important step to understanding the hazardous and meteorite feeding population.  
For example, NEOs that have recently evolved from the belt can still collide with main-belt asteroids while their aphelion remains inside the belt ($Q>$ 1.8~au; \citealp{Nesvorny:2024NEOMOD2}).
Collisions are therefore a dominant, strongly size-dependent driver of the main belt and near-Earth asteroid populations \citep{Bottke:2005_1stpaper_Fossilized...}.
For main-belt fragments 1 -- 10~cm in size, modern collisional-evolution models yield mean catastrophic lifetimes of $\sim$3–6 Myr (\citealp{Bottke:2005_2ndpaper_Linking..., EliaBrunini:2007}; see also \citealp{Farinella:1998}). 
Because NEOs encounter the main-belt fragments at higher relative velocities ($\approx$ 8–11 km/s instead of $\approx$ 5 km/s), their collisional lifetimes are shortened by roughly a factor of three to 1–3 Myr for the same size range.
This estimate agrees with dynamical fits to the orbital distribution of fireballs \citep{MorbidelliGladman:1998} and is broadly consistent with more recent Öpik-based calculations which find 5 -- 10~Myr for 1 -- 10~m NEOs \citep{Nesvorny:2024NEOMOD2}.

Another size-dependent feature to consider is the orbital inclination distribution of asteroids within the main belt.
The initial particle distributions used in the G18 and NEOMOD simulations were designed to mimic the inclinations of the currently observed ($>$100~m) main-belt asteroids.
However, much of the smaller asteroidal debris is not expected to follow this same distribution.
Younger asteroid families still contain large numbers of small fragments that have yet to disperse through Yarkovsky drift \citep{Bottke:2006}, keeping them dynamically concentrated and actively undergoing collisional cascades.
Several of these families are also associated with zodiacal dust bands — for example, the Veritas and Karin families — indicating that they produce debris down to micrometeoroid sizes \citep{Nesvorny:2006}.
Many such young families, including Massalia, Karin, Koronis, and Themis, have low orbital inclinations \citep{Nesvorny:2025catalogue}, where small meteoroids originating from them would share similarly low-inclination orbits before escaping the main belt.
As a result, the orbital evolution of small meteoroids in near-Earth space per source region may differ from that of the larger asteroids used in previous models.

Investigating the hazardous decametre NEOs can be accomplished by studying objects both an order of magnitude smaller and larger in diameter and inferring the size-dependent characteristics between the populations. 
The NEOMOD models extend to an absolute magnitude as high as H=28, or NEOs $\approx$ 10~m. 
In this paper, we present a model for smaller objects.
We use meteoroid observations from the Global Fireball Observatory (GFO) to analyse size-dependent main-belt migration and physical processing for meteoroids of sizes approximately 1~cm to 0.5~m.

We make the assumption that all of our observed meteoroids were liberated from their parent bodies while residing in the main asteroid belt.
Several mechanisms exist which remove meteoroids from a body in near-Earth space, such as sublimation-driven activity \citep{Jewitt:2012}, collisions and cratering \citep{Bottke:2025}, thermal fatigue or rotational breakup from close encounters with the sun \citep{Granvik:2016}, and tidal forces during close approaches with the terrestrial planets \citep{Schunova:2014, Granvik:2024} -- the latter of which is known to contribute to the H$<$25 NEO population.
We do not expect any of these sources to be a major contribution to the meteorite-precursor population that we model. 
This is because bodies in near-Earth space only remain on such orbits for on average $<$10 Myr \citep{Gladman:1997,Granvik:2018}, while the cosmic-ray exposure (CRE) ages of meteorites, which measure how long material has been isolated to a body $<$2~m in diameter, are generally much longer than these short dynamical lifetimes \citep{marti1992cosmic, Eugster:2003}. 
This implies that most meteorites must be liberated from parent bodies in the main asteroid belt, where delivery timescales match the measured CRE ages \citep{MorbidelliGladman:1998}. 
An exception are the rare CI/CM chondrites, whose short CRE ages (mostly $<$2 Myr) require an alternative delivery mechanism.
CI/CM chondrites, however, make up $<$2\% of falls\footnote{The Meteoritical Bulletin www.lpi.usra.edu/meteor/}.
Like the first debiased NEO model of \citet{Bottke:2002} that only incorporated 138 NEOs, our study is based on relatively few observations, at least at the larger sizes, with $\sim$200 meteoroids $\gtrsim$10~cm in diameter. %
Thus, we restrict our modelling to the dominant evolutionary pathway for meteorite precursors and do not attempt to include alternative sources.

The paper is divided as follows: Section \ref{sec:dataset} introduces the GFO dataset, the meteoroid selection criteria, and debiasing techniques. Section \ref{sec:model} describes the parameter fitting routines to create the NEO model. In Section \ref{sec:results} we evaluate the model's performance and limitations. Finally, we relate back to the meteoroid population in Section \ref{sec:discussion}.

\section{Data: Earth Impactors and Debiasing} \label{sec:dataset}
The Global Fireball Observatory (GFO) \citep{Devillepoix:2020} is an international collaborative project, born from the original efforts of the Australian Desert Fireball Network \citep{Bland:2012} to image bright fireballs and recover meteorites using long-exposure digital images from a distributed network of cameras.
The data that we use in this study was captured over several iterations of camera hardware \citep{Howie:2017:howtobuild, Howie:2017:debruijn}. 
Images are astrometrically calibrated \citep{Devillepoix:2018}, and the fireballs captured within them are detected \citep{Towner:2020}, triangulated using the straight line least squares method \citep{Borovicka:1990}, and dynamically modelled using an extended Kalman filter \citep{Sansom:2015}.
The pre-impact orbit and its uncertainty is determined by sampling within the measurement error of the highest recorded velocity vector with a Monte Carlo approach to create 1,000 clones \citep{Jansen-Sturgeon:2019}. 
The clones are individually propagated backward to a position outside the sphere of Earth's influence and converted into heliocentric orbital elements, where the mean and 1-sigma value in each orbital element distribution provides the final orbit and uncertainty.
Since the data collection began in 2014, upwards of 2,000 fireballs have been detected and triangulated over 10 years.

To create a debiased model of the cm to m-sized near-Earth object population, we selected fireball events for the calibration data set.
The quality cuts and selection criteria included: high quality observing geometry, lower limits on the meteoroid's initial mass, and no association with a known meteor shower (Appendix \ref{sec:subset}).
At the end of this process we are left with 1,202 sporadic meteoroids captured between 2014 and 2024.

\begin{figure}
    \centering
    \includegraphics[width=\linewidth]{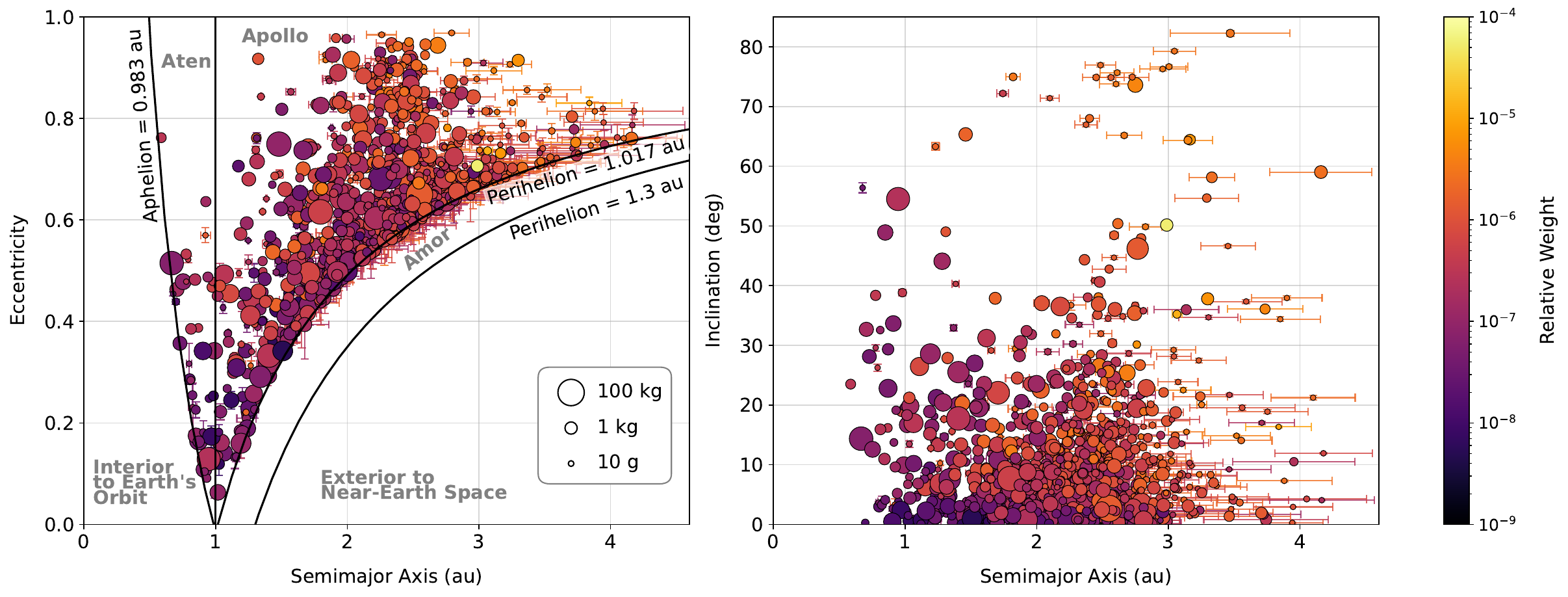}
    \caption{The GFO observed meteoroids used as calibration data plotted by their semi-major axis, eccentricity, and inclination. The sizes represent their masses at the top of the atmosphere, and the colours represent their relative weighting used for debiasing, derived from their Earth impact probability and detectability as a function of mass and speed. }
    \label{fig:GFOaei}
\end{figure}

Figure~\ref{fig:GFOaei} shows the debiased fireball data after the appropriate weightings were applied and it was normalised to 1202, the effective number of meteoroids it represents.
The slope of the size frequency distribution of the debiased meteoroids follows a broken power law very similar to that measured in the debiased fireball survey of \citet{Halliday:1996} covering the same size range as our data. 
For GFO meteoroids with masses $M$ above 2.0$\pm$0.5 kg, we measure the cumulative number $N$ as $log(N)\propto-1.059 log(M)$.
This matches, within reason, to the constant power law of the extrapolated main asteroid belt and NEO population for asteroids with scale-free disruption laws under collisional equilibrium (\citealp{Dohnanyi:1969}; slope -1.167; \citealp{OBrienGreenburg:2003}; slope -1.2), that of bolides (\citealp{Brown:2002}; slope -0.90), and that of fireballs (\citealp{Halliday:1996}; slope -1.06).

\section{Methods: Model Building} \label{sec:model}
Following the approach of previous NEO models (\citet{Bottke:2002}, G18, and NEOMOD), we have fitted a maximum likelihood function between our debiased orbital data and distributions of simulated NEOs originating from the either the main asteroid belt or Jupiter-family Comet region.
The aim of the model is to determine the relative flux of each NEO source required to reproduce the true orbital distribution of NEOs.
We describe below where our methodology is tailored towards the meteoroid population, with the remaining details of the statistical framework and parameter fitting routine discussed in Appendix \ref{sec:modelparameters}.

\subsection{Data Binning} \label{sec:bin}
The model calibration data were distributed into cells according to the meteoroid size and orbital elements: semi-major axis $a$, eccentricity $e$, and inclination $i$ (Table~\ref{tab:bins}). 
These bin sizes encompass the orbital uncertainties of the fireballs, which are typically less than 0.1~au, 0.02, and 0.5$^{\circ}$ for $a$, $e$, and $i$ respectively (see Figure~\ref{fig:GFOaei}), but do increase for higher semi-major axis orbits because of the greater speeds with which they impact the Earth (see \S \ref{sec:dataset} for more information on calculating meteoroid orbits from fireball observations).
This created the 4-dimensional distribution $N_{GFO}(a,e,i,H)$ containing 21$\times$12$\times$15$\times$28 = 105,840 cells. 
The absolute magnitude axis, $H$, was used as a proxy for meteoroid size. 
This was chosen so the model works at the same scale, $\propto log(m^{1/3})$, as other NEO models using telescopic survey data which intrinsically measure $H$. 
The meteoroid entry mass calculated by the GFO pipeline was transformed into $H$ assuming a density of 3,500~kg m$^{-3}$ and geometric albedo of 0.147 (typical values for an ordinary chondrite). 
Ideally, one would want to develop a model using a directly observed measure of size such as mass, diameter, or cross-sectional area.
In the case of fireballs, however, we do not directly observe any of those quantities without making some assumption about the meteoroid's physical properties.
Hence, using H as a proxy for size in the model is no worse than the alternatives; we will convert between H, mass, and diameter for the purpose of discussion throughout the paper, using the assumed quantities stated above.
To assess whether converting from mass to H without knowing individual meteoroid properties significantly impacts the model fit or interpretation, we consider the conversion across various meteorite types. The meteorite type most visibly distinct from ordinary chondrites are aubrite and eucrite, which are the bright E-/Xe- and V- type asteroids respectively. Using nominal albedo and bulk densities \citep{Usui:2013, Macke:2011}, the difference in H, for any mass, between an ordinary chondrite and an aubrite or eucrite is -2.0 or -1.8 respectively. This is significant compared to the H bin size, however, the model flexibility along the H axis is constrained to at maximum 3 steps which are least $\Delta$H=2 apart. Additionally, these meteoroid types are not significant among NEOs \citep{Marsset:2022} or meteorites\footnote{The Meteoritical Bulletin www.lpi.usra.edu/meteor/}, and thus we expect no more than 5\% of the 1202 dataset to have a greatly overestimated H i.e. they should be in much `brighter' H bins. The second most common meteoroid we expect after ordinary chondrites, carbonaceous chondrites, have a nominal $\Delta$H of -0.8. A reasonable portion of our data (somewhere between 4 and $\sim$32\% based on meteorite falls and NEO statistics \citep{Marsset:2022}) will have H underestimated by at maximum 3 bins. We would not expect these shifts to alter the calibration data enough to create significantly different model results, as the uncertainties on the final posterior distributions of the free parameters as they stand are rather large.

\begin{deluxetable}{ccc} \label{tab:bins}
\tablecaption{The cells used to divide the model calibration data -- the GFO meteoroids -- and residence time distributions.} 
\tablewidth{1.0\textwidth}
\tablehead{
\colhead{Dimension} & \colhead{Range} & \colhead{Step Size}}
\startdata
$a$ & 0~au -- 4.2~au & 0.2~au \\
$e$ & 0.04 -- 1.00 & 0.08 \\
$i$ & 0$^{\circ}$ -- 60$^{\circ}$ & 4$^{\circ}$\\
$H$ & 34.5 -- 41.5 & 0.25 
\enddata  
\end{deluxetable}

\subsection{Residence Time Distributions \Rs} \label{sec:residencetimes}
We call the distribution of orbits originating from a source $s$ the residence time distribution, \Rs, which is a function of ($a, e, i$).
We use the same as those created for the NEOMOD models \citep{Nesvorny:2023}; to which we direct the reader for a full description of their derivation, but describe briefly below. 
Eleven \Rs were created by populating the resonance gaps in the main asteroid belt with massless test-particles and tracking their motion within near-Earth space ($q<$1.3~au) during 500~Myr numerical integrations of the Solar System. The particles' initial locations within the mean-motion resonances with Jupiter (3:1, 5:2, 7:3, 8:3, 9:4, 11:5, and 2:1), and the secular resonance with Saturn (\nusix), were distributed bordering the resonance gaps according to the trends of main-belt asteroids ($>$100~m in size). The initial positions for the Hungaria population, Phocaeas population, and the particles expected to evolve along weak inner resonances (2.1$<a<$2.5~au, $i<$18$^{\circ}$, $q>$1.66~au), were taken from real asteroid positions in the respective populations. 
The time the particles spent with certain $(a,e,i)$ after being perturbed from the main-belt and reaching near-Earth space was recorded and averaged to create the residence time distributions as a function of those orbital element, $R_s(a,e,i)$. 
The 12th and final \Rs representing the Jupiter Family Comet (JFC) population was developed in \citep{Nesvorny:2017}. 

When we produced the \Rs to account for physical processes in the inner Solar System (see \S \ref{sec:alteredsources}), the finite number of particles in the original orbital integrations resulted in many empty bins. 
To accommodate this, the resolution of the 12 new \Rs were reduced from that used in NEOMOD to the resolution outlined in Table~\ref{tab:bins}.
Namely, the bin widths along the semi-major axis and eccentricity were doubled, and the inclination range was reduced to below 60 degrees. 
We also uniformly filled any empty bins with a small flux of 1$e^{-8}$, which is at the tail end of the flux distribution of the filled bins, and renormalised \Rs to 1 again. 
This downsampling and bin filling allowed for smooth log-likelihood calculations with minimal influence on the final model fit.

\subsection{Meteoroid Residence Time Distributions} \label{sec:alteredsources}
We now introduce a range of meteoroid specific \Rs distributions.
These were created because, upon reviewing the initial model fits with the NEOMOD \Rs, 
the distribution of the GFO meteoroids and could not be recreated with the original 12 sources. 
This is not surprising as the NEOMOD \Rs were created from orbital simulations designed to replicate the passage of decametre to kilometre asteroids from the main-belt. 
The additional \Rs, described below, replicate the physical effects applicable to smaller asteroids as they evolve from the main belt. 

\begin{figure}
    \centering
    \includegraphics[width=1\textwidth]{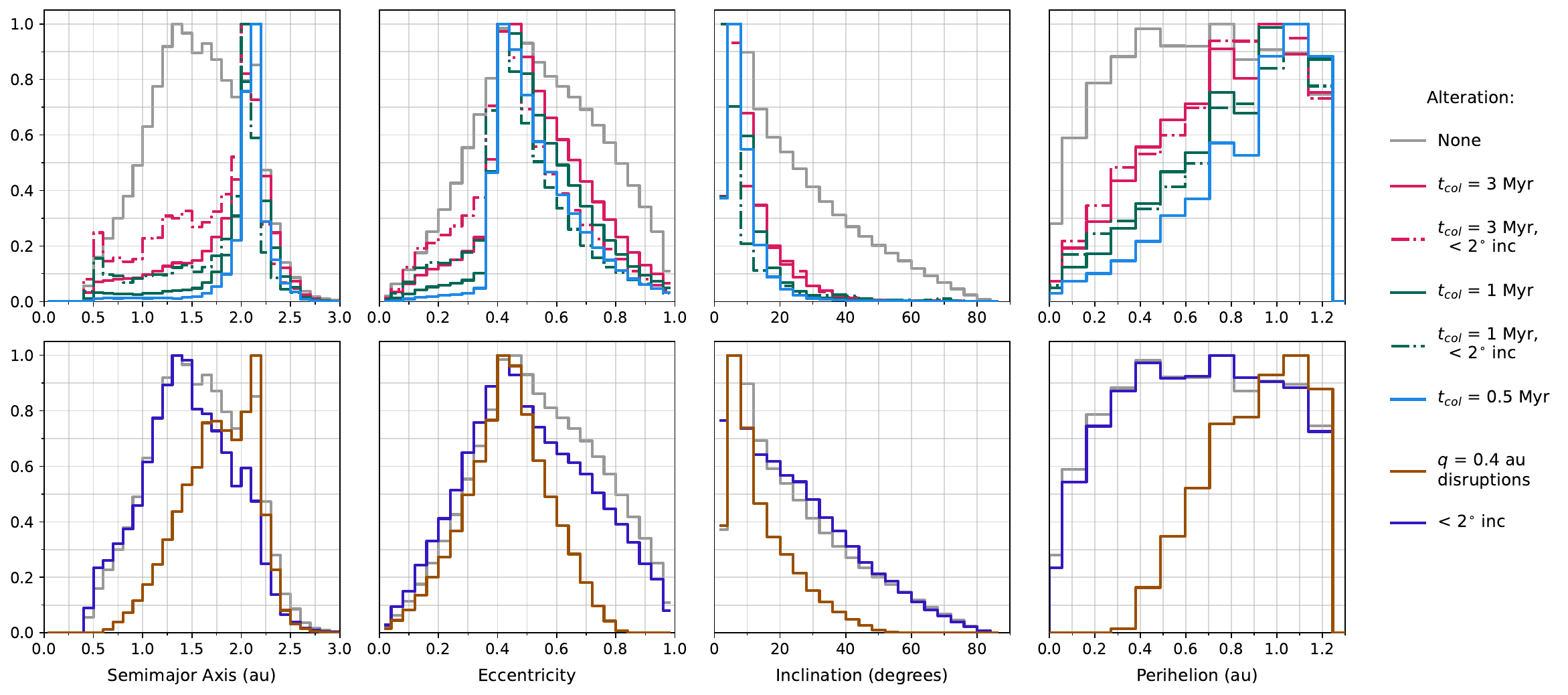}
    \caption{A comparison of orbital distributions for particles that have escaped from the \nusix resonance and evolved under various physical processes. The top panel shows the result for simulations including different collisional lifetimes (\tcol), while the bottom panel shows the output of low perihelion ($q<$0.4~au) disruptions or main-belt asteroids restricted to low ($<$2$^{\circ}$) inclinations. The top panel also shows the combination of collisions and low inclination input with the dot-dashed line. The simulation which created these distributions was integrated initially for 100~Myr to capture particle escape from the main-belt, and a further 400~Myr for particles within $q<$1.3~au to track further evolution (see \citet{Nesvorny:2023} for further details).}
    \label{fig:altered_sources}
\end{figure}

\subsubsection{Collisions $R_{s,Myr}$}\label{sec:collisionallifetimes}
Asteroids in the main asteroid belt are subject to disruption from a collision with another asteroid, and near-Earth asteroids are no exception. 
The collisional lifetimes of large NEAs are typically longer than the relatively short dynamical lifetimes of objects within the inner Solar System, so there has been no need to include them in previous NEO models (discussed in Section 7.6 of \citet{Nesvorny:2024NEOMOD2}).
As we are modelling meteoroids with diameters of less than 1 metre, however, the expected collisional lifetimes are now on the order of, or less than, the dynamical lifetimes; altering the expected orbital distribution of such a near-Earth meteoroid population.
We mimicked collisions in the numerical integrations to produce \textit{collisionally evolved} \Rs.
For each particle, the probability of a collision $P_{col}$ was modelled by an exponential cumulative distribution function, $P_{col} = 1 - e^{h/tcol}$, for the simulation interval $h$ = 1,000~years. 
When $P_{col} < x $ for some randomly generated value $0 < x < 1$, the particle was removed from the simulation. 
This was calculated at each simulation interval for particles with an aphelion of $Q>$1.8~au still coupled to the main-asteroid belt, where collisions are estimated to occur with the half life of \tcol. 
This aphelion condition was also used by \citet{MorbidelliGladman:1998} in their analysis of the collisional lifetimes of decimetre NEOs, representing only collisions between NEOs and the comparatively concentrated main-belt asteroid population.
We also tested collisions without any condition on $Q$ for comparison; however, the aphelion condition was preferred by the model and is used throughout the rest of the modelling procedure. 

The most appropriate collisional lifetime to include depends on the size of the objects being investigated (introduced in Section \ref{sec:intro}).
The sizes of the GFO meteoroids used to calibrate this model are approximately 1~cm to 0.5~m in diameter, so we created \Rs distributions of NEOs with $t_{col}$ = 3~Myr, 1~Myr, and 0.5~Myr. 
As only the \Rs with a mean dynamical lifetime the same as \tcol or less will be influenced significantly by collisions, for $t_{col}$ = 3~Myr, we only created $R_{s,3Myr}$ for the \nusix and inner weak sources.
Without introducing any depletion (e.g. collisions or low-$q$ disruptions), $R_{\nu_6}$ and $R_{Inner}$ have mean dynamical lifetimes of 9.5~Myr and 7.0~Myr respectively, while the rest of the sources employed in the model have much shorter dynamical lifetimes on the order of 1~Myr or less\footnote{The mean dynamical lifetimes of the base \Rs are different to \Rs with low-perihelion disruptions, which are the values reported in Table 5 of \citet{Nesvorny:2023}. \Rs without low-perihelion disruptions have a longer mean dynamical lifetime.}. 
In our 3~Myr collisional models we used $R_{\nu_6,3Myr}$ and $R_{Inner,3Myr}$ along with the NEOMOD \Rs for the other sources.
We created \Rs with \tcol = 1~Myr and 0.5~Myr for all 12 sources.

Figure~\ref{fig:altered_sources} shows an example of the meteoroid \Rs for the \nusix source region.
The dominant trend for the collisionally evolved \Rs is a decreasing contribution from particles on highly evolved orbits with decreasing collisional lifetime. 
This is shown by a significantly reduced signal from semi-major axis values $<$ 1.6~au. 
The eccentricity peak at 0.44 is also narrower, and the inclination distribution changed from a gradual decrease in inclinations from ecliptic towards 90 degrees to a sharp peak at 8 degrees and sudden decay to very little signal from orbits $>$ 40 degrees.
The perihelion distribution also shifted from a flat distribution ranging 0~au to 1~au to an increasing slope peaking at 1~au.

\subsubsection{Low-perihelion Disruption $R_{s,0.4au}$}\label{sec:lowperiheliondisruption}
We created \Rs including catastrophic disruptions for particles with $q\leq$0.4~au.
If meteoroids were to disrupt near the Sun in the same way as larger asteroids, then we would not expect them to survive within about 0.4~au. This distance corresponds to extrapolating the disruption law of \citet{Granvik:2016} down to objects of 1 m in diameter.
Such disruption was invoked by monitoring each simulated particle's perihelion distance and removing it when it reached $q\leq$0.4~au for the first time. 
This primarily removes highly eccentric orbits (Figure~\ref{fig:altered_sources}).
Other prominent features are a swap in the semi-major axis peak from $\sim$ 1.2~au to $\sim$ 2.1~au, a sharpening of the eccentricity peak at 0.4, a decrease in high inclination orbits, and a removal of low perihelion orbits to create a steep slope from 0.4~au to 1~au.

\subsubsection{Isolating Low Inclination Sources $R_{s,^{\circ}}$} \label{sec:lowincsources}
In final set of meteoroid \Rs distributions, we restricted the simulation input to low-inclination particles.
This was to mimic young asteroid families, often located at low inclinations, which predominantly feed neighbouring main-belt escape regions due to their greater amount of small debris (\citet{Broz:2024Nature} and references therein).
We restricted the input particles for $R_{\nu 6}$, $R_{3:1J}$, $R_{2:1J}$ to those below 2$^{\circ}$, 2$^{\circ}$, and 5$^{\circ}$ respectively. 
It is worth noting that the particles below these inclinations still follow the same input distributions originally developed for NEOMOD, which were derived from the positions of the largest asteroids in the main-belt. 
This created subtle differences for $R_{\nu_6,2^{\circ}}$ (Figure~\ref{fig:altered_sources}), the most significant being the reduction of the second semi-major axis peak at 2.2~au.

\section{Model Fitting and Results} \label{sec:results}

With both the NEOMOD residence time distributions and the new meteoroid-appropriate residence time distributions (\Rs and $R_{s,Myr}$, $R_{s,0.4au}$, $R_{s,^{\circ}}$ respectively), we followed the methods of NEOMOD and fitted such distributions to the calibration fireballs using a maximum likelihood estimate according to Equations \ref{eq:alpha(H)}--\ref{eq:priors2} in Appendix \ref{sec:modelparameters} to find the optimal number of sources $n$, strength parameters \alphas per source as a function of size, number of linear segments or ``slopes", $m$, that $\alpha_s(H)$ is divided into to create the model flexibility with size, and crossover index $\delta$, which describes the transition between the types of \Rs along the $H$ axis.

\subsection{Best Fitting Model} \label{sec:best_fit_model}

\begin{figure} \label{fig:buildingLL}
    \centering
    \includegraphics[width=0.5\textwidth]{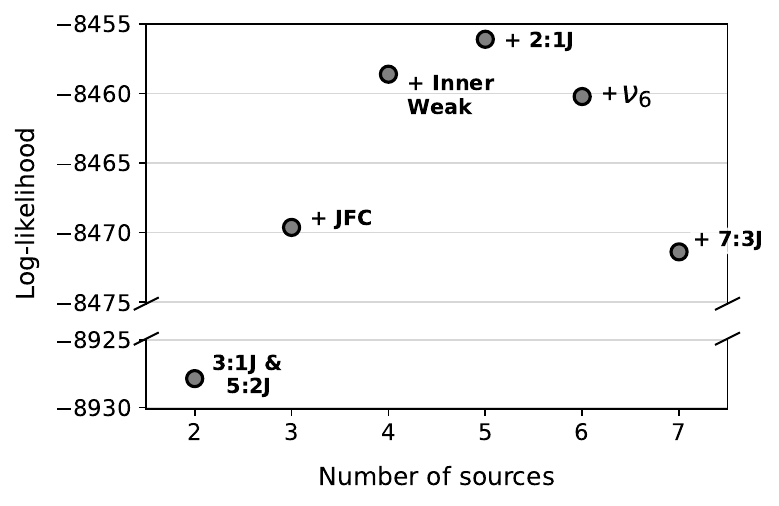}
    \caption{The log-likelihood values of the best performing source region combinations in a three-slope model. The log-likelihood has the highest value for the 5 sources 3:1J, 5:2J, Inner Weak, JFC, \& 2:1J. The addition of more sources decreases the log-likelihood; only marginally for the 6th (\nusix) but more significantly for the 7th (7:3J).}
\end{figure}

The optimal number of sources to include in the model was $n$=6, including: Inner Weak, \nusix, 3:1J, 5:2J, 2:1J, 
and JFC.
We tested this by adding the asteroidal \Rs (without any physical processes, using $m$=3) incrementally to the model until the highest log-likelihood value was reached (Figure~\ref{fig:buildingLL}).
One inner- and one outer-belt source, $R_{3:1J}$ and $R_{5:2J}$, were used as the starting input, from which the remaining sources were added.
\multinest penalises models with too many parameters to prevent over-fitting, which is why a turnover is seen in the log-likelihood for more than 5 sources.
We decided a 6-source model was the most appropriate, even though 5 sources produced the highest log-likelihood.
This is because $R_{\nu6}$ and $R_{3:1J}$ are very similar, so the degeneracy between them is important to evaluate.
Also, the difference in log-likelihood between 4, 5, and 6 sources is small, with Bayes factors ($\Delta \mathcal{L} $) of $< 5$.
When a 7th source was added, the best performing additional source was 7:3J; however, \multinest allocated that source 0\% contribution, i.e. $\alpha_{7:3J}(H)=0$.
This indicates that the calibration data set cannot distinguish between any more sources, and from this, we are certain these 6 sources are sufficient to describe the general trends of the cm to m NEO population. 

\startlongtable
\begin{deluxetable*}{ccccccccccccc}

\tablecaption{The parameters for the strength function, $\alpha_{s,m+1}$ for the best fitting $m$=3, $n$=6 model, with the crossover from a collisional lifetime of 3~Myr to 1~Myr at H=38.5 -- about 0.6~kg, between $\alpha_2$ and $\alpha_3$. $\alpha_1$ and $\alpha_4$ are the outer bounds of the model at H=34.5 and 41.5, and $\alpha_2$ and $\alpha_3$ are anchor points defining the breaks in the slope at H=39.75 and 39.25, respectively. The values shown are the medians and 1$\sigma$ bounds of the marginal posterior distributions. As each parameter may be drawn from a non-normal distribution and normalisation to 1 applies only to each posterior draw, the medians of the marginals do not always sum to 1 -- the normalised value is provided in brackets where it differed significantly.%
}
\tablehead{\colhead{Source} & \colhead{$\alpha_1$} & \colhead{$\alpha_2$} & \colhead{$\alpha_3$} & \colhead{$\alpha_4$} & \multicolumn{2}{c}{$\sigma(\alpha_1)$} & \multicolumn{2}{c}{$\sigma(\alpha_2)$} & \multicolumn{2}{c}{$\sigma(\alpha_3)$} & \multicolumn{2}{c}{$\sigma(\alpha_4)$} %
} 
\startdata
    \nusix & 0.203 (0.235)& 0.332  & 0.207 & 0.264 & 0.075 & 0.394 & 0.272 & 0.395 & 0.177 & 0.239 & 0.228 & 0.301 \\ 
        Inner & 0.143 (0.166) & 0.033  & 0.008  & 0.02  & 0.050 & 0.304 & 0.009 & 0.077 & 0.002 & 0.019 & 0.006 & 0.045 \\ 
        3:1J & 0.285 (0.330) & 0.158  & 0.512  & 0.177  & 0.128 & 0.489 & 0.087 & 0.241 & 0.473 & 0.553 & 0.139 & 0.217 \\ 
        5:2J & 0.145 (0.168) & 0.192  & 0.082  & 0.025  & 0.05 & 0.299 & 0.112 & 0.282 & 0.053 & 0.115 & 0.008 & 0.048 \\ 
        2:1J & 0.042 (0.049) & 0.104  & 0.043  & 0.007  & 0.013 & 0.102 & 0.061 & 0.148 & 0.026 & 0.061 & 0.002 & 0.016 \\ 
        JFC & 0.045 (0.052) & 0.156  & 0.141  & 0.496  & 0.013 & 0.108 & 0.097 & 0.221 & 0.111 & 0.171 & 0.465 & 0.528 \\ 
\enddata

\label{tab:alphas}
\end{deluxetable*}

The best performing number of slopes in the strength function (Eq. \ref{eq:alpha(H)}) for this dataset was $m$=3. 
We computed model fits with $m$ ranging 1 -- 4.
The log-likelihood continued to increase with increasing numbers of slopes and freedom given to the model.
We therefore used our intuition to determine that one or two slopes weren't enough to provide the model with sufficient freedom on the $H$ axis, and that four slopes produced rapid oscillations between sources for different sizes, indicating that the model was overfitting to the data.
Three slopes created a smooth transition between sources and were the most appropriate to describe the general trends of the data without over-fitting.

The best performing meteoroid \Rs had collisions for lifetimes of 3~Myr and 1~Myr.
Compared to the models using the asteroidal \Rs, these models have Bayes factors of 144 and 63, respectively.
What became clear when investigating the model performance over distinct size regions (Figure~\ref{fig:heatmap}), is that 3~Myr was dominant over the largest sizes and 1~Myr at smaller sizes. 
This was confirmed by fitting a model with both $R_{s,3Myr}$ and $R_{s,1Myr}$ and \multinest fitting the crossover at $\delta$=38.5 ($\sim$0.6~kg).
This crossover model has a Bayes Factor of 190 compared to a model with just the original \Rs. 
Figure~\ref{fig:heatmap} also shows the model including low-inclination inputs alongside collisions yields comparable performance to the collision only model, however we chose to adopt the collisions-only model moving forward for several reasons. 
First, the \Rs values from combined physical process which depleted meteoroids suffered most from empty bins (see Section \ref{sec:alteredsources}). 
Second, the restriction to low inclinations provides only a simple approximation of the true spatial distribution of small main-belt asteroids. 
Finally, the differences in performance and output distributions between the two model types are only small. Taken the performance of this model from these limitations, we demonstrate a strong case for future work on this framework.
Overall, we choose the crossover model transitioning from 3~Myr collisions to 1~Myr at around $\sim$0.6~kg as our best description of the cm to m NEO population.

\begin{figure} 
    \includegraphics[width=\textwidth]{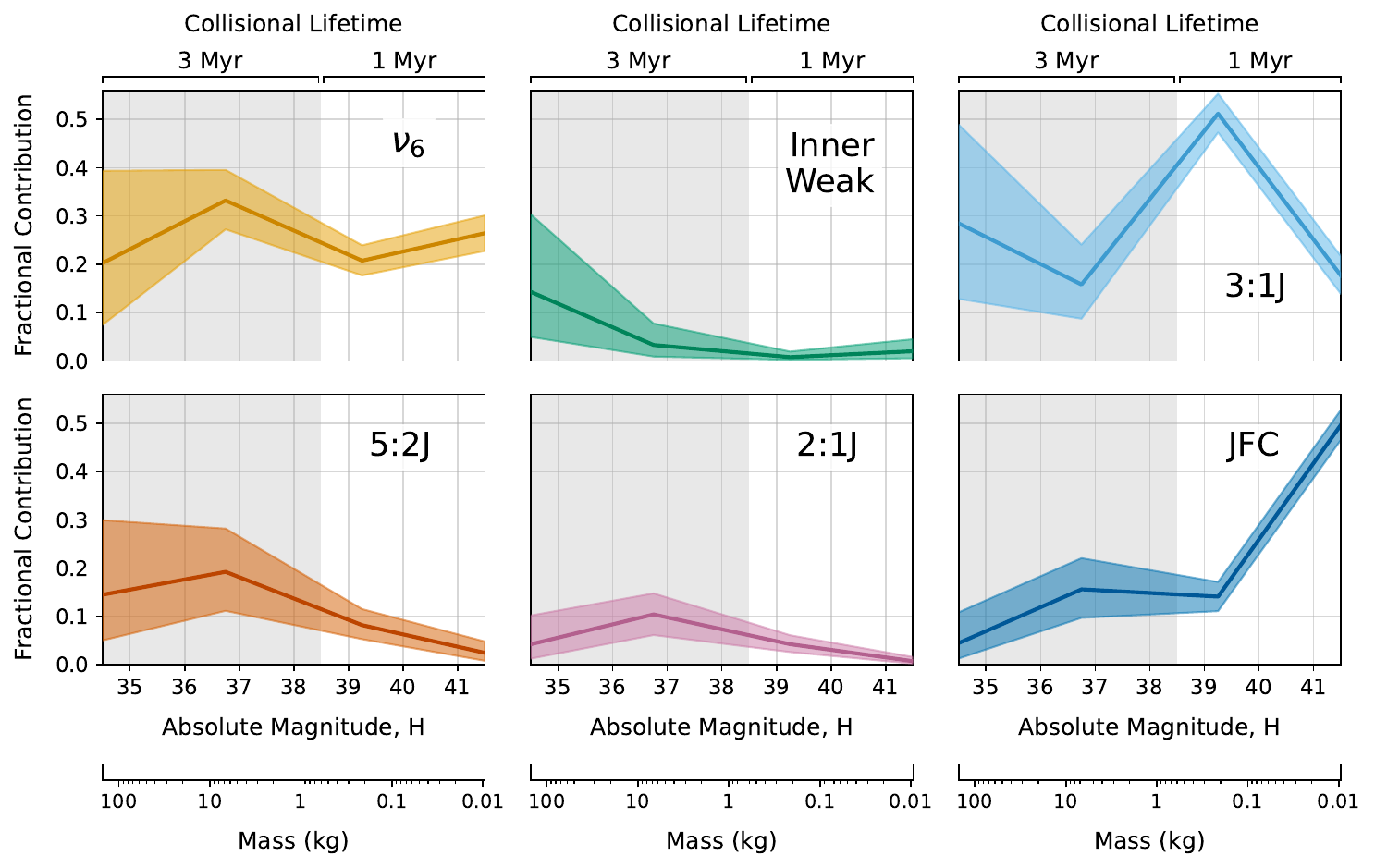}
    \caption{The relative contributions of the 6 sources, Inner Weak, \nusix, 3:1J, 5:2J, 2:1J, and JFC, for the optimal three slope model as per Table \ref{tab:alphas}. There is a transition between input sources with collisions at 3~My and 1~Myr lifetimes as indicated by the grey background.}
    \label{fig:alphas}
\end{figure}

The contributions from the sources for the best-fitting model, as determined by \multinest, are displayed in Table~\ref{tab:alphas} and shown in Figure~\ref{fig:alphas}, along with the corresponding 1-sigma uncertainties.
The largest sizes probed by this model, 150~kg – 10~kg, are supplied by a combination of the Inner Weak, \nusix, 3:1J, and 5:2J sources.
Looking across the size distribution, the \nusix contribution stays relatively constant, while the contributions from the inner weak, 5:2J, and 2:1J decrease.
Interestingly, the 3:1J source peaks around %
0.25~kg, with approximately a 50\% contribution; the next greatest contribution at this size being \nusix $\sim$20\%. 
The smallest size range 10~g -- 250~g is reproduced by JFC type orbits as it has the greatest contribution at ~50\%, with $<$ 10\% contribution from the Inner Weak, 5:2J, and 2:1J sources, and around 20\% from both \nusix and 3:1J. 
Overall, we have found the relative contributions between dominant main-belt sources that feed the cm to m NEO population.

The $\alpha$ parameter uncertainties are the greatest at the largest sizes of the model because of the smaller number of fireballs to calibrate the model and the degenerate nature of the inner main-belt sources. 
Figure~\ref{fig:corner_x4} shows examples of the marginal posterior distributions (full plot in Appendix \ref{sec:full_corner_plot}). 
The parameters for the largest sizes, $\alpha_{s,1}$, show major degeneracy for the sources \nusix, Inner Weak, 3:1J, and 5:2J, while 2:1J and JFC appear independent.
Unfortunately, we do not have enough data in this size region to constrain each inner-belt source individually. 
A further complication is that, while each individual sample of \multinest fulfils $\sum_{s=1}^{n}\alpha_{s}=1$, the median values of each posterior distribution do not always sum to 1, so we use the normalised values (bracketed column in Table \ref{tab:alphas}) when comparing source region contribution.
We demonstrate that the best fitting model is robust i.e. it is not fitting to specific data points, by dividing the calibration data set in half and refitting our $s$=6, $m$=3 model (Appendix \ref{sec:half_data_results}) and seeing that the relative strengths of the source regions not change significantly.

\begin{figure}
    \centering
    \includegraphics[width=0.5\textwidth]{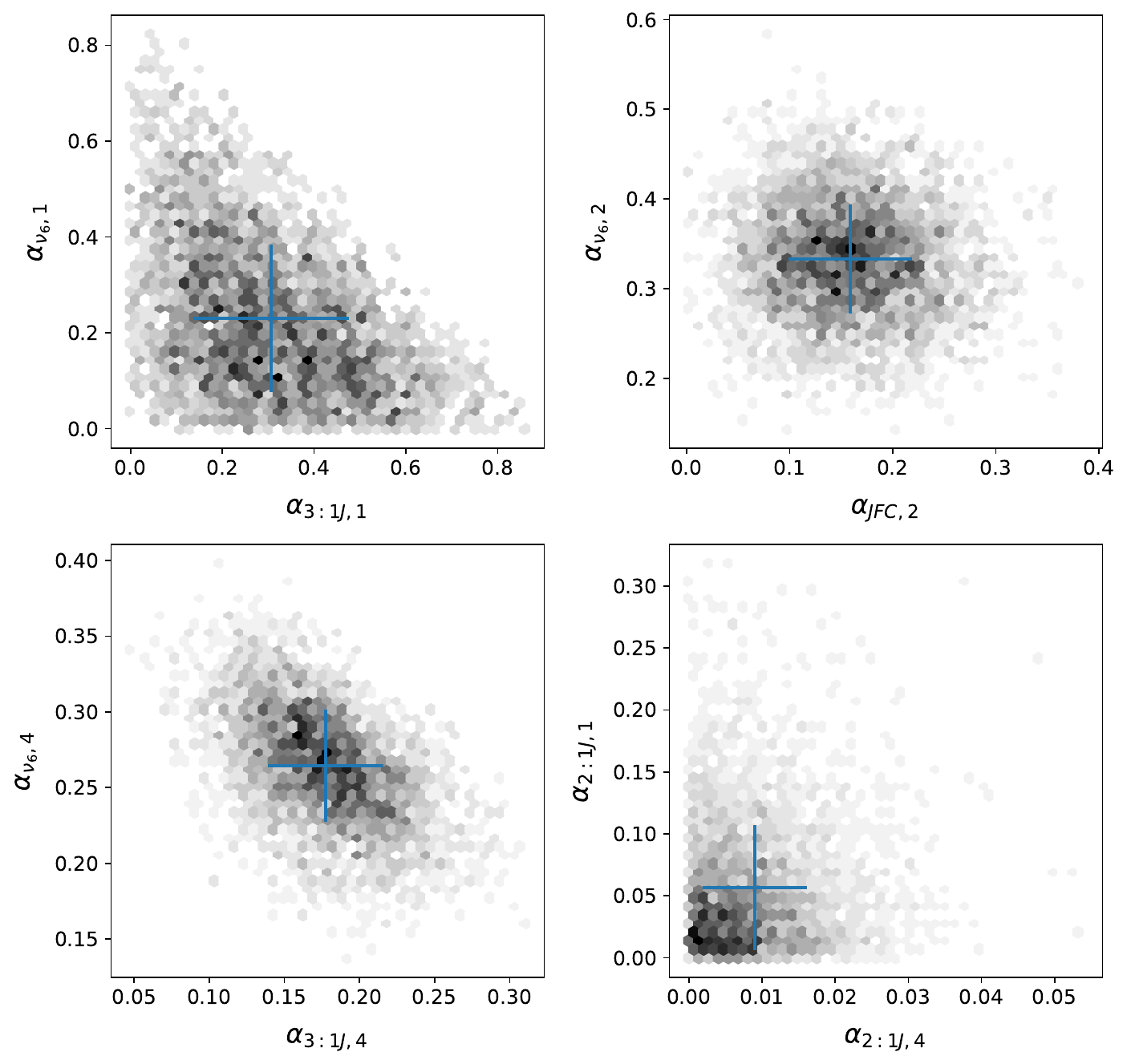}
    \caption{Four examples of posterior distributions from \multinest for the best fitting model showing major degeneracy (top left), a well constrained parameter (top right), minor degeneracy (bottom left), and low- to zero-contribution (bottom right).}
    \label{fig:corner_x4}
\end{figure}

\begin{figure}
    \centering
    \includegraphics[width=1\textwidth]{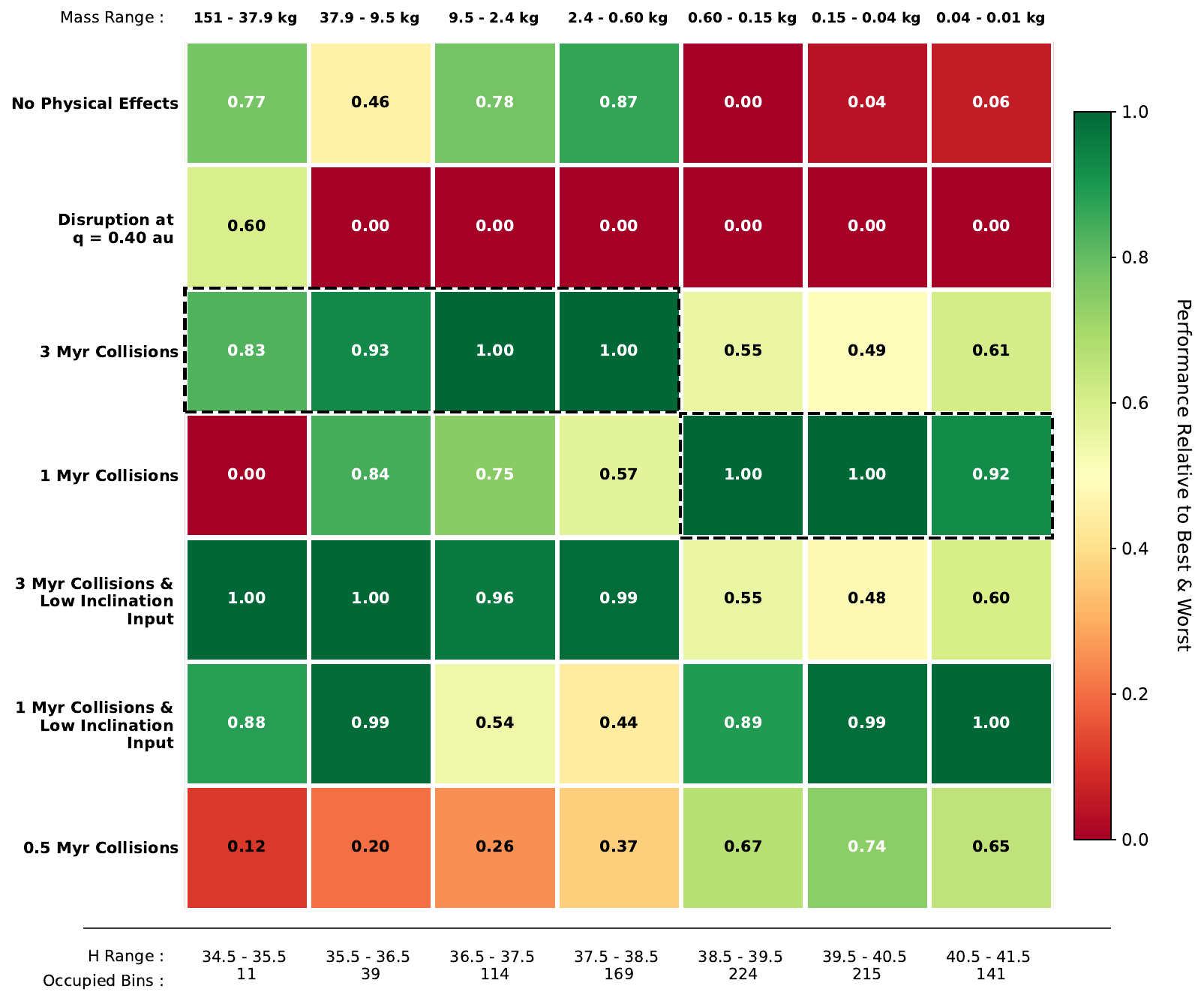}
    \caption{The relative performance of all the models over brackets of decreasing size. Performance is measured by taking the log-likelihood values of the model fit and the data in each size bracket (vertical column) and rescaling them to the range 0 to 1, between the lowest and highest log-likelihood values. Our choice of best-fit model is shown with the dashed line, switching physical process types at H=38.5 (0.6~kg). The number of bins in $N(a,e,i,H)$ occupied by meteoroids in each size bracket varies, as indicated below each column, and indicates the reliability of the performance measure. The less data, the more sensitive the model is to the presence of individual data as opposed to general trends. Note this does not add to 1,202 as multiple calibration meteoroids can occupy the same bin.}
    \label{fig:heatmap}
\end{figure}

\subsection{Evaluation of Fit} \label{sec:eval_of_fit}

\begin{figure}
    \centering
    \gridline{\fig{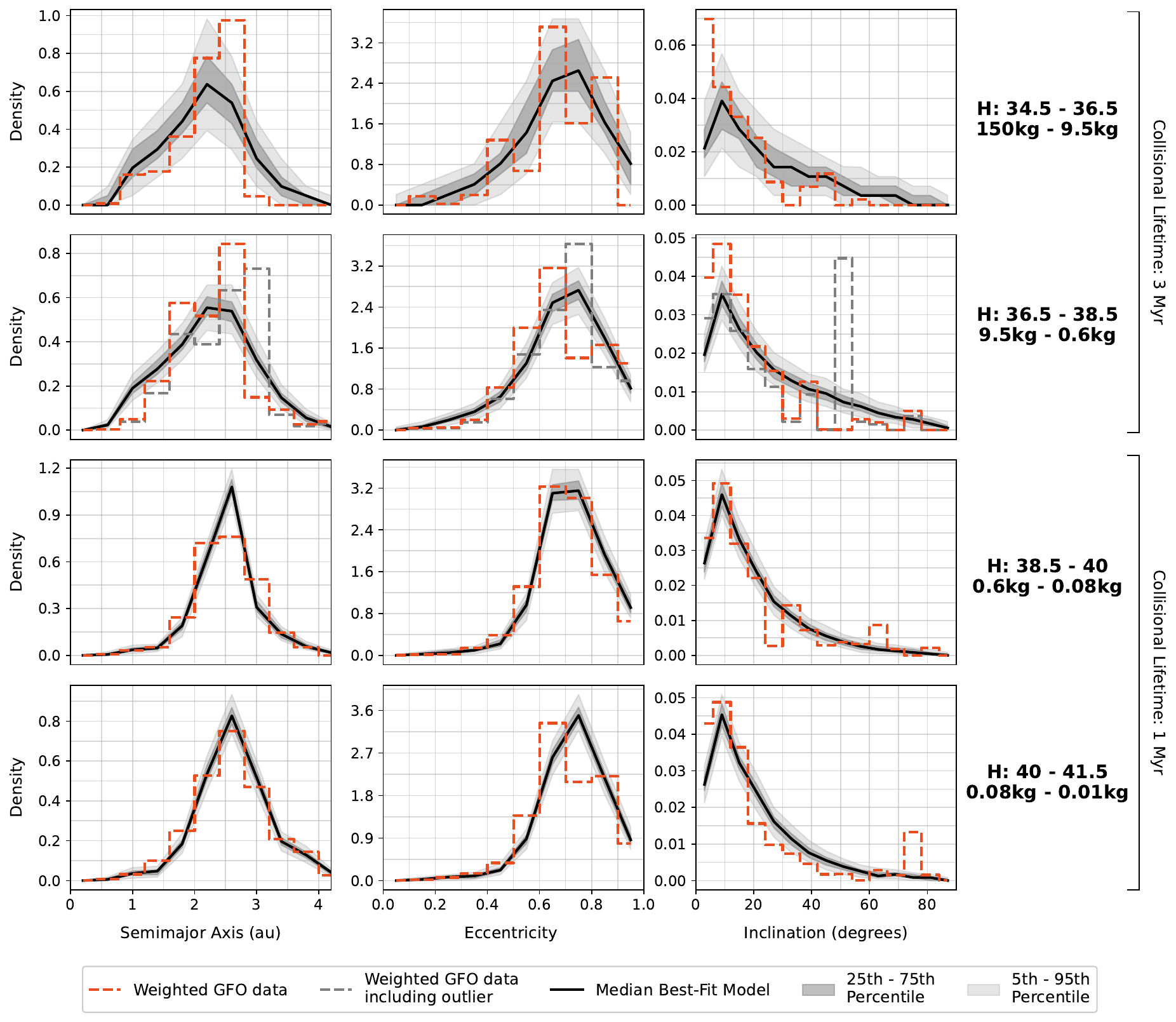}{\textwidth}{(a)}}
    \gridline{\fig{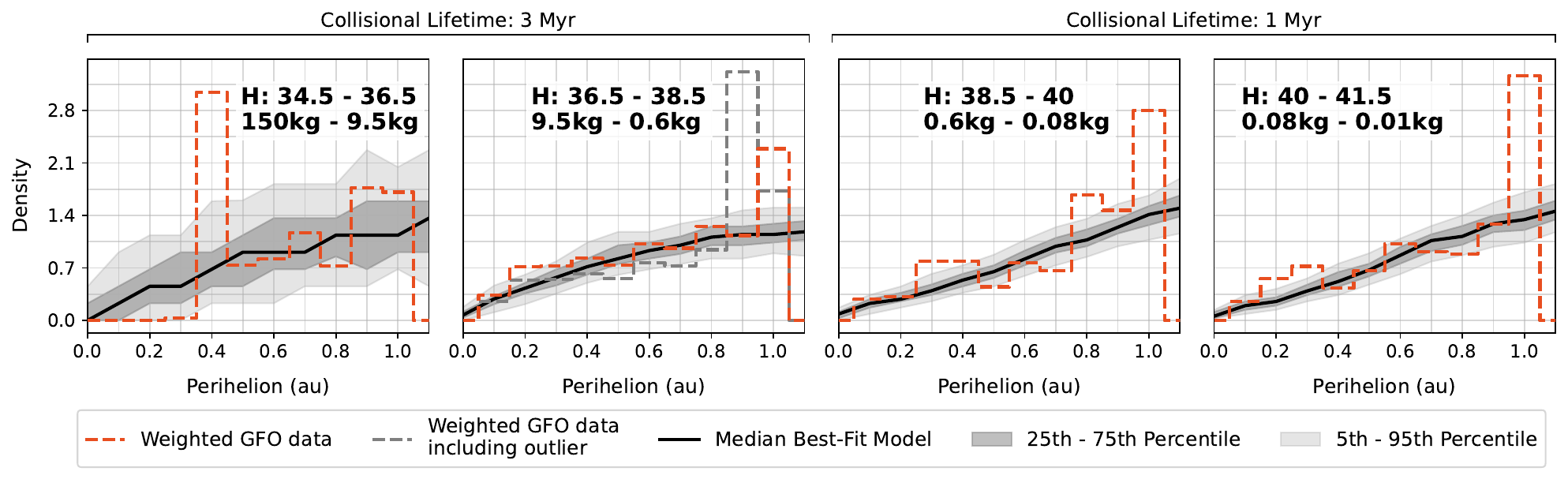}{\textwidth}{(b)}}

    \caption{Normalised orbital element distributions of the debiased Earth-impacting meteoroids and best-fit model for (a) Earth-crossing orbits ($q<$1.016~au and $Q>$0.98~au) and (b) orbits truncated to $q\leq$ 1.1~au. 200 model outputs were created sampling the parameter uncertainties as described in Section \ref{sec:eval_of_fit}, and sampled 53, 320, 425, and 425 times for the ranges H=34.5-36.5, 36.5-38.5, 38.5-40, 40-41.5 respectively. Though sampled as continuous distributions, the histograms here are plotted with 0.4~au, 0.1, 6$^{\circ}$, and 0.1~au resolution over the semi-major axis, eccentricity, inclination, and perihelion axes.}
    \label{fig:orbitaldistributions}
\end{figure}

The best-fit model comprehensively matches the debiased fireball population.
To compare the two distributions, 200 model outputs were created by taking a set of the $\alpha_{s,m+1}$ parameters from within the \multinest marginal posterior distributions. 
These $\alpha$ parameters all fall within the 1-sigma uncertainties reported in Table~\ref{tab:alphas} and combine to create models of comparable performance to the best-fit combination (Bayes Factors of $<$ 2).
Using parameters from the marginal posterior distribution ensures the degeneracy between the source region strength is preserved in each model created, and the parameter 1-$\sigma$ uncertainties are sampled effectively. 
For each combination, we used $t_1$=3~Myr, $t_2$=1~Myr and $\delta$=38.5 in Eq. \ref{eq:model} to create $N(a,e,i,H)$.
Each $N(a,e,i,H)$ was reduced in H range and sampled the same number of times as the number of meteoroids within that size range of the GFO calibration dataset to create an equivalent resolution distribution for comparison.
Finally, to change $N(a,e,i,H)$ from the model resolution (Table~\ref{tab:bins}) to a continuous distribution, $(a,e,i)$ was randomly resampled within each bin range. 
We extract the median and various percentiles of the orbital element distributions from the 200 model outputs for comparison against the weighted GFO calibration data in Figure~\ref{fig:orbitaldistributions}.
The general trends of each size range are well matched by the model outputs, particularly the reduction in flux at a semi-major axis of 1~au for decreasing sizes, and the narrowing of the eccentricity peak.
We show the distribution of the H=36.5-38.5 range both with and without one, strongly weighted, high inclination meteoroid which was used in the calibration data set. 
Note that the inclusion or exclusion of this event from the model fitting did not change which model performed best, nor the resulting $\alpha$ parameters.
The mismatch at the smallest sizes is a curiosity as it is not likely a result of a small quantity of data.
For H = 38.5–40, the model's semi-major axis distribution peaks above the GFO data at 2.6~au while the eccentricity shows excellent agreement. In contrast, for H = 40–41.5, the model's eccentricity peaks above the data at 0.75, while the semi-major axis aligns well.

This could be explained by a consistent peak at a perihelion of 1~au in the meteoroid data, not present in the model output, for data H$>$36.5 (Figure~\ref{fig:orbitaldistributions} (b)). 
There is also a spike at $q$=0.4~au for the largest objects, but this is an artefact of the small number of meteoroids in this size range.
As our calibration data is made up of only Earth-impactors (the majority of which have perihelion values of $q$=1~au), it is difficult to know, despite our debiasing procedures, whether or not we have created a true representation of near-Earth meteoroids to compare to main-belt derived material.
Overall, the strong match in the semi-major axis distribution between the debiased meteoroids and the model output is strong support for the inclusion of collisions in the cm- to m-sized NEOs, a reduction in collisional lifetime as sizes decrease, and the use of this model to represent the meteoroid population.

\section{Model Interpretation and Discussion} \label{sec:discussion}

\subsection{Inner-belt Dominance}

Placing our results in the context of meteoroid delivery to the inner Solar System, we can reaffirm the inner-belt dominance feeding the near-Earth region through the \nusix and 3:1J resonance.
The parameters $\alpha_{s}(H=34.5)$ and $\alpha_{s}(H=39.75)$ for the sources inner weak, \nusix, and 3:1J, shows that our model predicts a decreasing, 73\% to 53\% contribution from the inner-belt sources over the size range 150~kg to 6~kg (approximately 0.5~m to 15~cm assuming a density of 3,500~kgm$^{-3}$). 
These parameters, however, are correlated with those of the 5:2J resonance (Fig \ref{fig:corner_full}), which have median values 16 -- 19\% over this size range.
Only an increase in the number of observed fireballs over this size range could resolve the degeneracy between sources. 
In comparison to the predictions from other NEO models, our inner-belt contribution is generally less, though consistent within uncertainties.
NEOMOD2 predicts 91\% of H=28 (10~m diameter for an albedo of $p_v=0.14$) NEOs originate from the \nusix and 3:1J alone (Inner Weak $<$1\%).
\citet{Deienno:2025} estimate a 77\% contribution from \nusix and 3:1J and 11\% from the inner weak resonances, predicting an 88\% dominance from the inner main belt. 
Returning to submetre sizes, \citet{MorbidelliGladman:1998} proposed that Earth-impacting meteoroids can be fed primarily by the \nusix and 3:1J resonances in roughly equal proportions, which we agree with despite subtle differences in the dataset and methods.
Their analysis used 63 fireballs 
with initial masses ranging 30~g -- 25~kg.
While this matches the mass range of our calibration meteoroids, they only analysed potential meteorite-dropping fireballs, whereas we were inclusive of any meteorite fall potential.
They also used an \"{O}pik-style impact probability calculation
which has been shown to produce unexpected population analyses (e.g., the P.M. ratio calculation \citep{Wisdom:2020}).
Overall, our model reinforces the notion of inner-belt dominance across the extended size range of 100s of metres to centimetre diameter NEOs.

The generally agreed upon explanation for the inner-belt dominance for sub-kilometre asteroids is the greater mobility of material in the main asteroid belt and strength of these escape routes.
The primary mechanism for this mobility is the Yarkovsky effect, a thermophysical effect producing a net force either inwards or outwards from the Sun for prograde or retrograde rotating asteroids, respectively \citep{Farinella:1998, MorbidelliGladman:1998, Bottke:2002_asteroids3}.
The magnitude of the effect depends on the asteroid's size, distance from the Sun, and physical properties of the body.
The change in semi-major axis over time for a 1~m diameter body in the main belt ranges from 0.001 -- 0.01~au My$^{-1}$ \citep{Bottke:2002_asteroids3,Bottke:2006}.
This drift rate causes asteroids to `jump' across weak resonances until they reach the stronger \nusix and 3:1J regions, which are very efficient at evolving asteroids onto Earth-crossing orbits in comparison to outer-belt resonances - a factor of 20 in efficiency between \nusix, and 5:2J for a collisional lifetime of 3 Myr \citep{MorbidelliGladman:1998}. 
This phenomenon is evident by the change from outer- to inner-belt dominance among km and sub-kilometre asteroids in G18 and NEOMOD models, and it appears to continue regulating the metre population.

\subsection{Outer-belt Contribution}

Our model predicts a non-negligible contribution from the 5:2J and 2:1J outer-belt sources, which is not predicted for NEOs of 10~m diameter by previous models.
The only outer-belt contribution required for G18 at H=25 is JFC, only 8:3J by NEOMOD, and NEOMOD2 find no significant ($>$2\%) outer-belt source contribution for H=28. 
We propose several explanations for this.

Firstly, though Yarkovsky mobility still explains the overall inner-belt dominance, we see a greater outer-belt signal than comparative models because of a \textit{reduced} mobility of sub-metre meteoroids relative to their 10~m counterparts, leading to a lower proportion of meteoroids jumping the outer-belt resonances. 
1~m meteoroids can have a reduced average Yarkovsky drift over 1~Myr compared to 10~m asteroids, as much as a factor of 10, because these objects either collide or disrupt on short timescales which interrupts their migration towards the strong \nusix and 3:1J resonances \citep{Bottke:2006}.
The effectiveness of the Yarkovsky force also diminishes when the thermal penetration depth of solar heating exceeds the body’s diameter. 
Such reductions apply only to high-conductivity surfaces however, i.e., bare rock, while for dusty or porous bodies the strength of Yarkovsky, and therefore the semi-major axis drift rate, is roughly constant for asteroids 10~m and 1~m in size.

Secondly, the Yarkovsky-O’Keefe-Radzievskii-Paddack (YORP) effect is another thermal re-radiation interaction which reduces the flux of small asteroids entering the escape routes \citep{MorbidelliVokrouhlicky:2003, Granvik:2017} by altering an asteroid's obliquity and spin rate; interrupting the semi-major axis drift at shortening timescales $\propto$ radius$^2$.
YORP combined with Yarkovsky could reduce asteroid drift rates as a function of decreasing size for asteroids $\leq$1~m and decrease their ability to jump over the weak resonances, thereby reintroducing the contribution from the weaker, outer-belt resonances 5:2J and 2:1J.

A third explanation is a greater quantity of cm-m debris near outer-belt resonances, as this region is dominated by hydrated, friable, carbonaceous bodies \citep{DeMeoCarry:2013,Demeo:2014} which are more prone to fracturing during impacts or thermally fragmenting \citep{Molaro:2020,Delbo:2022,Shober:2025}. 

Our model interestingly predicts a negligible contribution from these sources for much smaller debris ($<$ 200~g or $\lesssim$ 5~cm), which we interpret as a relative increase in material coming from other sources (Fig.~\ref{fig:alphas}) rather than a decrease in outer-belt material.
Additional studies on the strength of the Yarkovsky and YORP effects, as well as sub-metre asteroid strength, would clarify the drivers of the outer-belt signal.

\subsection{Increase in Jupiter Family Comet Source Contribution}

As we examine the smallest meteoroid sizes in the dataset, we have a much greater contribution from $R_{JFC}$ within our model (Fig.~\ref{fig:alphas}). The interpretation of the JFC source here, however, must be taken with caution. JFCs are short-period, low-inclination comets largely deriving from the Kuiper belt and scattered disk as a result of a series of chaotic close encounters with Neptune and Jupiter \citep{Fernandez:1980,Levison:1994,Duncan:1997,Duncan:2004,DiSisto:2009,Nesvorny:2017}. 
Their orbital evolution is characterised by their close encounters with the gas giant, which cause large chaotic changes in the orbits over thousand-year timescales \citep{Tancredi:1995,Levison:1997,DiSisto:2009,Nesvorny:2017}.
$R_{JFC}$ was created by augmenting the statistics of comets within the model of \citet{Nesvorny:2017} that reach below $q<$1.3~au and $a<$4.5~au (see \citealp{Nesvorny:2023} for a full description of $R_{JFC}$). 
In this framework, any $R_{\rm JFC}$ contribution at cm–m sizes need not imply a purely cometary (volatile-rich) physical origin, nor a dynamical coupling with Jupiter; it primarily reflects where meteoroids reside in $(a,e,i)$ space.

Meteoroids on JFC-like orbits are predominantly stable \citep{Shober:2021,Shober:2024}, despite the fact that non-gravitational forces (PR-drag, Yarkovsky, etc.) are not strong enough to decouple these meteoroids from Jupiter. 
We therefore expect our model to be predicting a source of \textit{stable} JFC meteoroids, for which we present two plausible explanations: (1) cm-m JFC debris is produced during cometary splitting events of low-$q$ comets that are already dynamically stable, or (2) asteroidal material from an outer main-belt source has diffused onto JFC-like orbits.

Since the Earth is located within the inner solar system, the JFC population it samples is not very representative. Only about one third of JFCs become part of the `visible population' ($q<2.5$\,au), and only one fifth reach Earth-crossing orbits \citep{Levison:1997,Fernandez:2002, Horner:2004}. Despite the average dynamical lifetimes of JFCs being $\sim1.5\times10^{5}$\,yr, ones that reach the inner solar system only spend $\sim10^3$\,yr of that time $<$2.5\,au \citep{Duncan:1988,Levison:1994}. 
This limited dynamic lifetime results from both the frequent close encounters with Jupiter and the predicted short physical lifetimes of small JFCs in the inner solar system
($10^3-10^4$\,yrs; \citealp{Levison:1997,Fernandez:2002,DiSisto:2009}). 
Consequently, a JFC which entered into some stable state could break down, split, and produce cm-m sized debris \citep{Boehnhardt:2004, Jenniskens:2008_rev, Jenniskens:2008, Fernandez:2009} along the stable orbit.
This hypothesis is potentially supported by the observation that nearly all JFC-like (2$<T_J<$3) meteor showers (e.g., $\alpha$-Capricornids, Southern $\delta$-Aquariids, etc.) are more stable than the local sporadic background (Fig.~17 within \citealp{Shober:2024}) and the specific example of the Taurid complex and its primary parent comet 2P/Encke which is decoupled from Jupiter.
The only cometary shower that suffers frequent close encounters with Jupiter is the October Draconids - associated with JFC 21P/Giacobini-Zinner - which is consistent with their observed low-density, fragile meteoroids \citep{Borovicka:2007,Shober:2024}. 
Although possible, we consider this explanation unlikely, as JFCs entering onto more stable trajectories decoupled from Jupiter are very rare \citep{Fernandez:2002}.
\citet{Hsieh:2016} found 0.1–1\% of objects transferred from the JFC region are expected to evolve onto detached orbits, a value 4–40 times lower than the observed rate of stable near-Earth comets \citep{Fernandez:2015, Shober:2024}.
A large supply of cm-m cometary debris would also be inconsistent with Zodiacal Cloud (ZC) models. They show JFCs supply the majority of zodiacal dust through cometary splitting events \citep{Nesvorny:2010,Nesvorny:2011,Rigley:2022}, indicating the physical lifetimes of this cm-m JFC debris must be limited such that it eventually produces $<100\,\mu$\,m dust.

This widespread stability amongst JFC-like cometary showers and the uniqueness of the dynamic and physical properties of the October Draconids also points towards another explanation, asteroidal cross-contamination. 
The main asteroid belt is a much larger, closer, and more stable source that could be feeding these comet-like orbits with carbonaceous material.
Several studies of asteroidal contamination have found that outward diffusion of outer main-belt asteroids is capable of explaining the asteroids on cometary orbits (ACOs) that move on stable orbits \citep{Fernandez:2002,Fernandez:2015,Shober:2020,Hsieh:2020,Shober:2024}. 
Infrared catalogues show $\sim$20\% of ACOs have asteroid-like low albedos ($p_V>0.1$) \citep{Kim:2014}. 
Also, comet-like spectra dominate only for $T_J\lesssim2.8$; the $2.8<T_J\le3.1$ interval is a mixed regime with primitive and silicate objects in comparable proportions, with the comet-like fraction increasing toward fainter absolute magnitudes \citep{Simion:2021}. 
Based on the arguments above, the main-belt origin seems to be the most likely of these two explanations for the three-fold $R_{JFC}$ increase from H=39.125 ($\sim$5 cm) to H=41.5 ($\sim$1 cm). 
Further work examining the physical breakdown of material from different sources (JFC, carbonaceous, etc.) using thermophysical models is needed to better understand the sources of small cm-scale meteoroids.

\subsection{Physical Processes} \label{sec:phyical_processes}

We indirectly observe the decreasing collisional lifetime with size, which was only possible with the size-dependent analysis used here.
We reaffirm the range 1 -- 3~Myr is indeed correct for these sizes, with 3~Myr applying down to objects as small as approximately 0.6~kg (7~cm assuming a density of 3500~kgm$^{-3}$) and transitioning to 1~Myr for smaller objects.
This is evident in the relative performance of the collisionally depleted models over various size brackets (Figure~\ref{fig:heatmap}), as well as the match in slope of the semi-major axis distributions between calibration fireballs and the model output at low ($\leq$1\,au) values, which visibly changes with decreasing meteoroid size/collisional lifetime (Figure~\ref{fig:orbitaldistributions}).
Previous application of main-belt collision models to near-Earth space estimates a collisional half-life of 1 -- 3~Myr for 1~cm -- 10~cm \citep{Nesvorny:2024NEOMOD2}, which matches chondritic fireball entry velocities (3~Myr; \citealp{MorbidelliGladman:1998}) though is below the 6~Myr predicted for 10~g meteoroids of both asteroidal and cometary origin (\citealp{Soja:2019}; Fig 1).
As meteoroid sizes decrease, the physical lifetime is expected to continue decreasing until a minimum at around 0.1~g \citep{Jenniskens:2024}.
A different implementation of this type of modelling would have to be considered for $<$10~g. Collisions with the Zodiacal Cloud begin to dominate at smaller masses, and these have a different orbit dependency than main-belt collisions \citep{Pokorny:2024}.
Our estimates, and the model fit, are global averages of a diverse set of objects, so we do not make strong claims as to the exact meteoroid size of the transition from 3\,Myr to 1\,Myr.
This was made clear when the crossover index shifted between models fitted to the calibration data divided into two groups 
(see Appendix~\ref{sec:half_data_results}). 
Our model demonstrates NEOs across many sizes are compliant with our current understanding of collisions in the Solar System.

Thermal effects are not necessary to describe the orbital distribution of Earth-impacting meteoroids ranging from 1~cm to 0.5~m to the resolution of the data in this study.
The low perihelion disruption model was not preferred over the collisional model (Fig.~\ref{fig:heatmap}), and in some instances performed worse than the base model.
It is clear the removal of highly eccentric orbits (Fig.~\ref{fig:altered_sources}; the result of removing simulation particles reaching $q\leq$0.4\,au) provides a poor fit to the fireball data.
It is likely the extrapolation of the critical distance to $q$=0.4~au following the trend in \citet{Granvik:2016}, which has only been measured for NEOs $\ge$50~m, was not appropriate.
While telescopic surveys have found meteoroid orbits to be consistent with the $q$=0.4~au disruption \citep{Lue:2019}, studies of meteorites and near-Sun dwell times indicate some meteoroids do spend time near the Sun, possibly hinting that the meteorite precursors are fragments of low-$q$ super-catastrophic destruction \citep{Toliou:2021}.
We also conclude that near-Sun approaches do not appear to destroy the meteorite precursors, consistent with the notion that meteoroids are often monolithic \citep{Borovicka:2016, Kareta:2024_2022WJ1} and withstand greater thermal stresses than rubble pile asteroids.
New thermal effects would need to be considered for objects smaller than those modelled here, as it has been observed that lifetimes of primarily comet-derived material $\leq$1~cm have a $q$ dependency \citep{Jenniskens:2024}.
Further modelling of the combination of collisional and thermal processes on the cm- to m-sized NEO population is required to determine the extent to which these phenomena affect the smallest NEOs and connect to the depletion and creation of asteroids, meteoroids, sporadic cometary material, and dust particles.

\subsection{Lack of Hungarias}

The Hungaria region ($a\sim1.8$\,au and $i\sim15-20\degr$) is a significant source of large NEOs in the G18 model ($\sim$20\% for H=25), and to a lesser extent in NEOMOD2 ($<$2.9\% for H=28).
The proportion of NEOs originating from the Hungaria region in the G18 model varies considerably with asteroid size, but tends to be magnified by their long dynamical lifetimes, particularly at both the large (H=17, $\sim12$\%) and small (H=24, $\sim$20\%) ends of the populations.
Hungarians are made up mainly of Xe-type, S-type, and some C-type asteroids \citep{Lucas:2017}.
The largest of the group -- 434 Hungaria -- is an Xe-type \citep{BusBinzel:2002}, and also is the leading member of a collisional family \citep{Williams:1992, Lemaitre:1994}.

Based on G18 and NEOMOD2, one would expect that Xe-type material would represent a prominent source of meteorites on Earth.
It has been proposed that Xe-type asteroids are associated with enstatite achondrites (aka aubrites), based on both albedo and spectrum indications \citep{Gaffey:1992} and dynamics matching meteorite ages \citep{Cuk:2014}.
However, these enstatite achondrites only represent $\sim1$\% of meteorite falls (cf. Metbull), so there is a mismatch.
Our model confirms this mismatch: the Hungaria source region is not a preferred contributor of Earth-impacting NEOs in our size range (see Sec. \ref{sec:best_fit_model}).

The Xe-type Hungaria collisional family is a relatively old asteroid family; $207\pm65$\,Myr old \citep{Spoto:2015}.
As the Yarkovsky drift rates are high at meteoroid sizes $\pm [0.01–0.001]$\,AU\,My$^{-1}$, these objects can reach nearby resonances as early as a few million years after the collision, rapidly 
depleting the region of small debris and feeding near-Earth space.
Following the Hungaria family forming event, it is probable that small Xe-type impactors could have been a major source of meteorites then.
However, this increase would have dissipated a long time ago, within tens to a hundred million years of the family-forming event.
We still occasionally get small Xe-type meteorites landing on Earth, but these are likely the result of more recent collisions.
Interestingly, asteroid 2024 BX1, which delivered the Ribbeck enstatite achondrite \citep{Bischoff:2024}, has a negligible chance of having come from the Hungaria source region when using NEOMOD (\nusix 88.1\%, 3:1J 10.9\%, 5:2J 0.2\%, Inner Weak 0.7\%, \textbf{Hungarias 0.1\%}).
Assuming the Xe-type = enstatite achondrite connection holds, this could mean enstatite achondrites have another source in the belt, distinct from the Hungarian collisional family.

\subsection{Earth-Impacting Population Predictions}\label{sec:prediction}
\begin{figure}
    \centering
    \includegraphics[width=1\textwidth]{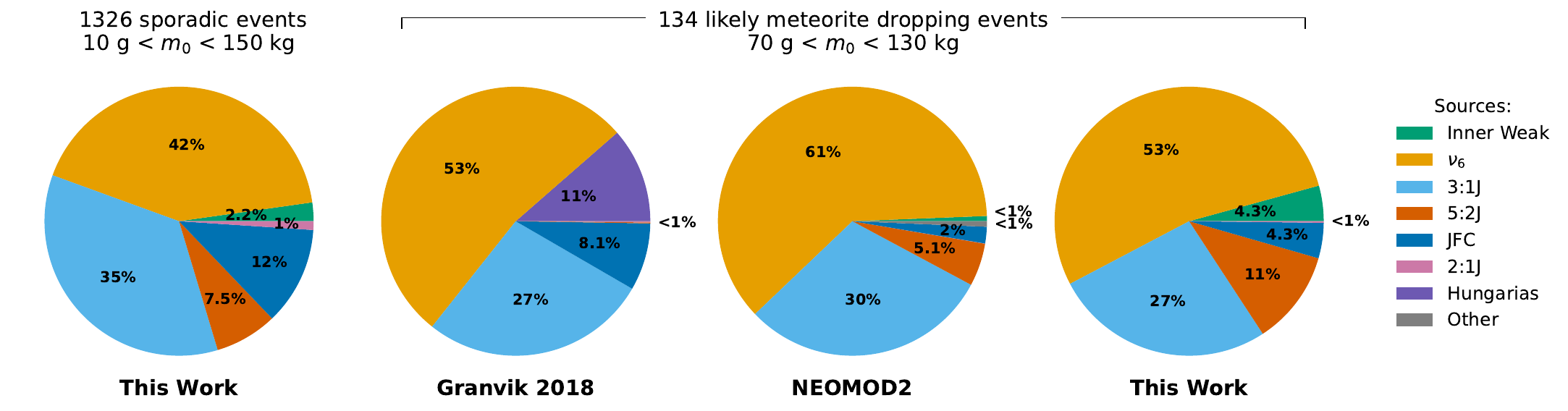}
    \caption{The sources of GFO meteoroids predicted by our nominal, best-fit model the Granvik 2018 model and NEOMOD2. Each event contributes its fractional source probabilities (e.g., 80\% \nusix, 10\% 3:1J, 10\% Inner Weak), and the chart shows the sum of these fractions, normalised over all events in order to evaluate the population as a whole. The first subgroups contains all sporadic GFO events with initial meteoroid masses $m_0>$10~g from 2014-2024. The second subgroup restricts to possible meteorite droppers following the likely meteorite dropper criteria of \citet{Sansom:2019}. The lowest initial mass of 70~g for this subset is a consequence of the dropper selection criteria.}
    \label{fig:piecharts}
\end{figure}

For each GFO event's orbital elements $(a,e,i,H)$, we calculated the fractional contribution to those coordinates from each of the sources as predicted by our model using the median $\alpha_s$ values reported in Table~\ref{tab:alphas} and the original \Rs binning.
We show the sum of the fractional probabilities for all events (as opposed to counting the maximum probability per event) to have a population-wide overview of source probabilities in Figure~\ref{fig:piecharts}.
The sources that predominantly feed the Earth-impacting population are the \nusix and 3:1J, contributing to 74\% of the sporadic GFO fireballs caused by meteoroid $>$10~grams.
When we restrict our sample to likely meteorite-dropping events, described in Appendix~\ref{sec:droppers}, we expect 83\% of the meteoroids to originate from the inner main belt (\nusix: 53\%, 3:1J: 26\%, Inner Weak: 4\%).
This is slightly less than NEOMOD's 91\% prediction for simulated Earth impactors (\nusix: 57.6\%, 3:1J: 29.7\%, Inner Weak: 3.8\% -- calculated from their $\alpha_j(H=25)$ and $f_{imp}$ parameters) and on par with the 88\% prediction for metre-scale Earth impactors from \citet{Brown:2016} using the \citet{Bottke:2002} model (intermediate source Mars crossing: 26\%, \nusix: 48\%, and 3:1J: 14\%).
When feeding our likely meteorite-dropping GFO events to the NEOMOD2 model with a pre-set size H=28, we can see the predictions are similar to this model's (Fig.~\ref{fig:piecharts}). 
Both models have twice as many bodies arriving from the \nusix compared to the 3:1J resonance.
Our model predicts a greater chance of arriving from each of the remaining sources -- Inner Weak, 5:2J, and JFC -- by about a factor of two. 
\citet{Brown:2016} using \citet{Bottke:2002} predict 8\% outer-belt and 4\% JFC contribution to metre-scale impactors, where 10/55 analysed events are also confirmed meteorite droppers.
All models allocate a negligible chance of the likely droppers arriving from the 2:1J resonance, or any other minor sources.
These predictions are not necessarily an indication of the amount of pristine material that is reaching us, however.
These NEO models do not provide any information about the type of material arriving from each source (besides from NEOMOD3, which includes asteroid albedo).
As a result, when using them to predict which source meteorite-droppers come from, we are assuming that material from all sources has an equal probability of surviving atmospheric entry.
In reality, this is not the case: if a likely meteorite-dropper has an equal chance of \textit{dynamically} arriving from an inner- and outer-belt source, it is more likely to be from the inner-belt, which is the primary source of generally stronger ordinary chondrite meteorites, whereas the outer-belt is the main source of weaker carbonaceous chondrites \citep{Binzel:2019, Jenniskens:2025, Broz:2024_A&A, Broz:2024Nature}. 
Until we can quantify for survivability as a function of source region, we must acknowledge these probabilities are biased.

When we compare the maximum source probability for each event, our model's results are nearly identical to the NEOMOD2 predictions for the GFO likely meteorite-droppers.
This is somewhat expected considering we are using the \Rs derived from the same numerical simulations, and the source strength for our dominant sources at H=28 and H=34.5 are nearly the same.
Even though collisions were included in our model, it was implemented in the same way for all sources, so they would not significantly change the source ratios per orbital bins in comparison to a collisionless model.
This speaks to the strength of the extrapolation of the NEOMOD models and its application to the near-Earth meteoroid environment.

\subsection{Link to Asteroid Families}
The attempts to connect meteorite types to asteroid families is a complex task; one informed through asteroid observation, classification \citep{DeMeo:2022} and association with meteorite analogues \citep{Battle:2025, McGraw:2025}, modelling of NEO delivery \citep{Broz:2024Nature, Marsset:2024, Broz:2024_A&A}, and observing meteorite pre-impact orbits to associate them with specific escape routes \citep{Jenniskens:2025} or with known NEOs \citep{Borovicka:2015,Shober:2025_streams,Shober:2025_Dn_meteorite_NEA_pairs}. 

From the study of asteroid family ages and expected size-frequency distributions, \citet{Broz:2024Nature} model the relative contribution of the Phocaea family to NEOs as decreasing with size, from 8.7\% at 1-km sized NEOs to a negligible contribution (0.7\%) for 1-m sized impacting meteoroids.
We also see no significant contribution from Phocaea in our model for near-Earth meteoroids ($<$1~m).
\citet{Jenniskens:2025} also do not think Phocaeas supplies many H-type chondrites - the meteorite type associated with the family - from the lack of orbital meteorites seen to originate from highly inclined inner-belt orbits still found in the 3:1 resonance. 
The residing orbits of small H-type candidate small ($<$300~m) NEOs are also dynamically consistent with \nusix or 3:1J origin, with none observed thus far connected to the Phocaeas \citep{Sanchez:2024}.

Additionally, there is a debate as to whether L-type chondrites come from the Massalia family \citep{Marsset:2024} or the Nysa/Hertha family \citep{Jenniskens:2025}.
Both are young asteroid families at low inclination (1.4$^{\circ}$ and 2.4$^{\circ}$) between the \nusix and 3:1J resonances.
High resolution simulations of these families evolving to Earth-crossing orbits could show distinctive orbital distributions.
By comparing the resulting near-Earth orbital distributions with observed data - as demonstrated here by our low inclination testing - it could become possible to assess which family is the major contributor of L-type chondrites, i.e., alike to METEOMOD\footnote{https://sirrah.troja.mff.cuni.cz/~mira/meteomod/meteomod.html}.

\subsection{Implications for the Decameter NEOs}
Work to investigate the population of decameter NEOs ($\sim$10~m diameter) is most difficult due to the lack of observations in this size range \citep{ChowBrown:2025}, however there are several inferences we can make using this model.
1) The inner main-belt, which predominantly supplies NEOs $>$10~m in diameter \citep{Bottke:2002, Granvik:2018, Nesvorny:2023} and NEO meteoroids $<$ 1~m (this work), is the primary source of the decametre NEO population.
2) The low-perihelion disruption process, as currently implemented in \citet{Granvik:2018, Nesvorny:2023} and extrapolated from \citet{Granvik:2016}, ceases to apply for NEOs $\leq$10 m.
Such a change could alter the orbital distribution and size-frequency distribution of decametre NEOs in ways that are yet understood.
Recent investigation into the perihelion history of NEOs shows that meteorite precursors likely dwell at low perihelion distances \citep{Toliou:2021}, and that low perihelia preferentially remove weaker material from the Earth-impacting $<$1~m population \citep{Shober:2025}. 
Since previous studies have already demonstrated the role of solar processing in NEO evolution, continued investigation of size-dependent thermal alteration processes is particularly relevant for understanding the hazardous Earth-impacting population.

\section{Model Limitations}\label{sec:limitations}
We highlight several limitations of this model for future users.
First, the model is calibrated only to meteoroids of the dynamical classes Aten and Apollo, as no Atira or Amor objects cross Earth's orbit. 
Second, the Jupiter Family Comet source region is used \textit{as is} from NEOMOD. This does not account for cometary fragmentation or dispersion of cm-sized particles. Improved input distributions bridging between those for mm-sized particles \citep{Nesvorny:2010, Nesvorny:2011} and that for comets \citep{Nesvorny:2017} would help confirm our JFC contribution interpretation. 
Third, the model is fitted in $log(m)$ rather than absolute magnitude H. We used a single conversion to H based on ordinary chondrites/S-complex asteroids. Fitting a model using informed physical properties per meteoroid, such as using fireballs with measured spectroscopy \citep{Vojavek:2015}, would be more accurate when translating the results into H.
Finally, the model uses meteoroids biased in size as we do not yet have a method to debias the GFO data in terms of absolute numbers. Model interpretation must be performed over discrete size brackets, as done here (Figure \ref{fig:orbitaldistributions}), and meteoroid orbit generation must also be in discrete size bins unless the user imposes their own size-frequency distribution for the flux of Earth-impactors.

\section{Model Application}\label{sec:application}
We promote the use of this model to investigate near-Earth objects in the size range 1~cm to 1~m.
We provide the resources to a) query the source region probability of orbits and b) generate meteoroid orbital distributions \citep{modelcode:2025}.
We suggest that previous calculations of the origin of meteoroids using NEOMOD do not need to be repeated, as the fractional contributions of sources are the same for the most dominant sources (Section \ref{sec:prediction}).
However, we do note a difference between predictions by this model/NEOMOD and predictions by the G18 model, such as in \citet{GranvikBrown:2018} or \citet{ChowBrown:2025}.
The greatest strength of sampling orbits from this model is the difference in orbital element distributions between this model (1~cm - 1~m) and all other NEO models ($>$ 10~m) due to the inclusion of size-dependent collisions, removing highly evolved orbits.

\section{Conclusion} \label{sec:conclusion}
This work utilised the unique data set of 1,202 sporadic meteoroids of mass $>$10 grams observed by the Global Fireball Observatory to create a debiased model for near-Earth objects. 
This model covers meteoroids 10~g -- 150~kg (approximately 1~cm -- 0.5~m) in size dynamically evolving from the main asteroid belt onto Earth-crossing orbits, experiencing collisions while still coupled to the main belt to deplete the population of highly evolved orbits, e.g., low semi-major axis and high inclination orbits.
The major findings of this model are:

\begin{itemize}
    \item  We observe the global decrease in collisional half-life from 3~Myr to 1~Myr for the meteoroid population, with the transition at around 0.6~kg/7~cm. This aligns with current understanding and estimates of collision processes.
    \item The model shows an uptick in Jupiter Family Comet type orbits feeding the near-Earth meteoroids for sizes smaller than 5~cm (250~g). 
\end{itemize}

This model, along with previous NEO models, borders the decameter size range - a population that is difficult to observe and potentially hazardous to civilians from airbursts in the atmosphere. 
Inferences we can make about the decameter population are:

\begin{itemize}
    \item The inner main belt (\nusix secular resonance, 3:1J, and various weak mean motion resonances 2.1 -- 2.5~au) continues to dominate feeding the near-Earth meteoroid population across 10~m to 1~m in diameter objects. 
    \item There is a change in the near-Sun disruption trends for asteroids smaller than 10~m, as \citet{Granvik:2016}'s low perihelion disruption law extrapolated to 0.4~au does not directly apply to 1~m diameter meteoroids. 
\end{itemize}

Overall, we find strong motivation to continue investigating and modelling decimetre meteoroids, particularly around the physical processing of these bodies near the Sun or the low inclination delivery into resonances by young asteroid families.
We therefore encourage the community to continue observing fireballs in the night sky.

\section*{Acknowledgements}
The Desert Fireball Network and Global Fireball Observatory programs have been funded by the Australian Research Council as part of the Australian Discovery Project scheme (DP170102529, DP200102073, DP230100301), the Linkage Infrastructure, Equipment and Facilities scheme (LE170100106), and receives institutional support from Curtin University. 
S. E. Deam acknowledges the support of an Australian Government Research Training Program Scholarship.
Luke Daly thanks STFC (ST/Y004817/1, ST/T002328/1, ST/W001128/1, and ST/V000799/1) for support.
C D K Herd acknowledges the support of the Canadian Space Agency FAST Grants 18FAALBB20 and 21FAALBB17, and Natural Sciences and Engineering Research Council of Canada Grant RGPIN-2018-04902.
PB was supported by NASA co-operative agreement 80NSSC24M0060.


\vspace{5mm}
\facilities{}

\software{Rebound \citep{REBOUND}, pymultinest \citep{Buchner:2014}, Matplotlib \citep{matplotlib}, NumPy \citep{numpy}, pandas \citep{pandas:2025}, SciPy \citep{scipy} 
          }

\appendix

\section{Data: Quality Cuts and Fireball Selection} \label{sec:subset}

\textbf{Capture geometry:} Meteoroids were only included in the model calibration if the best convergence angle of the data captured was above $10^{\circ}$. 
The convergence angle describes the angle between the image planes of two camera observations. 
As a single image can only localise a fireball within the two-dimensional image plane, the observation of a second camera provides a constraint on the distance to the fireball in the dimension perpendicular to the first image plane. 
This has the highest precision when the two image planes are perpendicular: a convergence angle of 90$^{\circ}$. 
Alternatively, if the convergence angle is low, then the two image uncertainties compound and the triangulation is less precise. 
The location of a meteoroid's entry over a local group of cameras is random, so the convergence angles are not correlated with any particular meteoroid orbit type.
Cameras within the GFO network in Australia are typically spaced 100 -- 150 kilometres apart for optimal sky coverage and triangulation. 
As a result, a minority ($\sim$8\%) of the detections have a convergence angle $<10^{\circ}$ and are disregarded here.

\textbf{Initial mass:} Only meteoroids with a pre-atmospheric mass of at least 10 grams were included in this analysis. 
This is to ensure we are working with meteoroids with a sufficient mass to be reliably detected by the GFO cameras, which have a limiting magnitude of $\sim$0.5 \citep{Howie:2017:howtobuild}.
The lower limit of each meteoroid's masses were determined from the dynamics of the luminous flight of the fireball \citep{Sansom:2015}. 
For the small number ($<$20) of low-deceleration, non-converging events, the mass was instead estimated using photometry by integrating the ablation equations of \citet{Ceplecha:1998} for the GFO cameras' green channel light curves using the luminous efficiency of \citet{Revelle&Ceplecha:2001}, as done for the European Fireball Network in \citet{Borovicka:2022paper2}.

\textbf{Meteor shower association:} As our intention is to model the sporadic meteoroid population, fireballs created from meteor shower material were not included when calibrating the model. 
To identify meteor stream events in the GFO dataset, an orbit similarity criterion was computed between all GFO events and the established showers or single shower entries (status flags 1 and 2 in the database) of the IAU Meteor Data Centre\footnote{downloaded 23 August 2023 from https://www.iaumeteordatacenter.org/} \citep{Hajdukova:2023}.
An orbit similarity criterion is a metric of orbit similarity, first described by \citet{SouthworthHawkins:1963}, and used to associate meteors together as showers or link them to parent bodies.
Here we used the \DN criterion \citep{Valsecchi:1999} as it is the most appropriate for our on-sky, geocentric measurements \citep{Moorhead:2016}. 
We associated any GFO fireball with a stream if \DN was below the cut-off retrieval percentages of \citet{Galligan:2001} which we show in Table~\ref{tab:DN}. 
The more accommodating 90\% retrieval cut-off was chosen for 14 major showers (Perseids, Geminids, Orionids, Northern Taurids, Southern Taurids, Northern $\delta$-Aquariids, Southern $\delta$-Aquariids, Quadrantids, $\alpha$-Capricornids, Leonids, $\sigma$-Hydrids, $\eta$-Aquariids, Comae Berenicids, November Orionids) as we were expecting to have many fireball events from these showers in our dataset which we wished to identify and remove, while the more constrained 70\% retrieval was used for all other showers.

\begin{deluxetable}{ccc}
\tablecaption{The \DN cut-off values for fractional recovery of stream members from \citet{Galligan:2001} used to associate GFO events with known meteor showers} \label{tab:DN}
\tablewidth{1.0\textwidth}
\tablehead{
\multicolumn2c{\DN cut-off} & \colhead{Inclination}  \\
\colhead{90\%} & \colhead{70\%}  & \colhead{} }
\startdata
0.11 & 0.08 & $i<10^{\circ}$ \\
0.14 & 0.09 & $10^{\circ}\leq i <90^{\circ}$ \\
0.26 & 0.17 & $i\geq90^{\circ}$ \\
\enddata
\end{deluxetable}

\section{Data: Correcting for Observational Biases} \label{sec:bias}

\textbf{Detection Efficiency:} First, we address the in-atmospheric detection bias.
Several factors influence the likelihood of a meteor or fireball being detected, including the signal's strength, duration, and the event's location in the sky as viewed by the detector.
Several works have independently addressed these biases for both radar and photometric systems, presenting bias-corrected meteor populations \citep{Galligan&Baggaley:2005, Campbell-Brown:2008, Jenniskens:2016}.
Here, we aim to correct for our instrument limiting magnitude, which manifests as a bias against detecting smaller, slower meteoroids.
The light from a fireball is created by the conversion of kinetic energy to luminous energy; described analytically in \citet{Ceplecha:1998}. As we do not know the meteoroid's composition, shape, or cross-sectional area etc., we combined their equations 1, 2, and 28 to isolate the dependence of luminosity on mass and velocity

\begin{equation}
    I \propto v^5 m^{2/3} \label{eq:detecteff}
\end{equation}

where 
$I$ represents the relative luminosity of a fireball, $v$ is the initial velocity of a meteoroid in km/s, and $m$ is the initial mass of a meteoroid in kg.
We expect every fireball above some brightness to be detected by the GFO cameras and pipeline, while fireballs dimmer than this to have a lower probability of being detected.
We assume that such probability scales linearly with the decrease in luminosity. 
We therefore weight each event by a detection efficiency factor $\epsilon$ (Eq. \ref{eq:epsilon}) if the luminosity $I$ falls below some threshold $\kappa$, while fireballs brighter than $\kappa$ have no weighting applied i.e. $\epsilon$ is 1.

\begin{equation} \label{eq:epsilon}
\epsilon(m, v) =
\begin{cases} 
  \frac{I}{\kappa}  & \text{if } I < \kappa, \\
  1               & \text{if } I \geq \kappa.
\end{cases}
\end{equation}

\begin{figure}
    \centering
    \includegraphics[width=0.5\linewidth]{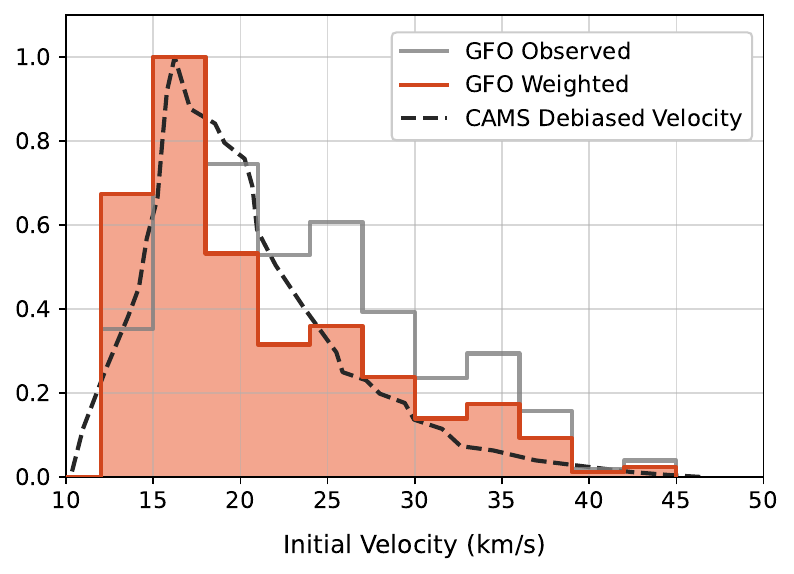}
    \caption{The top-of-the-atmosphere velocity distributions for anti-helion, sporadic meteors from GFO and CAMS (\citealp{Jenniskens:2016}; digitised from Fig 16) for initial masses of $<$ 30 grams, normalised by their maximum value. The $1/\epsilon$ weighting from Equation \ref{eq:epsilon} with $\kappa=9.5e4$ was applied to the GFO events to best reproduce the CAMS distribution.}
    \label{fig:CAMSvsGFO}
\end{figure}

\begin{figure}
    \centering
    \includegraphics[width=0.5\linewidth]{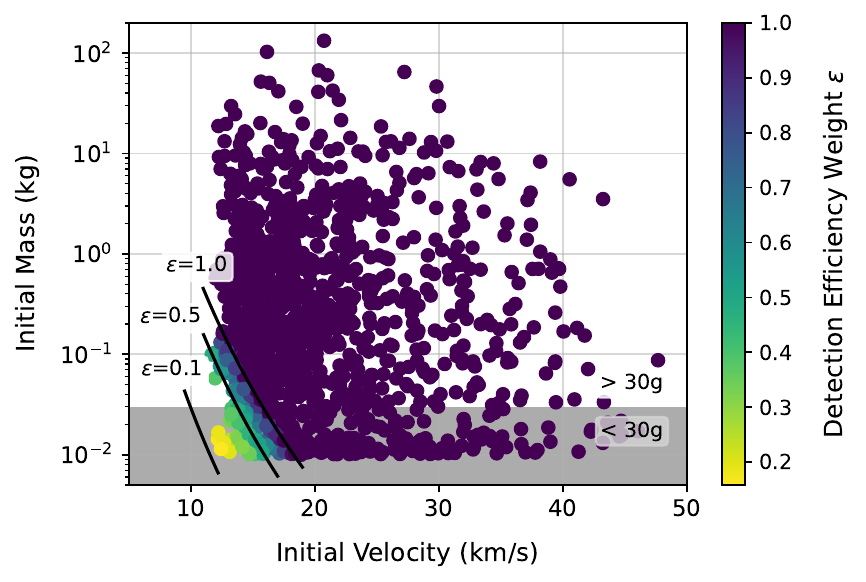}
    \caption{The velocity-mass dependent detection efficiency weighting. The 1,202 calibration meteoroids are coloured by $\epsilon$ (Eq. \ref{eq:epsilon}). Solid black lines show lines of constant weight. The majority (89\%) of events have a weighting of 1. The efficiency decreases with both mass and velocity, indicating that the smallest and slowest events will receive the greatest weighting. The grey background shows the $<$30~g mass fireballs which were used in the Figure~\ref{fig:CAMSvsGFO} comparison with CAMS data to determine $\kappa$.}
    \label{fig:detecteffsurface}
\end{figure}

The absolute value of the weighting is not important here, only the relative values between fireballs.
$\kappa$ was determined by comparing the entry velocities of the smallest GFO fireballs (10~g $<m_0<$ 30~g), weighted by the detection efficiency $\epsilon$ of Equation \ref{eq:epsilon}, against the all-data mass-weighted (up to 30~g) CAMS meteors \citep{Jenniskens:2016}. 
The match between the distributions, particularly for velocities $>$30~km/s, is best with $\kappa=9.5e4$ (Figure~\ref{fig:CAMSvsGFO}). 
The resulting $\epsilon(m,v)$ weighting, plotted in Figure~\ref{fig:detecteffsurface}, shows that 1073 (89\%) of the GFO meteoroids used in the model calibration are above the $\kappa$ threshold and therefore have $\epsilon$=1 (i.e. no weighting) and the remaining 129 events (11\%) are weighted by $1/\epsilon$.

\textbf{Earth Impact Probability:} The following section describes the efforts to debias the GFO meteoroids from the limitation of only observing objects that collide with the Earth, and not uniformly sampling the inner Solar System. 
We account for the meteoroid’s probability of impacting Earth and the gravitational focusing of the Earth. 
After comparing several probability calculation techniques, we decided to use the semi-analytical method of \citet{FuentesMunoz:2023} to calculate the relative Earth-impact probabilities for each GFO meteoroid. This method is preferred because, a) we know the osculating orbital elements with uncertainties for each meteoroid before impact, including a mean anomaly, b) the dynamical lifetimes of near-Earth objects are less than the long-term dynamics perturbations implemented in other impact probability calculations (e.g. \citet{VokrouhlickyPokornyNesvorny2012, PokornyVokrouhlicky2013}), and c) the numerical integrations of this method reduce the assumptions about the previous orbital propagation per meteoroid.

Orbital histories of GFO meteoroids were numerically integrated to evaluate their previous proximity to the Earth and calculate relative Earth impact probabilities.
For each of the 1,202 events, a cluster of 200 particles was injected into a \texttt{REBOUND} simulation populated with the Sun and the planets from Mercury to Neptune as well as Earth's Moon \citep{REBOUND}. 
Each meteoroid's latitude, longitude, altitude, and velocity vector (as measured in the Earth-Centered Earth Fixed coordinate frame) from the highest point of the triangulated fireball were used as initial conditions for the simulated particle.
The velocities for the 200 particles were sampled within the fireball measurement uncertainties.   
With appropriate coordinate transforms, the simulation was populated with these 200 particles 
, and integrated backwards for 100,000 years using the IAS15 integrator \citep{IAS15integrator} while recording the particles' orbits every 10 Earth years.

The Earth impacting probability of each particle was calculated from Equations 1 and 4 of \citet{FuentesMunoz:2023} (Equations \ref{eq:PMA} and \ref{eq:P} below). 
Equation \ref{eq:PMA} here calculates the probability, $P_{MA}$, of the mean anomaly of the Earth and the simulated particle aligning such that there is a close encounter, from the relative velocity between the Earth and particle, $U$, the orbital periods of the Earth and particle, $T_{\oplus}$ and $T_{p}$, the velocity vectors of the Earth and particle, $\textbf{v}_{\oplus}$ and $\textbf{v}_{p}$, and the Minimum Orbital Intersect Distance or MOID between the particle and Earth. 
The MOID was calculated for each recorded orbital configuration using the algorithm of \citet{Wisniowski2013}\footnote{accessed from \url{https://github.com/mkretlow/MOID.jl}}. 
$P_{MA}$ was then calculated for intervals when the MOID was below $d$ = 0.01 au, approximately Earth's Hill radius. 
The impact probability of a particle over the entire integration period $T$ was then calculated as $P$ in equation \ref{eq:P}, where $K_{\oplus}$ and $K_p$ are the orbital parameters of the Earth and particle respectively at time $t$. 
The average probability of the 200 particles was used as the impact probability for the fireball event. 
The final $P$ was relatively insensitive to both the $P_{MA}$ sampling interval (both 0.25 and 10 years were tested), and the simulation integration length (most particles per event converge toward the final $P$ within $T$=30,000 years). 
For a single GFO meteoroid, we report the relative Earth impact probability as the median $P$ value of the 200 simulated particles with a non-zero probability. 
Within a simulation, stochastic processes could result in the ejection of particles from the inner Solar System, resulting in some particles with $P=0$.

\begin{equation} \label{eq:PMA}
    P_{MA} = \frac{2Ud}{T_{\oplus} T_p | \textbf{v}_{\oplus} \times \textbf{v}_p|}\sqrt{1 - \frac{MOID^2}{d^2}}
\end{equation}

\begin{equation} \label{eq:P}
    P = \frac{1}{T} \int_{T} P_{MA}(d, K_{\oplus}, K_p) dt 
\end{equation}

For meteoroids that approach the Earth's Hill sphere, the effective collisional radius of the Earth depends on the meteoroid’s relative velocity because of the mutual gravity between the two objects.
Objects with a higher relative velocity have a smaller effective collisional radius and therefore a lower chance of impact \citep{Opik:1951}. 
We account for such relative gravitational focusing by including an additional weighting to each GFO meteoroid through Equation \ref{eq:gravfocus}, where $V_{\infty}$ is the velocity of the projectile as it first enters Earth’s gravitational influence.

\begin{equation} \label{eq:gravfocus}
    \left(1+\frac{1}{V_{\infty}^{2}}\right)
\end{equation}

\section{Methods: Model Parameters and Optimisation} \label{sec:modelparameters}

To create a model of the observed meteoroid population, each source region is multiplied by a strength factor, $\alpha_s(H)$, to scale the relative contributions of each meteoroid source. 
This implements the size-dependent contribution of a source to the NEO population, as previously established in the G18 and NEOMOD models. 
We model this size dependence with $m$ continuous, piecewise, linear functions.
For $m$ linear functions (or slopes), there are $m$+1 free parameters.
For example, $m$ = 1 is a single slope between $H$ = 34.5 and 41.5 described by two parameters $\alpha_1$ and $\alpha_2$, while $m$ = 2 contains 3 $\alpha$ parameters with an apex at $H_2 = 38$.
We varied the number of slopes to best fit the calibration dataset. 
The strength function for source $s$ is therefore a collection of piecewise functions outlined in equation \ref{eq:alpha(H)}, where $H_1$ and $H_{m+1}$ are the bounds of the $H$ axis.

\begin{equation}\label{eq:alpha(H)}
\alpha_s(H) =
\begin{cases}
    \alpha_{s,1}+(\alpha_{s,2}-\alpha_{s,1})(H-H_1) & \text{for } H_1 < H \leq H_2, \\
    \alpha_{s,2}+(\alpha_{s,3}-\alpha_{s,2})(H-H_2) & \text{for } H_2 < H \leq H_3, \\
    \vdots & \vdots \\
    \alpha_{s,m-1}+(\alpha_{s,m}-\alpha_{s,m-1})(H-H_{m-1}) & \text{for } H_{m-1} < H \leq H_m, \\
    \alpha_{s,m}+(\alpha_{s,m+1}-\alpha_{s,m})(H-H_m) & \text{for } H_{m} < H \leq H_{m+1}.
\end{cases}
\end{equation}

The contributions from the $n$ sources are added together under the constraint that the sum of the strength function for any size $H$ equals 1.

\begin{align} \label{eq:sum_a(H)}
    \sum_{s=1}^{n}\alpha_s(H)=1
\end{align}

The introduction of physical processes, e.g., collisional lifetimes, provided the opportunity to include multiple process types, $t$, within a single model.
This allowed for more freedom, for example, to adjust the collisional lifetime as a function of size.
The residence time distribution for source $s$ and process type $t$, $R_{s,t}$ was selected as a function of size, $H$, with a simple crossover at size $\delta$.
Note that it is possible for $\delta$ to equal the boundaries of the H axis, 34.5 or 41.5, which simply allocates a single type of process over the entire size range.

\begin{equation} \label{eq:crossover}
    t(H;\delta) = 
    \begin{cases}
    t_1 & \text{for } H < \delta, \\
    t_2 & \text{for } H \geq \delta.
    \end{cases}
\end{equation}

To summarise, the model fits for the strength factors $\alpha_{s,m+1}$ for $n$ number of main-belt sources $s$ and a chosen number of slopes $m$, with the crossover between process type $t_1$ and $t_2$ at index $\delta$. 
These parameters combine with the residence time distributions to make the model distribution $N(a,e,i,H)$ (Equation \ref{eq:model}), where $\alpha_s(H)$ is defined in Eq. \ref{eq:alpha(H)} and $t(H;\delta)$ is defined in Eq. \ref{eq:crossover}.
To be explicit, this model does not fit for the absolute flux of meteoroids in either near-Earth space or as Earth-impactors. 
The flux of the weighted GFO meteoroids was renormalised to the number of meteoroids it represented i.e. $\int_a\int_e\int_i\int_H N_{GFO}(a,e,i,H)=1202$ while $\int_a\int_e\int_i\int_H N(a,e,i,H)=28$ (the \Rs sum to 1 for each of he 28 $H$ bins).
We chose this because 1) scaling both $N$ and $N_{GFO}$ to match some size-frequency distribution $N(H)$ would not provide any new information, and 2) we do not yet have a method to debias our GFO data in absolute terms (also see Section \ref{sec:bias}).
We are therefore cautious to only investigate the model output over discrete size brackets in our interpretation and discussion, since the size frequency of the GFO meteoroids does not match that of the interplanetary space.
Despite this, the model is still perfectly capable of fitting for the best orbital distribution per size bracket, and diagnosing the size-dependent trends of the meteoroid population.

\begin{align} \label{eq:model}
    N(a,e,i,H) = \sum_{s=1}^{n}\alpha_s(H)R_{s, t(H;\delta)}(a,e,i)
\end{align}

The log-likelihood calculation uses a joint probability of a Poisson distribution, following the methods used in NEOMOD. For each bin $j$, the joint probability $P$ of predicting $n_j$ objects from the binned model output $N(a,e,i,H)$, for the number of expected objects $\lambda_j$ in the debiased dataset $N_{GFO}(a,e,i,H)$, is 

\begin{align} \label{eq:jointprob}
    P = \prod_j = \frac{\lambda_j^{n_j}exp(-\lambda_j)}{n_j!}
\end{align}

The log-likelihood is therefore

\begin{align} \label{eq:LL}
    \mathcal{L}=lnP=-\sum_j\lambda_j+\sum_jn_jln\lambda_j
\end{align}

where the first term is the same for every model run, and the second term is evaluated only over non-zero bins: $n_j\neq 0$ and $\lambda_j\neq 0$.
We also adopt the same prior function used in NEOMOD for our $\alpha_{s,m+1}$ to facilitate the constraint that they are not independent priors and must sum to 1 following Equation \ref{eq:sum_a(H)}. For sources $s=1$ to $s=n-1$, uniformly random parameters $X_s$ are generated and transformed using

\begin{align} \label{eq:priors1}
    \alpha_{s,m+1}=[1-(1-X_s)^{1/(n-s)}]\left(1-\sum_{k=1}^{s-1}\alpha_{k,m+1}\right)
\end{align}

and

\begin{align} \label{eq:priors2}
    \alpha_{n,m+1}=1-\sum_{k=1}^{n-1}\alpha_{k,m+1}
\end{align}

This ensures Equation \ref{eq:sum_a(H)} is fulfilled and \multinest fits each $\alpha$ independently. 
The prior for the crossover parameter, $\delta$, is uniform over the H axis (Tab.~\ref{tab:bins}). 
It ranges between 34.625 and 41.375 with the constrain of being an increment of 0.25 ($H$ axis bin width). 
Both $N(a,e,i,H)$ and $N_{GFO}(a,e,i,H)$ are given to \multinest along with the log-likelihood and prior functions to find the maximum log-likelihood between the two distributions. 
\multinest is a Bayesian inference technique sampler \citep{Feroz:2019}, and we specifically used the python interface, \texttt{pymultinest} \citep{Buchner:2014}.
The parameters $\alpha$ and $\delta$ are free parameters given to \multinest, whereas $n$, $s$, $m$, and $t$ are chosen before running a model fit.
The number of parameters \multinest fits for is therefore $n \times (m+1) + 1$.

\section{Results: Marginal Posterior Distributions}\label{sec:full_corner_plot}
\begin{figure}
    \centering
    \includegraphics[width=1\textwidth]{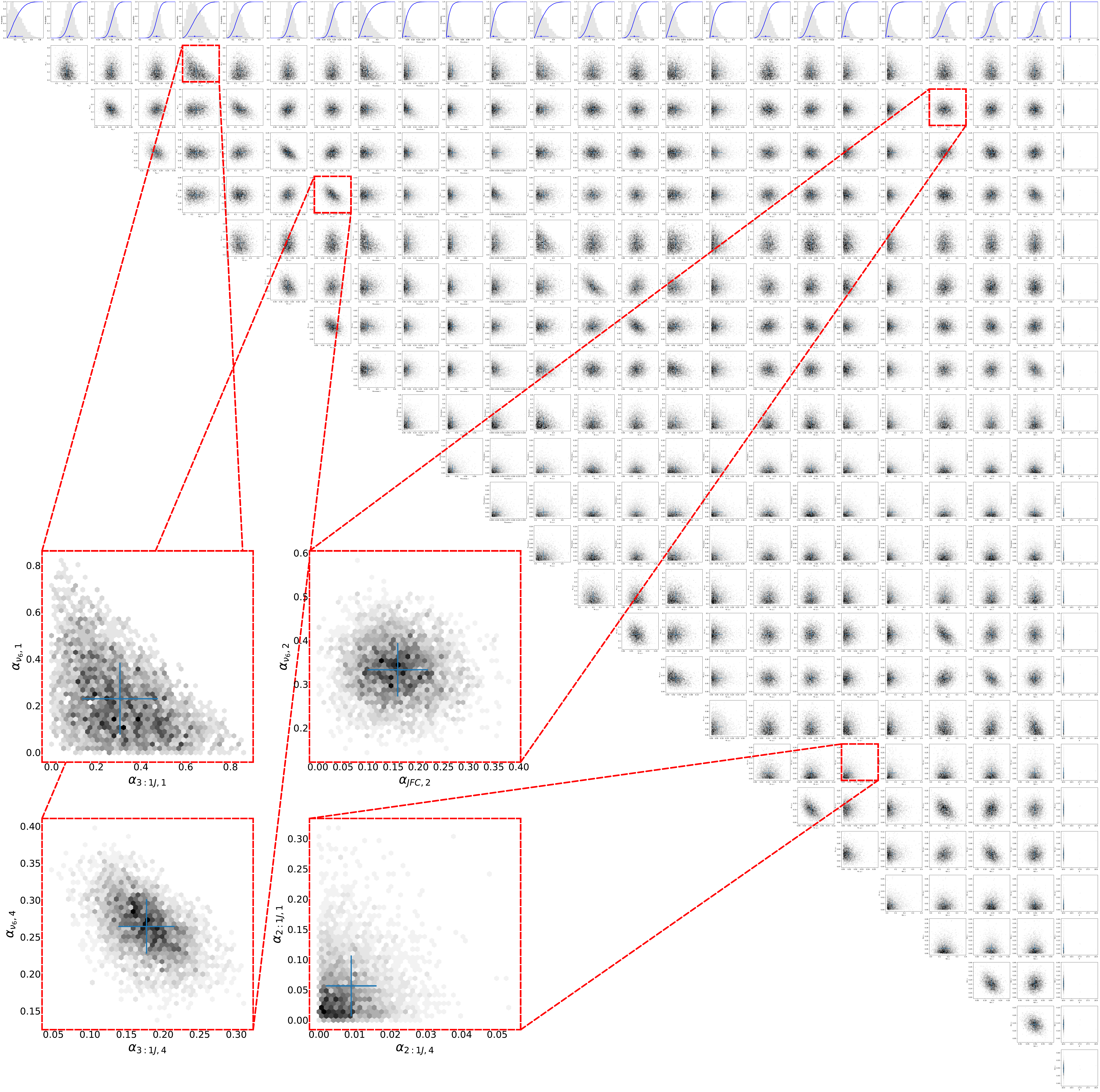}
    \caption{The marginal posterior distributions for the best fitting model from \multinest. It contains the sources for the Inner Weak resonances, \nusix, 3:1J, 5:2J, 2:1J, and Jupiter Family Comets (JFC). The zoomed insets show four examples of posterior distributions; major degeneracy (top left), well constrained (top right), minor degeneracy (bottom left), and low to zero-contribution (bottom right). At the largest sizes of the model, the \multinest samples $\alpha_s(H=34.5)$ for the InnerWeak, \nusix, 3:1J and 5:2J all occupy the lower left of the parameter space between pairs of parameters, indicating major degeneracy between the sources. Other minor degeneracies are seen between numerous sources (e.g. $\alpha_{\nu_6,4}$ and $\alpha_{3:1J,4}$). The crossover parameter, restricted to integer values, on the far right column shows significant preference for an index of 16 (H = 38.5).}
    \label{fig:corner_full}
\end{figure}

\section{Testing Model Stability} \label{sec:half_data_results}
\begin{figure}
    \centering
    \includegraphics[width=1\linewidth]{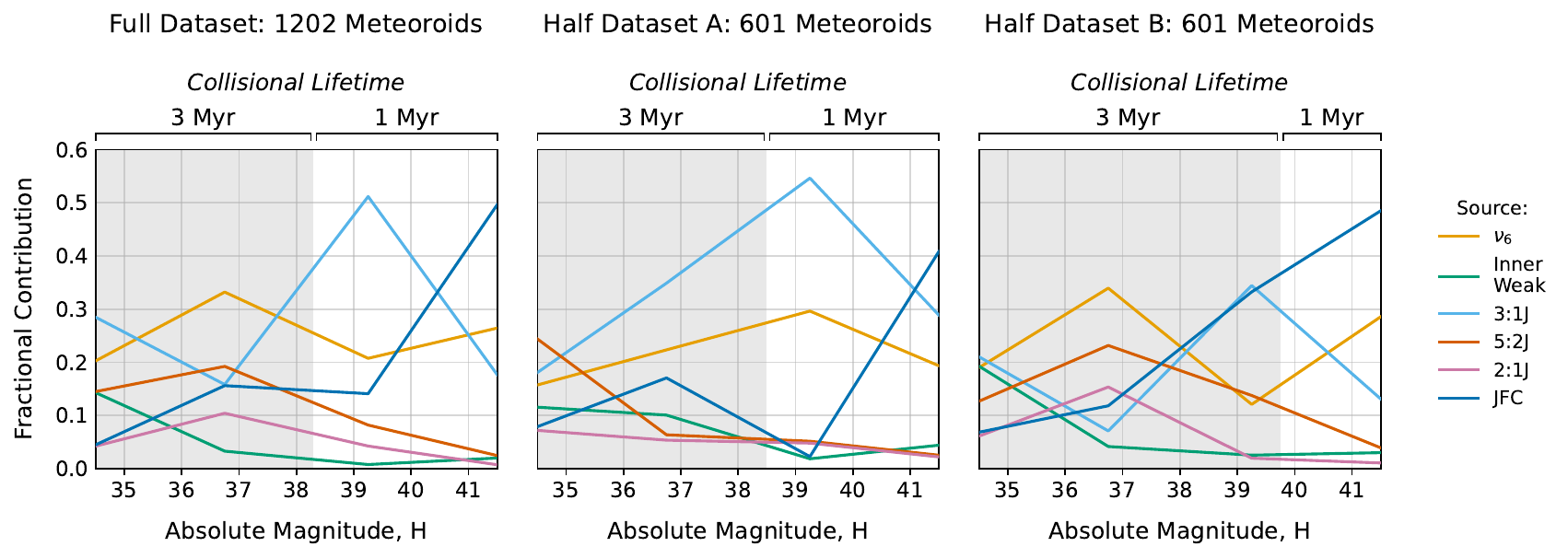}
    \caption{The median alpha parameters output from \multinest for both the full dataset and half datasets A and B. The uncertainty bands for the half data sets are slightly wider than the full dataset (plotted in Figure~\ref{fig:alphas}) but omitted here for clarity. The crossover index for the models run on datasets A and B are H=38.5 ($\sim$0.6~kg) and H=39.75 ($\sim$0.1~kg) respectively.}
    \label{fig:half_alphas}
\end{figure}

To investigate the stability of the model fit and ensure the results aren't specific to any single meteoroid in the calibration data set, we fitted the model to half the data set at a time. 
The data was randomly split into two subsets, A and B, and the model fitted. 
The $N_{GFO}(a,e,i)$ was normalised to 601 instead of 1202 after weighting. 
Figure~\ref{fig:half_alphas} shows the general trends are consistent between subsets A and B; \nusix is consistent, the Inner Weak, 2:1J, and 5:2J have minimal and decreasing contributions. 
The ratio of 3:1J and the JFC uptick with decreasing size is not consistent. 
This model is working with a small amount of data, so this is not unexpected.

\section{Identifying the Meteorite Dropping Population} \label{sec:droppers}
We used several criteria to identify which fireballs captured by the GFO likely resulted in a meteorite falling to the ground. 
The fireball must have exhibited significant ($\geq$ 80\%) overall deceleration and fall within the `likely dropper' parameter space of \citet{Sansom:2019}. 
This criteria identifies events by their deceleration profile, specifically $ln(\beta) \leq ln(4.4 - ln(\alpha sin(\gamma)))$ for the mass loss and ablation parameters $\alpha$ and $\beta$ and slope of entry $\gamma$.
It is worth noting that this criteria were used generally for the entire dataset and does not take into account individual circumstances which would be certainly be considered when determining whether any particular even did in fact drop a meteorite.

\section{Author publication charges} \label{sec:pubcharge}

\bibliography{sophiesbib}{}

\begin{thebibliography}{}
\expandafter\ifx\csname natexlab\endcsname\relax\def\natexlab#1{#1}\fi
\providecommand{\url}[1]{\href{#1}{#1}}
\providecommand{\dodoi}[1]{doi:~\href{http://doi.org/#1}{\nolinkurl{#1}}}
\providecommand{\doeprint}[1]{\href{http://ascl.net/#1}{\nolinkurl{http://ascl.net/#1}}}
\providecommand{\doarXiv}[1]{\href{https://arxiv.org/abs/#1}{\nolinkurl{https://arxiv.org/abs/#1}}}

\bibitem[{{Battle} {et~al.}(2025){Battle}, {Reddy}, {Sanchez}, {Pearson}, {Sharkey}, \& {Kareta}}]{Battle:2025}
{Battle}, A., {Reddy}, V., {Sanchez}, J.~A., {et~al.} 2025, \psj, 6, 31, \dodoi{10.3847/PSJ/ada286}

\bibitem[{{Binzel} {et~al.}(2015){Binzel}, {Reddy}, \& {Dunn}}]{Binzel:2015}
{Binzel}, R.~P., {Reddy}, V., \& {Dunn}, T.~L. 2015, in Asteroids IV, ed. P.~{Michel}, F.~E. {DeMeo}, \& W.~F. {Bottke}, 243--256, \dodoi{10.2458/azu_uapress_9780816532131-ch013}

\bibitem[{{Binzel} {et~al.}(2019){Binzel}, {DeMeo}, {Turtelboom}, {Bus}, {Tokunaga}, {Burbine}, {Lantz}, {Polishook}, {Carry}, {Morbidelli}, {Birlan}, {Vernazza}, {Burt}, {Moskovitz}, {Slivan}, {Thomas}, {Rivkin}, {Hicks}, {Dunn}, {Reddy}, {Sanchez}, {Granvik}, \& {Kohout}}]{Binzel:2019}
{Binzel}, R.~P., {DeMeo}, F.~E., {Turtelboom}, E.~V., {et~al.} 2019, \icarus, 324, 41, \dodoi{10.1016/j.icarus.2018.12.035}

\bibitem[{{Bischoff} {et~al.}(2024){Bischoff}, {Patzek}, {Barrat}, {Berndt}, {Busemann}, {Degering}, {Di Rocco}, {Ek}, {Harries}, {Godinho}, {Heinlein}, {Kriele}, {Krietsch}, {Maden}, {Marchhart}, {Marshal}, {Martschini}, {Merchel}, {M{\"o}ller}, {Pack}, {Raab}, {Reitze}, {Rendtel}, {R{\"u}fenacht}, {Sachs}, {Sch{\"o}nb{\"a}chler}, {Schuppisser}, {Weber}, {Wieser}, \& {Wimmer}}]{Bischoff:2024}
{Bischoff}, A., {Patzek}, M., {Barrat}, J.-A., {et~al.} 2024, \maps, 59, 2660, \dodoi{10.1111/maps.14245}

\bibitem[{{Bland} {et~al.}(2012){Bland}, {Spurn{\'y}}, {Bevan}, {Howard}, {Towner}, {Benedix}, {Greenwood}, {Shrben{\'y}}, {Franchi}, {Deacon}, {Borovi{\v{c}}ka}, {Ceplecha}, {Vaughan}, \& {Hough}}]{Bland:2012}
{Bland}, P.~A., {Spurn{\'y}}, P., {Bevan}, A.~W.~R., {et~al.} 2012, Australian Journal of Earth Sciences, 59, 177, \dodoi{10.1080/08120099.2011.595428}

\bibitem[{Boehnhardt(2004)}]{Boehnhardt:2004}
Boehnhardt, H. 2004, Comets II, 301

\bibitem[{Borovi{\v c}ka {et~al.}(2015)Borovi{\v c}ka, Spurn{\'y}, \& Brown}]{Borovicka:2015}
Borovi{\v c}ka, J., Spurn{\'y}, P., \& Brown, P. 2015, in Asteroids {IV} (University of Arizona Press), 257--280

\bibitem[{Borovi{\v{c}}ka {et~al.}(2007)Borovi{\v{c}}ka, Spurn{\`y}, \& Koten}]{Borovicka:2007}
Borovi{\v{c}}ka, J., Spurn{\`y}, P., \& Koten, P. 2007, Astronomy \& Astrophysics, 473, 661

\bibitem[{{Borovi{\v{c}}ka}(1990)}]{Borovicka:1990}
{Borovi{\v{c}}ka}, J. 1990, Bulletin of the Astronomical Institutes of Czechoslovakia, 41, 391

\bibitem[{{Borovi{\v{c}}ka}(2016)}]{Borovicka:2016}
{Borovi{\v{c}}ka}, J. 2016, in IAU Symposium, Vol. 318, Asteroids: New Observations, New Models, ed. S.~R. {Chesley}, A.~{Morbidelli}, R.~{Jedicke}, \& D.~{Farnocchia}, 80--85, \dodoi{10.1017/S174392131500873X}

\bibitem[{{Borovi{\v{c}}ka} {et~al.}(2013){Borovi{\v{c}}ka}, {Spurn{\'y}}, {Brown}, {Wiegert}, {Kalenda}, {Clark}, \& {Shrben{\'y}}}]{Borovicka:2013}
{Borovi{\v{c}}ka}, J., {Spurn{\'y}}, P., {Brown}, P., {et~al.} 2013, \nat, 503, 235, \dodoi{10.1038/nature12671}

\bibitem[{{Borovi{\v{c}}ka} {et~al.}(2022{\natexlab{a}}){Borovi{\v{c}}ka}, {Spurn{\'y}}, \& {Shrben{\'y}}}]{Borovicka:2022paper1}
{Borovi{\v{c}}ka}, J., {Spurn{\'y}}, P., \& {Shrben{\'y}}, L. 2022{\natexlab{a}}, \aap, 667, A158, \dodoi{10.1051/0004-6361/202244197}

\bibitem[{{Borovi{\v{c}}ka} {et~al.}(2022{\natexlab{b}}){Borovi{\v{c}}ka}, {Spurn{\'y}}, \& {Shrben{\'y}}}]{Borovicka:2022paper2}
---. 2022{\natexlab{b}}, \aap, 667, A158, \dodoi{10.1051/0004-6361/202244197}

\bibitem[{{Boslough} \& {Crawford}(2008)}]{Boslough:2008}
{Boslough}, M.~B.~E., \& {Crawford}, D.~A. 2008, International Journal of Impact Engineering, 35, 1441, \dodoi{10.1016/j.ijimpeng.2008.07.053}

\bibitem[{{Bottke} {et~al.}(2005{\natexlab{a}}){Bottke}, {Durda}, {Nesvorn{\'y}}, {Jedicke}, {Morbidelli}, {Vokrouhlick{\'y}}, \& {Levison}}]{Bottke:2005_1stpaper_Fossilized...}
{Bottke}, W.~F., {Durda}, D.~D., {Nesvorn{\'y}}, D., {et~al.} 2005{\natexlab{a}}, \icarus, 175, 111, \dodoi{10.1016/j.icarus.2004.10.026}

\bibitem[{{Bottke} {et~al.}(2005{\natexlab{b}}){Bottke}, {Durda}, {Nesvorn{\'y}}, {Jedicke}, {Morbidelli}, {Vokrouhlick{\'y}}, \& {Levison}}]{Bottke:2005_2ndpaper_Linking...}
---. 2005{\natexlab{b}}, \icarus, 179, 63, \dodoi{10.1016/j.icarus.2005.05.017}

\bibitem[{{Bottke} {et~al.}(2002{\natexlab{a}}){Bottke}, {Morbidelli}, {Jedicke}, {Petit}, {Levison}, {Michel}, \& {Metcalfe}}]{Bottke:2002}
{Bottke}, W.~F., {Morbidelli}, A., {Jedicke}, R., {et~al.} 2002{\natexlab{a}}, \icarus, 156, 399, \dodoi{10.1006/icar.2001.6788}

\bibitem[{{Bottke} {et~al.}(2025){Bottke}, {Meyer}, {Vokrouhlick{\'y}}, {Nesvorn{\'y}}, {Bierhaus}, {DellaGiustina}, {Hoover}, {Connolly}, \& {Lauretta}}]{Bottke:2025}
{Bottke}, W.~F., {Meyer}, A.~J., {Vokrouhlick{\'y}}, D., {et~al.} 2025, \psj, 6, 150, \dodoi{10.3847/PSJ/add46a}

\bibitem[{{Bottke} {et~al.}(2002{\natexlab{b}}){Bottke}, {Vokrouhlick{\'y}}, {Rubincam}, \& {Broz}}]{Bottke:2002_asteroids3}
{Bottke}, Jr., W.~F., {Vokrouhlick{\'y}}, D., {Rubincam}, D.~P., \& {Broz}, M. 2002{\natexlab{b}}, in Asteroids III, ed. W.~F. {Bottke}, Jr., A.~{Cellino}, P.~{Paolicchi}, \& R.~P. {Binzel} (University of Arizona Press), 395--408

\bibitem[{{Bottke} {et~al.}(2006){Bottke}, {Vokrouhlick{\'y}}, {Rubincam}, \& {Nesvorn{\'y}}}]{Bottke:2006}
{Bottke}, Jr., W.~F., {Vokrouhlick{\'y}}, D., {Rubincam}, D.~P., \& {Nesvorn{\'y}}, D. 2006, Annual Review of Earth and Planetary Sciences, 34, 157, \dodoi{10.1146/annurev.earth.34.031405.125154}

\bibitem[{{Bro{\v{z}}} {et~al.}(2024{\natexlab{a}}){Bro{\v{z}}}, {Vernazza}, {Marsset}, {DeMeo}, {Binzel}, {Vokrouhlick{\'y}}, \& {Nesvorn{\'y}}}]{Broz:2024Nature}
{Bro{\v{z}}}, M., {Vernazza}, P., {Marsset}, M., {et~al.} 2024{\natexlab{a}}, \nat, 634, 566, \dodoi{10.1038/s41586-024-08006-7}

\bibitem[{{Bro{\v{z}}} {et~al.}(2024{\natexlab{b}}){Bro{\v{z}}}, {Vernazza}, {Marsset}, {Binzel}, {DeMeo}, {Birlan}, {Colas}, {Anghel}, {Bouley}, {Blanpain}, {Gattacceca}, {Jeanne}, {Jorda}, {Lecubin}, {Malgoyre}, {Steinhausser}, {Vaubaillon}, \& {Zanda}}]{Broz:2024_A&A}
---. 2024{\natexlab{b}}, \aap, 689, A183, \dodoi{10.1051/0004-6361/202450532}

\bibitem[{{Brown} {et~al.}(2002){Brown}, {Spalding}, {ReVelle}, {Tagliaferri}, \& {Worden}}]{Brown:2002}
{Brown}, P., {Spalding}, R.~E., {ReVelle}, D.~O., {Tagliaferri}, E., \& {Worden}, S.~P. 2002, \nat, 420, 294, \dodoi{10.1038/nature01238}

\bibitem[{{Brown} {et~al.}(2016){Brown}, {Wiegert}, {Clark}, \& {Tagliaferri}}]{Brown:2016}
{Brown}, P., {Wiegert}, P., {Clark}, D., \& {Tagliaferri}, E. 2016, \icarus, 266, 96, \dodoi{10.1016/j.icarus.2015.11.022}

\bibitem[{{Brown} {et~al.}(2013){Brown}, {Assink}, {Astiz}, {Blaauw}, {Boslough}, {Borovi{\v{c}}ka}, {Brachet}, {Brown}, {Campbell-Brown}, {Ceranna}, {Cooke}, {de Groot-Hedlin}, {Drob}, {Edwards}, {Evers}, {Garces}, {Gill}, {Hedlin}, {Kingery}, {Laske}, {Le Pichon}, {Mialle}, {Moser}, {Saffer}, {Silber}, {Smets}, {Spalding}, {Spurn{\'y}}, {Tagliaferri}, {Uren}, {Weryk}, {Whitaker}, \& {Krzeminski}}]{Brown:2013}
{Brown}, P.~G., {Assink}, J.~D., {Astiz}, L., {et~al.} 2013, \nat, 503, 238, \dodoi{10.1038/nature12741}

\bibitem[{{Buchner} {et~al.}(2014){Buchner}, {Georgakakis}, {Nandra}, {Hsu}, {Rangel}, {Brightman}, {Merloni}, {Salvato}, {Donley}, \& {Kocevski}}]{Buchner:2014}
{Buchner}, J., {Georgakakis}, A., {Nandra}, K., {et~al.} 2014, \aap, 564, A125, \dodoi{10.1051/0004-6361/201322971}

\bibitem[{{Burbine} \& {O'Brien}(2004)}]{BurbineObrien:2004}
{Burbine}, T.~H., \& {O'Brien}, K.~M. 2004, \maps, 39, 667, \dodoi{10.1111/j.1945-5100.2004.tb00110.x}

\bibitem[{{Bus} \& {Binzel}(2002)}]{BusBinzel:2002}
{Bus}, S.~J., \& {Binzel}, R.~P. 2002, \icarus, 158, 146, \dodoi{10.1006/icar.2002.6856}

\bibitem[{{Campbell-Brown}(2008)}]{Campbell-Brown:2008}
{Campbell-Brown}, M.~D. 2008, \icarus, 196, 144, \dodoi{10.1016/j.icarus.2008.02.022}

\bibitem[{{Ceplecha} {et~al.}(1998){Ceplecha}, {Borovi{\v{c}}ka}, {Elford}, {Revelle}, {Hawkes}, {Porub{\v{c}}an}, \& {{\v{S}}imek}}]{Ceplecha:1998}
{Ceplecha}, Z., {Borovi{\v{c}}ka}, J., {Elford}, W.~G., {et~al.} 1998, \ssr, 84, 327, \dodoi{10.1023/A:1005069928850}

\bibitem[{{Chow} \& {Brown}(2025)}]{ChowBrown:2025}
{Chow}, I., \& {Brown}, P.~G. 2025, \icarus, 429, 116444, \dodoi{10.1016/j.icarus.2024.116444}

\bibitem[{{Colas} {et~al.}(2020){Colas}, {Zanda}, {Bouley}, {Jeanne}, {Malgoyre}, {Birlan}, {Blanpain}, {Gattacceca}, {Jorda}, {Lecubin}, {Marmo}, {Rault}, {Vaubaillon}, {Vernazza}, {Yohia}, {Gardiol}, {Nedelcu}, {Poppe}, {Rowe}, {Forcier}, {Koschny}, {Trigo-Rodriguez}, {Lamy}, {Behrend}, {Ferri{\`e}re}, {Barghini}, {Buzzoni}, {Carbognani}, {Di Carlo}, {Di Martino}, {Knapic}, {Londero}, {Pratesi}, {Rasetti}, {Riva}, {Stirpe}, {Valsecchi}, {Volpicelli}, {Zorba}, {Coward}, {Drolshagen}, {Drolshagen}, {Hernandez}, {Jehin}, {Jobin}, {King}, {Nitschelm}, {Ott}, {Sanchez-Lavega}, {Toni}, {Abraham}, {Affaticati}, {Albani}, {Andreis}, {Andrieu}, {Anghel}, {Antaluca}, {Antier}, {App{\'e}r{\'e}}, {Armand}, {Ascione}, {Audureau}, {Auxepaules}, {Avoscan}, {Baba Aissa}, {Bacci}, {B{\v{a}}descu}, {Baldini}, {Baldo}, {Balestrero}, {Baratoux}, {Barbotin}, {Bardy}, {Basso}, {Bautista}, {Bayle}, {Beck}, {Bellitto}, {Belluso}, {Benna}, {Benammi}, {Beneteau}, {Benkhaldoun}, {Bergamini}, {Bernardi}, {Bertaina}, {Bessin}, {Betti},
  {Bettonvil}, {Bihel}, {Birnbaum}, {Blagoi}, {Blouri}, {Boac{\u{a}}}, {Boat{\v{a}}}, {Bobiet}, {Bonino}, {Boros}, {Bouchet}, {Borgeot}, {Bouchez}, {Boust}, {Boudon}, {Bouman}, {Bourget}, {Brandenburg}, {Bramond}, {Braun}, {Bussi}, {Cacault}, {Caillier}, {Calegaro}, {Camargo}, {Caminade}, {Campana}, {Campbell-Burns}, {Canal-Domingo}, {Carell}, {Carreau}, {Cascone}, {Cattaneo}, {Cauhape}, {Cavier}, {Celestin}, {Cellino}, {Champenois}, {Chennaoui Aoudjehane}, {Chevrier}, {Cholvy}, {Chomier}, {Christou}, {Cricchio}, {Coadou}, {Cocaign}, {Cochard}, {Cointin}, {Colombi}, {Colque Saavedra}, {Corp}, {Costa}, {Costard}, {Cottier}, {Cournoyer}, {Coustal}, {Cremonese}, {Cristea}, {Cuzon}, {D'Agostino}, {Daiffallah}, {D{\v{a}}nescu}, {Dardon}, {Dasse}, {Davadan}, {Debs}, {Defaix}, {Deleflie}, {D'Elia}, {De Luca}, {De Maria}, {Deverch{\`e}re}, {Devillepoix}, {Dias}, {Di Dato}, {Di Luca}, {Dominici}, {Drouard}, {Dumont}, {Dupouy}, {Duvignac}, {Egal}, {Erasmus}, {Esseiva}, {Ebel}, {Eisengarten}, {Federici}, {Feral},
  {Ferrant}, {Ferreol}, {Finitzer}, {Foucault}, {Francois}, {Fr{\^\i}ncu}, {Froger}, {Gaborit}, {Gagliarducci}, {Galard}, {Gardavot}, {Garmier}, {Garnung}, {Gautier}, {Gendre}, {Gerard}, {Gerardi}, {Godet}, {Grandchamps}, {Grouiez}, {Groult}, {Guidetti}, {Giuli}, \& {Hello}}]{Colas:2020}
{Colas}, F., {Zanda}, B., {Bouley}, S., {et~al.} 2020, \aap, 644, A53, \dodoi{10.1051/0004-6361/202038649}

\bibitem[{{{\'C}uk} {et~al.}(2014){{\'C}uk}, {Gladman}, \& {Nesvorn{\'y}}}]{Cuk:2014}
{{\'C}uk}, M., {Gladman}, B.~J., \& {Nesvorn{\'y}}, D. 2014, \icarus, 239, 154, \dodoi{10.1016/j.icarus.2014.05.048}

\bibitem[{{de El{\'\i}a} \& {Brunini}(2007)}]{EliaBrunini:2007}
{de El{\'\i}a}, G.~C., \& {Brunini}, A. 2007, \aap, 466, 1159, \dodoi{10.1051/0004-6361:20066046}

\bibitem[{Deam \& Nesvorny(2025)}]{modelcode:2025}
Deam, S.~E., \& Nesvorny, D. 2025, Sampling and querying code for the Near-Earth Object Model Calibrated to Earth Impactors, 1.0 (in review),  Zenodo, \dodoi{10.5281/zenodo.17809512}

\bibitem[{{Deienno} {et~al.}(2025){Deienno}, {Denneau}, {Nesvorn{\'y}}, {Vokrouhlick{\'y}}, {Bottke}, {Jedicke}, {Naidu}, {Chesley}, {Farnocchia}, \& {Chodas}}]{Deienno:2025}
{Deienno}, R., {Denneau}, L., {Nesvorn{\'y}}, D., {et~al.} 2025, \icarus, 425, 116316, \dodoi{10.1016/j.icarus.2024.116316}

\bibitem[{Delbo {et~al.}(2022)Delbo, Walsh, Matonti, Wilkerson, Pajola, Al~Asad, Avdellidou, Ballouz, Bennett, Connolly~Jr, {et~al.}}]{Delbo:2022}
Delbo, M., Walsh, K.~J., Matonti, C., {et~al.} 2022, Nature Geoscience, 15, 453

\bibitem[{{DeMeo} \& {Carry}(2013)}]{DeMeoCarry:2013}
{DeMeo}, F.~E., \& {Carry}, B. 2013, \icarus, 226, 723, \dodoi{10.1016/j.icarus.2013.06.027}

\bibitem[{DeMeo \& Carry(2014)}]{Demeo:2014}
DeMeo, F.~E., \& Carry, B. 2014, Nature, 505, 629

\bibitem[{{DeMeo} {et~al.}(2022){DeMeo}, {Burt}, {Marsset}, {Polishook}, {Burbine}, {Carry}, {Binzel}, {Vernazza}, {Reddy}, {Tang}, {Thomas}, {Rivkin}, {Moskovitz}, {Slivan}, \& {Bus}}]{DeMeo:2022}
{DeMeo}, F.~E., {Burt}, B.~J., {Marsset}, M., {et~al.} 2022, \icarus, 380, 114971, \dodoi{10.1016/j.icarus.2022.114971}

\bibitem[{{Devillepoix} {et~al.}(2018){Devillepoix}, {Sansom}, {Bland}, {Towner}, {Cup{\'a}K}, {Howie}, {Jansen-Sturgeon}, {Cox}, {Hartig}, {Benedix}, \& {Paxman}}]{Devillepoix:2018}
{Devillepoix}, H. A.~R., {Sansom}, E.~K., {Bland}, P.~A., {et~al.} 2018, \maps, 53, 2212, \dodoi{10.1111/maps.13142}

\bibitem[{{Devillepoix} {et~al.}(2020){Devillepoix}, {Cup{\'a}k}, {Bland}, {Sansom}, {Towner}, {Howie}, {Hartig}, {Jansen-Sturgeon}, {Shober}, {Anderson}, {Benedix}, {Busan}, {Sayers}, {Jenniskens}, {Albers}, {Herd}, {Hill}, {Brown}, {Krzeminski}, {Osinski}, {Aoudjehane}, {Benkhaldoun}, {Jabiri}, {Guennoun}, {Barka}, {Darhmaoui}, {Daly}, {Collins}, {McMullan}, {Suttle}, {Ireland}, {Bonning}, {Baeza}, {Alrefay}, {Horner}, {Swindle}, {Hergenrother}, {Fries}, {Tomkins}, {Langendam}, {Rushmer}, {O'Neill}, {Janches}, {Hormaechea}, {Shaw}, {Young}, {Alexander}, {Mardon}, \& {Tate}}]{Devillepoix:2020}
{Devillepoix}, H.~A.~R., {Cup{\'a}k}, M., {Bland}, P.~A., {et~al.} 2020, \planss, 191, 105036, \dodoi{10.1016/j.pss.2020.105036}

\bibitem[{Di~Sisto {et~al.}(2009)Di~Sisto, Fern{\'a}ndez, \& Brunini}]{DiSisto:2009}
Di~Sisto, R.~P., Fern{\'a}ndez, J.~A., \& Brunini, A. 2009, Icarus, 203, 140

\bibitem[{{Dohnanyi}(1969)}]{Dohnanyi:1969}
{Dohnanyi}, J.~S. 1969, \jgr, 74, 2531, \dodoi{10.1029/JB074i010p02531}

\bibitem[{Duncan {et~al.}(2004)Duncan, Levison, \& Dones}]{Duncan:2004}
Duncan, M., Levison, H., \& Dones, L. 2004, Comets II, 193, 204

\bibitem[{Duncan {et~al.}(1988)Duncan, Quinn, \& Tremaine}]{Duncan:1988}
Duncan, M., Quinn, T., \& Tremaine, S. 1988, Astrophysical Journal, Part 2-Letters (ISSN 0004-637X), vol. 328, May 15, 1988, p. L69-L73. Research supported by the University of Toronto and NSERC., 328, L69

\bibitem[{Duncan \& Levison(1997)}]{Duncan:1997}
Duncan, M.~J., \& Levison, H.~F. 1997, Science, 276, 1670

\bibitem[{{Durda} {et~al.}(1998){Durda}, {Greenberg}, \& {Jedicke}}]{Durda:1998}
{Durda}, D.~D., {Greenberg}, R., \& {Jedicke}, R. 1998, \icarus, 135, 431, \dodoi{10.1006/icar.1998.5960}

\bibitem[{{Egal} {et~al.}(2025){Egal}, {Vida}, {Colas}, {Zanda}, {Bouley}, {Steinhausser}, {Vernazza}, {Ferri{\`e}re}, {Gattacceca}, {Birlan}, {Vaubaillon}, {Antier}, {Anghel}, {Desmars}, {Bailli{\'e}}, {Maquet}, {Bouquillon}, {Malgoyre}, {Jeanne}, {Fripon International Team}, {Trigo-Rodriguez}, {Herrero}, {Rowe}, {Smedley}, {King}, {Sylla}, {Gardiol}, {Barghini}, {Lamy}, {Jehin}, {Koschny}, {Poppe}, {Jord{\'a}n}, {Mendez}, {Vieira}, {Cremades}, {Aoudjehane}, {Benkhaldoun}, {Borovi{\v{c}}ka}, {Spurn{\'y}}, {Devillepoix}, {Micheli}, {Farnocchia}, {Naidu}, {Brown}, {Wiegert}, {S{\'a}rneczky}, {P{\'a}l}, {Moskovitz}, {Kareta}, {Santana-Ros}, {Le Pichon}, {Mazet-Roux}, {Vergoz}, {McFadden}, {Assink}, {Evers}, {Krietsch}, {Busemann}, {Maden}, {Eckart}, {Barrat}, {Povinec}, {S{\'y}kora}, {Kontul'}, {Marchhart}, {Martschini}, {Merchel}, {Wieser}, {Gounelle}, {Pont}, {Sans-Jofre}, {de Vet}, {Baziotis}, {Bro{\v{z}}}, {Marsset}, {Vergne}, {Hanu{\v{s}}}, {Devog{\`e}le}, {Conversi}, {Oca{\~n}a}, {Buzzi}, {Nedelcu},
  {Sonka}, {Losse}, {Dupouy}, {Korlevi{\'c}}, {Husar}, {Jahn}, {{\v{S}}egon}, {McIntyre}, {Neubert}, {Beck}, {Shober}, {Lagain}, {Hernandez}, {Robertson}, \& {Jenniskens}}]{Egal:2025_2023CX1}
{Egal}, A., {Vida}, D., {Colas}, F., {et~al.} 2025, Nature Astronomy, \dodoi{10.1038/s41550-025-02659-8}

\bibitem[{{Eugster}(2003)}]{Eugster:2003}
{Eugster}, O. 2003, Chemie der Erde / Geochemistry, 63, 3, \dodoi{10.1078/0009-2819-00021}

\bibitem[{{Farinella} {et~al.}(1998){Farinella}, {Vokrouhlick{\'y}}, \& {Hartmann}}]{Farinella:1998}
{Farinella}, P., {Vokrouhlick{\'y}}, D., \& {Hartmann}, W.~K. 1998, \icarus, 132, 378, \dodoi{10.1006/icar.1997.5872}

\bibitem[{{Farnocchia} {et~al.}(2016){Farnocchia}, {Chesley}, {Brown}, \& {Chodas}}]{Farnocchia:2016_2014AA}
{Farnocchia}, D., {Chesley}, S.~R., {Brown}, P.~G., \& {Chodas}, P.~W. 2016, \icarus, 274, 327, \dodoi{10.1016/j.icarus.2016.02.056}

\bibitem[{Fern{\'a}ndez(1980)}]{Fernandez:1980}
Fern{\'a}ndez, J.~A. 1980, Monthly Notices of the Royal Astronomical Society, 192, 481

\bibitem[{Fern{\'a}ndez {et~al.}(2002)Fern{\'a}ndez, Gallardo, \& Brunini}]{Fernandez:2002}
Fern{\'a}ndez, J.~A., Gallardo, T., \& Brunini, A. 2002, Icarus, 159, 358

\bibitem[{Fern{\'a}ndez \& Sosa(2015)}]{Fernandez:2015}
Fern{\'a}ndez, J.~A., \& Sosa, A. 2015, Planetary and Space Science, 118, 14

\bibitem[{Fern{\'a}ndez(2009)}]{Fernandez:2009}
Fern{\'a}ndez, Y.~R. 2009, Planetary and Space Science, 57, 1218

\bibitem[{{Feroz} {et~al.}(2019){Feroz}, {Hobson}, {Cameron}, \& {Pettitt}}]{Feroz:2019}
{Feroz}, F., {Hobson}, M.~P., {Cameron}, E., \& {Pettitt}, A.~N. 2019, The Open Journal of Astrophysics, 2, 10, \dodoi{10.21105/astro.1306.2144}

\bibitem[{{Fuentes-Mu{\~n}oz} {et~al.}(2023){Fuentes-Mu{\~n}oz}, {Scheeres}, {Farnocchia}, \& {Park}}]{FuentesMunoz:2023}
{Fuentes-Mu{\~n}oz}, O., {Scheeres}, D.~J., {Farnocchia}, D., \& {Park}, R.~S. 2023, \aj, 166, 10, \dodoi{10.3847/1538-3881/acd378}

\bibitem[{{Gaffey} {et~al.}(1992){Gaffey}, {Reed}, \& {Kelley}}]{Gaffey:1992}
{Gaffey}, M.~J., {Reed}, K.~L., \& {Kelley}, M.~S. 1992, \icarus, 100, 95, \dodoi{10.1016/0019-1035(92)90021-X}

\bibitem[{{Galligan}(2001)}]{Galligan:2001}
{Galligan}, D.~P. 2001, \mnras, 327, 623, \dodoi{10.1046/j.1365-8711.2001.04858.x}

\bibitem[{{Galligan} \& {Baggaley}(2005)}]{Galligan&Baggaley:2005}
{Galligan}, D.~P., \& {Baggaley}, W.~J. 2005, \mnras, 359, 551, \dodoi{10.1111/j.1365-2966.2005.08918.x}

\bibitem[{{Geng} {et~al.}(2023){Geng}, {Zhou}, \& {Li}}]{Geng:2023_2022EB5}
{Geng}, S., {Zhou}, B., \& {Li}, M. 2023, \aap, 670, A27, \dodoi{10.1051/0004-6361/202244084}

\bibitem[{{Gi} {et~al.}(2018){Gi}, {Brown}, \& {Aftosmis}}]{Gi:2018}
{Gi}, N., {Brown}, P., \& {Aftosmis}, M. 2018, \maps, 53, 1413, \dodoi{10.1111/maps.13085}

\bibitem[{{Gianotto} {et~al.}(2025){Gianotto}, {Carbognani}, {Fenucci}, {Devog{\`e}le}, {Ramirez-Moreta}, {Micheli}, {Salerno}, {Santana-Ros}, {Cano}, {Conversi}, {Drury}, {Faggioli}, {F{\"o}hring}, {Kresken}, {Machnitzky}, {Moissl}, {Oca{\~n}a}, {Oliviero}, {Alonso-Peleato}, {Revellino}, \& {Rudawska}}]{Gianotto:2025_2024XA1}
{Gianotto}, F., {Carbognani}, A., {Fenucci}, M., {et~al.} 2025, \icarus, 433, 116511, \dodoi{10.1016/j.icarus.2025.116511}

\bibitem[{{Gladman} {et~al.}(1997){Gladman}, {Migliorini}, {Morbidelli}, {Zappala}, {Michel}, {Cellino}, {Froeschle}, {Levison}, {Bailey}, \& {Duncan}}]{Gladman:1997}
{Gladman}, B.~J., {Migliorini}, F., {Morbidelli}, A., {et~al.} 1997, Science, 277, 197, \dodoi{10.1126/science.277.5323.197}

\bibitem[{{Granvik} \& {Brown}(2018)}]{GranvikBrown:2018}
{Granvik}, M., \& {Brown}, P. 2018, \icarus, 311, 271, \dodoi{10.1016/j.icarus.2018.04.012}

\bibitem[{{Granvik} {et~al.}(2017){Granvik}, {Morbidelli}, {Vokrouhlick{\'y}}, {Bottke}, {Nesvorn{\'y}}, \& {Jedicke}}]{Granvik:2017}
{Granvik}, M., {Morbidelli}, A., {Vokrouhlick{\'y}}, D., {et~al.} 2017, \aap, 598, A52, \dodoi{10.1051/0004-6361/201629252}

\bibitem[{{Granvik} \& {Walsh}(2024)}]{Granvik:2024}
{Granvik}, M., \& {Walsh}, K.~J. 2024, \apjl, 960, L9, \dodoi{10.3847/2041-8213/ad151b}

\bibitem[{{Granvik} {et~al.}(2016){Granvik}, {Morbidelli}, {Jedicke}, {Bolin}, {Bottke}, {Beshore}, {Vokrouhlick{\'y}}, {Delb{\`o}}, \& {Michel}}]{Granvik:2016}
{Granvik}, M., {Morbidelli}, A., {Jedicke}, R., {et~al.} 2016, \nat, 530, 303, \dodoi{10.1038/nature16934}

\bibitem[{{Granvik} {et~al.}(2018){Granvik}, {Morbidelli}, {Jedicke}, {Bolin}, {Bottke}, {Beshore}, {Vokrouhlick{\'y}}, {Nesvorn{\'y}}, \& {Michel}}]{Granvik:2018}
---. 2018, \icarus, 312, 181, \dodoi{10.1016/j.icarus.2018.04.018}

\bibitem[{{Greenstreet} {et~al.}(2012){Greenstreet}, {Ngo}, \& {Gladman}}]{Greenstreet:2012}
{Greenstreet}, S., {Ngo}, H., \& {Gladman}, B. 2012, \icarus, 217, 355, \dodoi{10.1016/j.icarus.2011.11.010}

\bibitem[{{Hajdukov{\'a}} {et~al.}(2023){Hajdukov{\'a}}, {Rudawska}, {Jopek}, {Koseki}, {Kokhirova}, \& {Neslu{\v{s}}an}}]{Hajdukova:2023}
{Hajdukov{\'a}}, M., {Rudawska}, R., {Jopek}, T.~J., {et~al.} 2023, \aap, 671, A155, \dodoi{10.1051/0004-6361/202244964}

\bibitem[{{Halliday} {et~al.}(1996){Halliday}, {Griffin}, \& {Blackwell}}]{Halliday:1996}
{Halliday}, I., {Griffin}, A.~A., \& {Blackwell}, A.~T. 1996, \maps, 31, 185, \dodoi{10.1111/j.1945-5100.1996.tb02014.x}

\bibitem[{Harris \& Chodas(2021)}]{HarrisChodas:2021}
Harris, A.~W., \& Chodas, P.~W. 2021, Icarus, 365, 114452, \dodoi{https://doi.org/10.1016/j.icarus.2021.114452}

\bibitem[{Harris {et~al.}(2020)Harris, Millman, van~der Walt, Gommers, Virtanen, Cournapeau, Wieser, Taylor, Berg, Smith, Kern, Picus, Hoyer, van Kerkwijk, Brett, Haldane, del R{\'{i}}o, Wiebe, Peterson, G{\'{e}}rard-Marchant, Sheppard, Reddy, Weckesser, Abbasi, Gohlke, \& Oliphant}]{numpy}
Harris, C.~R., Millman, K.~J., van~der Walt, S.~J., {et~al.} 2020, Nature, 585, 357, \dodoi{10.1038/s41586-020-2649-2}

\bibitem[{{Horner} {et~al.}(2004){Horner}, {Evans}, \& {Bailey}}]{Horner:2004}
{Horner}, J., {Evans}, N.~W., \& {Bailey}, M.~E. 2004, \mnras, 354, 798, \dodoi{10.1111/j.1365-2966.2004.08240.x}

\bibitem[{{Horner} {et~al.}(2020){Horner}, {Kane}, {Marshall}, {Dalba}, {Holt}, {Wood}, {Maynard-Casely}, {Wittenmyer}, {Lykawka}, {Hill}, {Salmeron}, {Bailey}, {L{\"o}hne}, {Agnew}, {Carter}, \& {Tylor}}]{Horner:2020}
{Horner}, J., {Kane}, S.~R., {Marshall}, J.~P., {et~al.} 2020, \pasp, 132, 102001, \dodoi{10.1088/1538-3873/ab8eb9}

\bibitem[{{Howie} {et~al.}(2017{\natexlab{a}}){Howie}, {Paxman}, {Bland}, {Towner}, {Cupak}, {Sansom}, \& {Devillepoix}}]{Howie:2017:howtobuild}
{Howie}, R.~M., {Paxman}, J., {Bland}, P.~A., {et~al.} 2017{\natexlab{a}}, Experimental Astronomy, 43, 237, \dodoi{10.1007/s10686-017-9532-7}

\bibitem[{{Howie} {et~al.}(2017{\natexlab{b}}){Howie}, {Paxman}, {Bland}, {Towner}, {Sansom}, \& {Devillepoix}}]{Howie:2017:debruijn}
---. 2017{\natexlab{b}}, \maps, 52, 1669, \dodoi{10.1111/maps.12878}

\bibitem[{Hsieh \& Haghighipour(2016)}]{Hsieh:2016}
Hsieh, H.~H., \& Haghighipour, N. 2016, Icarus, 277, 19

\bibitem[{Hsieh {et~al.}(2020)Hsieh, Novakovi{\'c}, Walsh, \& Sch{\"o}rghofer}]{Hsieh:2020}
Hsieh, H.~H., Novakovi{\'c}, B., Walsh, K.~J., \& Sch{\"o}rghofer, N. 2020, The Astronomical Journal, 159, 179

\bibitem[{Hunter(2007)}]{matplotlib}
Hunter, J.~D. 2007, Computing in Science \& Engineering, 9, 90, \dodoi{10.1109/MCSE.2007.55}

\bibitem[{{Ingebretsen} {et~al.}(2025){Ingebretsen}, {Bolin}, {Jedicke}, {Vere{\v{s}}}, {Chen}, {Lisse}, {McMillan}, {Sutherland}, \& {Townsend}}]{Ingebretsen:2025_2024RW1}
{Ingebretsen}, C., {Bolin}, B.~T., {Jedicke}, R., {et~al.} 2025, \aj, 170, 237, \dodoi{10.3847/1538-3881/adfb63}

\bibitem[{{Jansen-Sturgeon} {et~al.}(2019){Jansen-Sturgeon}, {Sansom}, \& {Bland}}]{Jansen-Sturgeon:2019}
{Jansen-Sturgeon}, T., {Sansom}, E.~K., \& {Bland}, P.~A. 2019, \maps, 54, 2149, \dodoi{10.1111/maps.13376}

\bibitem[{{Jenniskens}(2008{\natexlab{a}})}]{Jenniskens:2008_rev}
{Jenniskens}, P. 2008{\natexlab{a}}, Earth Moon and Planets, 102, 505, \dodoi{10.1007/s11038-007-9169-z}

\bibitem[{{Jenniskens}(2008{\natexlab{b}})}]{Jenniskens:2008}
---. 2008{\natexlab{b}}, \icarus, 194, 13, \dodoi{10.1016/j.icarus.2007.09.016}

\bibitem[{{Jenniskens} \& {Devillepoix}(2025)}]{Jenniskens:2025}
{Jenniskens}, P., \& {Devillepoix}, H. A.~R. 2025, \maps, 60, 928, \dodoi{10.1111/maps.14321}

\bibitem[{{Jenniskens} {et~al.}(2009){Jenniskens}, {Shaddad}, {Numan}, {Elsir}, {Kudoda}, {Zolensky}, {Le}, {Robinson}, {Friedrich}, {Rumble}, {Steele}, {Chesley}, {Fitzsimmons}, {Duddy}, {Hsieh}, {Ramsay}, {Brown}, {Edwards}, {Tagliaferri}, {Boslough}, {Spalding}, {Dantowitz}, {Kozubal}, {Pravec}, {Borovi{\v{c}}ka}, {Charvat}, {Vaubaillon}, {Kuiper}, {Albers}, {Bishop}, {Mancinelli}, {Sandford}, {Milam}, {Nuevo}, \& {Worden}}]{Jenniskens:2009}
{Jenniskens}, P., {Shaddad}, M.~H., {Numan}, D., {et~al.} 2009, \nat, 458, 485, \dodoi{10.1038/nature07920}

\bibitem[{{Jenniskens} {et~al.}(2016){Jenniskens}, {N{\'e}non}, {Gural}, {Albers}, {Haberman}, {Johnson}, {Morales}, {Grigsby}, {Samuels}, \& {Johannink}}]{Jenniskens:2016}
{Jenniskens}, P., {N{\'e}non}, Q., {Gural}, P.~S., {et~al.} 2016, \icarus, 266, 384, \dodoi{10.1016/j.icarus.2015.11.009}

\bibitem[{{Jenniskens} {et~al.}(2021){Jenniskens}, {Gabadirwe}, {Yin}, {Proyer}, {Moses}, {Kohout}, {Franchi}, {Gibson}, {Kowalski}, {Christensen}, {Gibbs}, {Heinze}, {Denneau}, {Farnocchia}, {Chodas}, {Gray}, {Micheli}, {Moskovitz}, {Onken}, {Wolf}, {Devillepoix}, {Ye}, {Robertson}, {Brown}, {Lyytinen}, {Moilanen}, {Albers}, {Cooper}, {Assink}, {Evers}, {Lahtinen}, {Seitshiro}, {Laubenstein}, {Wantlo}, {Moleje}, {Maritinkole}, {Suhonen}, {Zolensky}, {Ashwal}, {Hiroi}, {Sears}, {Sehlke}, {Maturilli}, {Sanborn}, {Huyskens}, {Dey}, {Ziegler}, {Busemann}, {Riebe}, {Meier}, {Welten}, {Caffee}, {Zhou}, {Li}, {Li}, {Liu}, {Tang}, {McLain}, {Dworkin}, {Glavin}, {Schmitt-Kopplin}, {Sabbah}, {Joblin}, {Granvik}, {Mosarwa}, \& {Botepe}}]{Jenniskens:2021_2018LA}
{Jenniskens}, P., {Gabadirwe}, M., {Yin}, Q.-Z., {et~al.} 2021, \maps, 56, 844, \dodoi{10.1111/maps.13653}

\bibitem[{{Jenniskens} {et~al.}(2024){Jenniskens}, {Pilorz}, {Gural}, {Samuels}, {Rau}, {Abbott}, {Albers}, {Austin}, {Avner}, {Baggaley}, {Beck}, {Blomquist}, {Boyukata}, {Breukers}, {Cooney}, {Cooper}, {De Cicco}, {Devillepoix}, {Egland}, {Fahl}, {Gialluca}, {Grigsby}, {Hanke}, {Harris}, {Heathcote}, {Hemmelgarn}, {Howell}, {Jehin}, {Johannink}, {Juneau}, {Kisvarsanyi}, {Mey}, {Moskovitz}, {Odeh}, {Rachford}, {Rollinson}, {Scott}, {Towner}, {Unsalan}, {van Wyk}, {Wood}, {Wray}, {Vaubaillon}, \& {Lauretta}}]{Jenniskens:2024}
{Jenniskens}, P., {Pilorz}, S., {Gural}, P.~S., {et~al.} 2024, \icarus, 415, 116034, \dodoi{10.1016/j.icarus.2024.116034}

\bibitem[{{Jewitt}(2012)}]{Jewitt:2012}
{Jewitt}, D. 2012, \aj, 143, 66, \dodoi{10.1088/0004-6256/143/3/66}

\bibitem[{{Kareta} {et~al.}(2024){Kareta}, {Vida}, {Micheli}, {Moskovitz}, {Wiegert}, {Brown}, {McCausland}, {Devillepoix}, {Male{\v{c}}i{\'c}}, {Prtenjak}, {{\v{S}}egon}, {Shafransky}, \& {Farnocchia}}]{Kareta:2024_2022WJ1}
{Kareta}, T., {Vida}, D., {Micheli}, M., {et~al.} 2024, \psj, 5, 253, \dodoi{10.3847/PSJ/ad8b22}

\bibitem[{Kim {et~al.}(2014)Kim, Ishiguro, \& Usui}]{Kim:2014}
Kim, Y., Ishiguro, M., \& Usui, F. 2014, The Astrophysical Journal, 789, 151

\bibitem[{{King} {et~al.}(2022){King}, {Daly}, {Rowe}, {Joy}, {Greenwood}, {Devillepoix}, {Suttle}, {Chan}, {Russell}, {Bates}, {Bryson}, {Clay}, {Vida}, {Lee}, {O'Brien}, {Hallis}, {Stephen}, {Tart{\`e}se}, {Sansom}, {Towner}, {Cupak}, {Shober}, {Bland}, {Findlay}, {Franchi}, {Verchovsky}, {Abernethy}, {Grady}, {Floyd}, {Van Ginneken}, {Bridges}, {Hicks}, {Jones}, {Mitchell}, {Genge}, {Jenkins}, {Martin}, {Sephton}, {Watson}, {Salge}, {Shirley}, {Curtis}, {Warren}, {Bowles}, {Stuart}, {Di Nicola}, {Gy{\"o}re}, {Boyce}, {Shaw}, {Elliott}, {Steele}, {Povinec}, {Laubenstein}, {Sanderson}, {Cresswell}, {Jull}, {S{\'y}kora}, {Sridhar}, {Harrison}, {Willcocks}, {Harrison}, {Hallatt}, {Wozniakiewicz}, {Burchell}, {Alesbrook}, {Dignam}, {Almeida}, {Smith}, {Clark}, {Humphreys-Williams}, {Schofield}, {Cornwell}, {Spathis}, {Morgan}, {Perkins}, {Kacerek}, {Campbell-Burns}, {Colas}, {Zanda}, {Vernazza}, {Bouley}, {Jeanne}, {Hankey}, {Collins}, {Young}, {Shaw}, {Horak}, {Jones}, {James}, {Bosley}, {Shuttleworth},
  {Dickinson}, {McMullan}, {Robson}, {Smedley}, {Stanley}, {Bassom}, {McIntyre}, {Suttle}, {Fleet}, {Bastiaens}, {Ih{\'a}sz}, {McMullan}, {Boazman}, {Dickeson}, {Grindrod}, {Pickersgill}, {Weir}, {Suttle}, {Farrelly}, {Spencer}, {Naqvi}, {Mayne}, {Skilton}, {Kirk}, {Mounsey}, {Mounsey}, {Mounsey}, {Godfrey}, {Bond}, {Bond}, {Wilcock}, {Wilcock}, \& {Wilcock}}]{King:2022}
{King}, A.~J., {Daly}, L., {Rowe}, J., {et~al.} 2022, Science Advances, 8, eabq3925, \dodoi{10.1126/sciadv.abq3925}

\bibitem[{{Kurlander} {et~al.}(2025){Kurlander}, {Bernardinelli}, {Schwamb}, {Juric}, {Murtagh}, {Chandler}, {Merritt}, {Nesvorny}, {Vokrouhlicky}, {Jones}, {Fedorets}, {Cornwall}, {Holman}, {Eggl}, {Oldag}, {West}, {Kubica}, {Yoachim}, {Moeyens}, {Kiker}, \& {Buchanan}}]{Kurlander:2025}
{Kurlander}, J.~A., {Bernardinelli}, P.~H., {Schwamb}, M.~E., {et~al.} 2025, arXiv e-prints, arXiv:2506.02487.
\newblock \doarXiv{2506.02487}

\bibitem[{{Lemaitre}(1994)}]{Lemaitre:1994}
{Lemaitre}, A. 1994, in Astronomical Society of the Pacific Conference Series, Vol.~63, 75 Years of Hirayama Asteroid Families: The Role of Collisions in the Solar System History, ed. Y.~{Kozai}, R.~P. {Binzel}, \& T.~{Hirayama}, 140

\bibitem[{Levison \& Duncan(1994)}]{Levison:1994}
Levison, H.~F., \& Duncan, M.~J. 1994, Icarus, 108, 18

\bibitem[{Levison \& Duncan(1997)}]{Levison:1997}
---. 1997, Icarus, 127, 13

\bibitem[{{Lucas} {et~al.}(2017){Lucas}, {Emery}, {Pinilla-Alonso}, {Lindsay}, \& {Lorenzi}}]{Lucas:2017}
{Lucas}, M.~P., {Emery}, J.~P., {Pinilla-Alonso}, N., {Lindsay}, S.~S., \& {Lorenzi}, V. 2017, \icarus, 291, 268, \dodoi{10.1016/j.icarus.2016.11.002}

\bibitem[{{Lue} {et~al.}(2019){Lue}, {Ruprecht}, {Varey}, {Czerwinski}, \& {Viggh}}]{Lue:2019}
{Lue}, A., {Ruprecht}, J.~D., {Varey}, J., {Czerwinski}, M., \& {Viggh}, H. E.~M. 2019, \icarus, 325, 105, \dodoi{10.1016/j.icarus.2019.02.019}

\bibitem[{{Macke} {et~al.}(2011){Macke}, {Britt}, \& {Consolmagno}}]{Macke:2011}
{Macke}, R.~J., {Britt}, D.~T., \& {Consolmagno}, G.~J. 2011, \maps, 46, 311, \dodoi{10.1111/j.1945-5100.2010.01155.x}

\bibitem[{{Mainzer} {et~al.}(2023){Mainzer}, {Masiero}, {Abell}, {Bauer}, {Bottke}, {Buratti}, {Carey}, {Cotto-Figueroa}, {Cutri}, {Dahlen}, {Eisenhardt}, {Fernandez}, {Furfaro}, {Grav}, {Hoffman}, {Kelley}, {Kim}, {Kirkpatrick}, {Lawler}, {Lilly}, {Liu}, {Marocco}, {Marsh}, {Masci}, {McMurtry}, {Pourrahmani}, {Reinhart}, {Ressler}, {Satpathy}, {Schambeau}, {Sonnett}, {Spahr}, {Surace}, {Vaquero}, {Wright}, {Zengilowski}, \& {NEO Surveyor Mission Team}}]{Mainzer:2023}
{Mainzer}, A.~K., {Masiero}, J.~R., {Abell}, P.~A., {et~al.} 2023, \psj, 4, 224, \dodoi{10.3847/PSJ/ad0468}

\bibitem[{{Marsset} {et~al.}(2022){Marsset}, {DeMeo}, {Burt}, {Polishook}, {Binzel}, {Granvik}, {Vernazza}, {Carry}, {Bus}, {Slivan}, {Thomas}, {Moskovitz}, \& {Rivkin}}]{Marsset:2022}
{Marsset}, M., {DeMeo}, F.~E., {Burt}, B., {et~al.} 2022, \aj, 163, 165, \dodoi{10.3847/1538-3881/ac532f}

\bibitem[{{Marsset} {et~al.}(2024){Marsset}, {Vernazza}, {Bro{\v{z}}}, {Thomas}, {DeMeo}, {Burt}, {Binzel}, {Reddy}, {McGraw}, {Avdellidou}, {Carry}, {Slivan}, \& {Polishook}}]{Marsset:2024}
{Marsset}, M., {Vernazza}, P., {Bro{\v{z}}}, M., {et~al.} 2024, \nat, 634, 561, \dodoi{10.1038/s41586-024-08007-6}

\bibitem[{Marti \& Graf(1992)}]{marti1992cosmic}
Marti, K., \& Graf, T. 1992, In: Annual review of earth and planetary sciences. Vol. 20 (A93-45370 18-46), p. 221-243., 20, 221

\bibitem[{{Marty} {et~al.}(2024){Marty}, {Bermingham}, {Nittler}, \& {Raymond}}]{Marty:2024}
{Marty}, B., {Bermingham}, K.~R., {Nittler}, L.~R., \& {Raymond}, S.~N. 2024, in Comets III, ed. K.~J. {Meech}, M.~R. {Combi}, D.~{Bockel{\'e}e-Morvan}, S.~N. {Raymodn}, \& M.~E. {Zolensky}, 681--730, \dodoi{10.48550/arXiv.2506.02721}

\bibitem[{{McGraw} {et~al.}(2025){McGraw}, {Reddy}, \& {Sanchez}}]{McGraw:2025}
{McGraw}, A., {Reddy}, V., \& {Sanchez}, J.~A. 2025, \mnras, 537, 3145, \dodoi{10.1093/mnras/staf061}

\bibitem[{Molaro {et~al.}(2020)Molaro, Walsh, Jawin, Ballouz, Bennett, DellaGiustina, Golish, Drouet~d’Aubigny, Rizk, Schwartz, {et~al.}}]{Molaro:2020}
Molaro, J., Walsh, K., Jawin, E., {et~al.} 2020, Nature Communications, 11, 2913

\bibitem[{{Moorhead}(2016)}]{Moorhead:2016}
{Moorhead}, A.~V. 2016, \mnras, 455, 4329, \dodoi{10.1093/mnras/stv2610}

\bibitem[{{Morbidelli} \& {Gladman}(1998)}]{MorbidelliGladman:1998}
{Morbidelli}, A., \& {Gladman}, B. 1998, \maps, 33, 999, \dodoi{10.1111/j.1945-5100.1998.tb01707.x}

\bibitem[{{Morbidelli} \& {Vokrouhlick{\'y}}(2003)}]{MorbidelliVokrouhlicky:2003}
{Morbidelli}, A., \& {Vokrouhlick{\'y}}, D. 2003, \icarus, 163, 120, \dodoi{10.1016/S0019-1035(03)00047-2}

\bibitem[{{Nesvorn{\'y}} {et~al.}(2011){Nesvorn{\'y}}, {Janches}, {Vokrouhlick{\'y}}, {Pokorn{\'y}}, {Bottke}, \& {Jenniskens}}]{Nesvorny:2011}
{Nesvorn{\'y}}, D., {Janches}, D., {Vokrouhlick{\'y}}, D., {et~al.} 2011, \apj, 743, 129, \dodoi{10.1088/0004-637X/743/2/129}

\bibitem[{{Nesvorn{\'y}} {et~al.}(2010){Nesvorn{\'y}}, {Jenniskens}, {Levison}, {Bottke}, {Vokrouhlick{\'y}}, \& {Gounelle}}]{Nesvorny:2010}
{Nesvorn{\'y}}, D., {Jenniskens}, P., {Levison}, H.~F., {et~al.} 2010, \apj, 713, 816, \dodoi{10.1088/0004-637X/713/2/816}

\bibitem[{{Nesvorny} {et~al.}(2025){Nesvorny}, {Roig}, {Vokrouhlicky}, \& {Broz}}]{Nesvorny:2025catalogue}
{Nesvorny}, D., {Roig}, F., {Vokrouhlicky}, D., \& {Broz}, M. 2025, {VizieR Online Data Catalog: Orbits for 1.25 million main-belt asteroids (Nesvorny+, 2024)}, VizieR On-line Data Catalog: J/ApJS/274/25. Originally published in: 2024ApJS..274...25N

\bibitem[{{Nesvorn{\'y}} {et~al.}(2023){Nesvorn{\'y}}, {Deienno}, {Bottke}, {Jedicke}, {Naidu}, {Chesley}, {Chodas}, {Granvik}, {Vokrouhlick{\'y}}, {Bro{\v{z}}}, {Morbidelli}, {Christensen}, {Shelly}, \& {Bolin}}]{Nesvorny:2023}
{Nesvorn{\'y}}, D., {Deienno}, R., {Bottke}, W.~F., {et~al.} 2023, \aj, 166, 55, \dodoi{10.3847/1538-3881/ace040}

\bibitem[{{Nesvorn{\'y}} {et~al.}(2024{\natexlab{a}}){Nesvorn{\'y}}, {Vokrouhlick{\'y}}, {Shelly}, {Deienno}, {Bottke}, {Christensen}, {Jedicke}, {Naidu}, {Chesley}, {Chodas}, {Farnocchia}, \& {Granvik}}]{Nesvorny:2024NEOMOD2}
{Nesvorn{\'y}}, D., {Vokrouhlick{\'y}}, D., {Shelly}, F., {et~al.} 2024{\natexlab{a}}, \icarus, 411, 115922, \dodoi{10.1016/j.icarus.2023.115922}

\bibitem[{{Nesvorn{\'y}} {et~al.}(2024{\natexlab{b}}){Nesvorn{\'y}}, {Vokrouhlick{\'y}}, {Shelly}, {Deienno}, {Bottke}, {Fuls}, {Jedicke}, {Naidu}, {Chesley}, {Chodas}, {Farnocchia}, \& {Delbo}}]{Nesvorny:2024NEOMOD3}
---. 2024{\natexlab{b}}, \icarus, 417, 116110, \dodoi{10.1016/j.icarus.2024.116110}

\bibitem[{Nesvorný {et~al.}(2006)Nesvorný, Vokrouhlický, Bottke, \& Sykes}]{Nesvorny:2006}
Nesvorný, D., Vokrouhlický, D., Bottke, W.~F., \& Sykes, M. 2006, Icarus, 181, 107, \dodoi{https://doi.org/10.1016/j.icarus.2005.10.022}

\bibitem[{{Nesvorný} {et~al.}(2017){Nesvorný}, {Vokrouhlický}, {Dones}, {Levison}, {Kaib}, \& {Morbidelli}}]{Nesvorny:2017}
{Nesvorný}, D., {Vokrouhlický}, D., {Dones}, L., {et~al.} 2017, The Astrophysical Journal, 845, 27, \dodoi{10.3847/1538-4357/aa7cf6}

\bibitem[{{O'Brien} \& {Greenberg}(2003)}]{OBrienGreenburg:2003}
{O'Brien}, D.~P., \& {Greenberg}, R. 2003, \icarus, 164, 334, \dodoi{10.1016/S0019-1035(03)00145-3}

\bibitem[{{{\"O}pik}(1951)}]{Opik:1951}
{{\"O}pik}, E.~J. 1951, Pattern Recognition and Image Analysis, 54, 165

\bibitem[{pandas~development team(2025)}]{pandas:2025}
pandas~development team, T. 2025, pandas-dev/pandas: Pandas, v3.0.0rc0,  Zenodo, \dodoi{10.5281/zenodo.17806077}

\bibitem[{{Piani} {et~al.}(2020){Piani}, {Marrocchi}, {Rigaudier}, {Vacher}, {Thomassin}, \& {Marty}}]{Piani:2020}
{Piani}, L., {Marrocchi}, Y., {Rigaudier}, T., {et~al.} 2020, Science, 369, 1110, \dodoi{10.1126/science.aba1948}

\bibitem[{{Pokorn{\'y}} {et~al.}(2024){Pokorn{\'y}}, {Moorhead}, {Kuchner}, {Szalay}, \& {Malaspina}}]{Pokorny:2024}
{Pokorn{\'y}}, P., {Moorhead}, A.~V., {Kuchner}, M.~J., {Szalay}, J.~R., \& {Malaspina}, D.~M. 2024, \psj, 5, 82, \dodoi{10.3847/PSJ/ad2de8}

\bibitem[{{Pokorn{\'y}} \& {Vokrouhlick{\'y}}(2013)}]{PokornyVokrouhlicky2013}
{Pokorn{\'y}}, P., \& {Vokrouhlick{\'y}}, D. 2013, \icarus, 226, 682, \dodoi{10.1016/j.icarus.2013.06.015}

\bibitem[{{Popova} {et~al.}(2013){Popova}, {Jenniskens}, {Emel'yanenko}, {Kartashova}, {Biryukov}, {Khaibrakhmanov}, {Shuvalov}, {Rybnov}, {Dudorov}, {Grokhovsky}, {Badyukov}, {Yin}, {Gural}, {Albers}, {Granvik}, {Evers}, {Kuiper}, {Kharlamov}, {Solovyov}, {Rusakov}, {Korotkiy}, {Serdyuk}, {Korochantsev}, {Larionov}, {Glazachev}, {Mayer}, {Gisler}, {Gladkovsky}, {Wimpenny}, {Sanborn}, {Yamakawa}, {Verosub}, {Rowland}, {Roeske}, {Botto}, {Friedrich}, {Zolensky}, {Le}, {Ross}, {Ziegler}, {Nakamura}, {Ahn}, {Lee}, {Zhou}, {Li}, {Li}, {Liu}, {Tang}, {Hiroi}, {Sears}, {Weinstein}, {Vokhmintsev}, {Ishchenko}, {Schmitt-Kopplin}, {Hertkorn}, {Nagao}, {Haba}, {Komatsu}, {Mikouchi}, \& {aff34}}]{Popova:2013}
{Popova}, O.~P., {Jenniskens}, P., {Emel'yanenko}, V., {et~al.} 2013, Science, 342, 1069, \dodoi{10.1126/science.1242642}

\bibitem[{{Rein} \& {Liu}(2012)}]{REBOUND}
{Rein}, H., \& {Liu}, S.~F. 2012, Astronomy \& Astrophysics, 537, A128

\bibitem[{{Rein} \& {Spiegel}(2015)}]{IAS15integrator}
{Rein}, H., \& {Spiegel}, D.~S. 2015, \mnras, 446, 1424, \dodoi{10.1093/mnras/stu2164}

\bibitem[{{Revelle} \& {Ceplecha}(2001)}]{Revelle&Ceplecha:2001}
{Revelle}, D.~O., \& {Ceplecha}, Z. 2001, in ESA Special Publication, Vol. 495, Meteoroids 2001 Conference, ed. B.~{Warmbein}, 507--512

\bibitem[{Rigley \& Wyatt(2022)}]{Rigley:2022}
Rigley, J.~K., \& Wyatt, M.~C. 2022, Monthly Notices of the Royal Astronomical Society, 510, 834

\bibitem[{{Ryabova} {et~al.}(2019){Ryabova}, {Asher}, \& {Campbell-Brown}}]{Ryabova2019_meteoroidsbook}
{Ryabova}, G.~O., {Asher}, D.~J., \& {Campbell-Brown}, M.~D. 2019, {Meteoroids: Sources of Meteors on Earth and Beyond}

\bibitem[{{Sanchez} {et~al.}(2024){Sanchez}, {Reddy}, {Thirouin}, {Bottke}, {Kareta}, {De Florio}, {Sharkey}, {Battle}, {Cantillo}, \& {Pearson}}]{Sanchez:2024}
{Sanchez}, J.~A., {Reddy}, V., {Thirouin}, A., {et~al.} 2024, \psj, 5, 131, \dodoi{10.3847/PSJ/ad445f}

\bibitem[{{Sansom} {et~al.}(2015){Sansom}, {Bland}, {Paxman}, \& {Towner}}]{Sansom:2015}
{Sansom}, E.~K., {Bland}, P., {Paxman}, J., \& {Towner}, M. 2015, \maps, 50, 1423, \dodoi{10.1111/maps.12478}

\bibitem[{{Sansom} {et~al.}(2019){Sansom}, {Gritsevich}, {Devillepoix}, {Jansen-Sturgeon}, {Shober}, {Bland}, {Towner}, {Cup{\'a}k}, {Howie}, \& {Hartig}}]{Sansom:2019}
{Sansom}, E.~K., {Gritsevich}, M., {Devillepoix}, H. A.~R., {et~al.} 2019, \apj, 885, 115, \dodoi{10.3847/1538-4357/ab4516}

\bibitem[{{Schunov{\'a}} {et~al.}(2014){Schunov{\'a}}, {Jedicke}, {Walsh}, {Granvik}, {Wainscoat}, \& {Haghighipour}}]{Schunova:2014}
{Schunov{\'a}}, E., {Jedicke}, R., {Walsh}, K.~J., {et~al.} 2014, \icarus, 238, 156, \dodoi{10.1016/j.icarus.2014.05.006}

\bibitem[{{Shober}(2025)}]{Shober:2025_Dn_meteorite_NEA_pairs}
{Shober}, P.~M. 2025, \aap, 702, A36, \dodoi{10.1051/0004-6361/202555857}

\bibitem[{{Shober} {et~al.}(2025{\natexlab{a}}){Shober}, {Courtot}, \& {Vaubaillon}}]{Shober:2025_streams}
{Shober}, P.~M., {Courtot}, A., \& {Vaubaillon}, J. 2025{\natexlab{a}}, \aap, 693, A23, \dodoi{10.1051/0004-6361/202452123}

\bibitem[{{Shober} {et~al.}(2020){Shober}, {Jansen-Sturgeon}, {Bland}, {Devillepoix}, {Sansom}, {Towner}, {Cup{\'a}k}, {Howie}, \& {Hartig}}]{Shober:2020}
{Shober}, P.~M., {Jansen-Sturgeon}, T., {Bland}, P.~A., {et~al.} 2020, \mnras, 498, 5240, \dodoi{10.1093/mnras/staa2559}

\bibitem[{Shober {et~al.}(2021)Shober, Sansom, Bland, Devillepoix, Towner, Cup{\'a}k, Howie, Hartig, \& Anderson}]{Shober:2021}
Shober, P.~M., Sansom, E.~K., Bland, P.~A., {et~al.} 2021, The Planetary Science Journal, 2, 98

\bibitem[{{Shober} {et~al.}(2024){Shober}, {Tancredi}, {Vaubaillon}, {Devillepoix}, {Deam}, {Anghel}, {Sansom}, {Colas}, \& {Martino}}]{Shober:2024}
{Shober}, P.~M., {Tancredi}, G., {Vaubaillon}, J., {et~al.} 2024, \aap, 687, A181, \dodoi{10.1051/0004-6361/202449635}

\bibitem[{{Shober} {et~al.}(2025{\natexlab{b}}){Shober}, {Devillepoix}, {Vaubaillon}, {Anghel}, {Deam}, {Sansom}, {Colas}, {Zanda}, {Vernazza}, \& {Bland}}]{Shober:2025}
{Shober}, P.~M., {Devillepoix}, H. A.~R., {Vaubaillon}, J., {et~al.} 2025{\natexlab{b}}, Nature Astronomy, \dodoi{10.1038/s41550-025-02526-6}

\bibitem[{{Shustov} {et~al.}(2017){Shustov}, {Naroenkov}, \& {Efremova}}]{Shustov:2017}
{Shustov}, B.~M., {Naroenkov}, S.~A., \& {Efremova}, E.~V. 2017, Solar System Research, 51, 38, \dodoi{10.1134/S0038094617010038}

\bibitem[{{Silber} {et~al.}(2018){Silber}, {Boslough}, {Hocking}, {Gritsevich}, \& {Whitaker}}]{Silber:2018}
{Silber}, E.~A., {Boslough}, M., {Hocking}, W.~K., {Gritsevich}, M., \& {Whitaker}, R.~W. 2018, Advances in Space Research, 62, 489, \dodoi{10.1016/j.asr.2018.05.010}

\bibitem[{Simion {et~al.}(2021)Simion, Popescu, Licandro, Vaduvescu, de~Le{\'o}n, \& Gherase}]{Simion:2021}
Simion, N., Popescu, M., Licandro, J., {et~al.} 2021, Monthly Notices of the Royal Astronomical Society, 508, 1128

\bibitem[{{Soja} {et~al.}(2019){Soja}, {Gr{\"u}n}, {Strub}, {Sommer}, {Millinger}, {Vaubaillon}, {Alius}, {Camodeca}, {Hein}, {Laskar}, {Gastineau}, {Fienga}, {Schwarzkopf}, {Herzog}, {Gutsche}, {Skuppin}, \& {Srama}}]{Soja:2019}
{Soja}, R.~H., {Gr{\"u}n}, E., {Strub}, P., {et~al.} 2019, \aap, 628, A109, \dodoi{10.1051/0004-6361/201834892}

\bibitem[{{Southworth} \& {Hawkins}(1963)}]{SouthworthHawkins:1963}
{Southworth}, R.~B., \& {Hawkins}, G.~S. 1963, Smithsonian Contributions to Astrophysics, 7, 261

\bibitem[{{Spoto} {et~al.}(2015){Spoto}, {Milani}, \& {Kne{\v{z}}evi{\'c}}}]{Spoto:2015}
{Spoto}, F., {Milani}, A., \& {Kne{\v{z}}evi{\'c}}, Z. 2015, \icarus, 257, 275, \dodoi{10.1016/j.icarus.2015.04.041}

\bibitem[{{Spurn{\'y}} {et~al.}(2024){Spurn{\'y}}, {Borovi{\v{c}}ka}, {Shrben{\'y}}, {Hankey}, \& {Neubert}}]{Spurny:2024_2024BX1}
{Spurn{\'y}}, P., {Borovi{\v{c}}ka}, J., {Shrben{\'y}}, L., {Hankey}, M., \& {Neubert}, R. 2024, \aap, 686, A67, \dodoi{10.1051/0004-6361/202449735}

\bibitem[{Svetsov {et~al.}(2019)Svetsov, Shuvalov, Collins, \& Popova}]{Svetsov:2019}
Svetsov, V., Shuvalov, V., Collins, G., \& Popova, O. 2019, in Meteoroids (Cambridge University Press), 275--298

\bibitem[{Tancredi(1995)}]{Tancredi:1995}
Tancredi, G. 1995, Astronomy and Astrophysics, v. 299, p. 288, 299, 288

\bibitem[{{Toliou} {et~al.}(2021){Toliou}, {Granvik}, \& {Tsirvoulis}}]{Toliou:2021}
{Toliou}, A., {Granvik}, M., \& {Tsirvoulis}, G. 2021, \mnras, 506, 3301, \dodoi{10.1093/mnras/stab1934}

\bibitem[{{Towner} {et~al.}(2020){Towner}, {Cupak}, {Deshayes}, {Howie}, {Hartig}, {Paxman}, {Sansom}, {Devillepoix}, {Jansen-Sturgeon}, \& {Bland}}]{Towner:2020}
{Towner}, M.~C., {Cupak}, M., {Deshayes}, J., {et~al.} 2020, \pasa, 37, e008, \dodoi{10.1017/pasa.2019.48}

\bibitem[{{Usui} {et~al.}(2013){Usui}, {Kasuga}, {Hasegawa}, {Ishiguro}, {Kuroda}, {M{\"u}ller}, {Ootsubo}, \& {Matsuhara}}]{Usui:2013}
{Usui}, F., {Kasuga}, T., {Hasegawa}, S., {et~al.} 2013, \apj, 762, 56, \dodoi{10.1088/0004-637X/762/1/56}

\bibitem[{{Valsecchi} {et~al.}(1999){Valsecchi}, {Jopek}, \& {Froeschle}}]{Valsecchi:1999}
{Valsecchi}, G.~B., {Jopek}, T.~J., \& {Froeschle}, C. 1999, \mnras, 304, 743, \dodoi{10.1046/j.1365-8711.1999.02264.x}

\bibitem[{{Vida} {et~al.}(2021){Vida}, {{\v{S}}egon}, {Gural}, {Brown}, {McIntyre}, {Dijkema}, {Pavleti{\'c}}, {Kuki{\'c}}, {Mazur}, {Eschman}, {Roggemans}, {Merlak}, \& {Zubovi{\'c}}}]{Vida:2021}
{Vida}, D., {{\v{S}}egon}, D., {Gural}, P.~S., {et~al.} 2021, \mnras, 506, 5046, \dodoi{10.1093/mnras/stab2008}

\bibitem[{Virtanen {et~al.}(2020)Virtanen, Gommers, Oliphant, Haberland, Reddy, Cournapeau, Burovski, Peterson, Weckesser, Bright, {van der Walt}, Brett, Wilson, Millman, Mayorov, Nelson, Jones, Kern, Larson, Carey, Polat, Feng, Moore, {VanderPlas}, Laxalde, Perktold, Cimrman, Henriksen, Quintero, Harris, Archibald, Ribeiro, Pedregosa, {van Mulbregt}, \& {SciPy 1.0 Contributors}}]{scipy}
Virtanen, P., Gommers, R., Oliphant, T.~E., {et~al.} 2020, Nature Methods, 17, 261, \dodoi{10.1038/s41592-019-0686-2}

\bibitem[{{Voj{\'a}{\v{c}}ek} {et~al.}(2015){Voj{\'a}{\v{c}}ek}, {Borovi{\v{c}}ka}, {Koten}, {Spurn{\'y}}, \& {{\v{S}}tork}}]{Vojavek:2015}
{Voj{\'a}{\v{c}}ek}, V., {Borovi{\v{c}}ka}, J., {Koten}, P., {Spurn{\'y}}, P., \& {{\v{S}}tork}, R. 2015, \aap, 580, A67, \dodoi{10.1051/0004-6361/201425047}

\bibitem[{{Vokrouhlick{\'y}} {et~al.}(2012){Vokrouhlick{\'y}}, {Pokorn{\'y}}, \& {Nesvorn{\'y}}}]{VokrouhlickyPokornyNesvorny2012}
{Vokrouhlick{\'y}}, D., {Pokorn{\'y}}, P., \& {Nesvorn{\'y}}, D. 2012, \icarus, 219, 150, \dodoi{10.1016/j.icarus.2012.02.021}

\bibitem[{{Wetherill}(1985)}]{Wetherill:1985}
{Wetherill}, G.~W. 1985, Meteoritics, 20, 1, \dodoi{10.1111/j.1945-5100.1985.tb00842.x}

\bibitem[{{Wiegert} {et~al.}(2020){Wiegert}, {Brown}, {Pokorn{\'y}}, {Ye}, {Gregg}, {Lenartowicz}, {Krzeminski}, \& {Clark}}]{Wiegert:2020}
{Wiegert}, P., {Brown}, P., {Pokorn{\'y}}, P., {et~al.} 2020, \aj, 159, 143, \dodoi{10.3847/1538-3881/ab700d}

\bibitem[{Williams(1992)}]{Williams:1992}
Williams, J.~G. 1992, Icarus, 96, 251, \dodoi{https://doi.org/10.1016/0019-1035(92)90079-M}

\bibitem[{{Wisdom}(2020)}]{Wisdom:2020}
{Wisdom}, J. 2020, \maps, 55, 766, \dodoi{10.1111/maps.13463}

\bibitem[{{Wi{\'s}niowski} \& {Rickman}(2013)}]{Wisniowski2013}
{Wi{\'s}niowski}, T., \& {Rickman}, H. 2013, \actaa, 63, 293

\end{thebibliography}
\bibliographystyle{aasjournal}

\listofchanges

\end{document}